\documentclass[a4paper,11pt]{article}
\usepackage{jheppub}
\usepackage{amsmath,amssymb,amsfonts,amsthm}
\usepackage{graphicx}
\usepackage{float}
\usepackage{tikz-cd}

\usepackage[]{enumitem}
\setlist[enumerate]{labelindent=\parindent,leftmargin=*,topsep=0.4ex,itemsep=0.1ex}
\setlist[itemize]{labelindent=\parindent,leftmargin=*,topsep=0.4ex,itemsep=-1ex,partopsep=1ex,parsep=1ex}
\setlist[enumerate,1]{labelindent=\parindent, leftmargin=*,label=\textup{(\arabic*)},ref=\textup{\arabic*}}

\theoremstyle{plain}
\newtheorem*{statement}{Statement}
\usepackage{caption,subcaption}
\usepackage{tikz}
\title{Mixed Hodge structure and  $\mathcal{N}=2$ Coulomb branch solution}

\author[a,b]{Dan Xie}
\author[a]{Dingxin Zhang}

\affiliation[a]{Yau Mathematics Sciences center, Tsinghua University, Beijing, 100084, China}
\affiliation[b]{Department of Mathematics, Tsinghua University, Beijing, 100084, China}
\abstract{%
  The Coulomb branch of four dimensional $\mathcal{N}=2$ theories can be solved
  by finding a Seiberg--Witten (SW) geometry and a SW differential.
  While lots of SW geometries are found, the extraction of low energy theory out
  of it is limited due to following reasons:
  (a) the difficulty of distinguishing electric-magnetic and flavor charges;
  (b) the difficulty of determining the low energy theory at singular point,
  (c) the lack of SW differential.
  We show that the mixed Hodge structure (MHS) can be used to fully solve the
  low energy physics of Coulomb branch at every vacuum.
  The MH{S} can be used to solve above three problems as
  follows:
  (a) The smooth fiber of SW geometry carries a Mixed Hodge Structure
  structure with two weights: one weight describes electric-magnetic charge and
  the other for the flavor charge;
  (b) for the singular fiber, there are three MHS which can used to determine the IR
  theory.
  (c) the MHS at $\infty$ is used to find the SW differential.
}

\begin{document}
\maketitle

\flushbottom
%%%%%%%%%%%%%%%%%%%%%%%%%%%%%%%%%%%%%%%%%%%%%%%%%%%%%%%%%%%

\section{Introduction}

The low energy effective theory at the Coulomb branch%
\footnote{%
  It is called Coulomb branch because the low energy theory of generic vacua on
  this branch is described by abelian gauge theory. }%
of a four dimensional $\mathcal{N}=2$ $SU(2)$ gauge theory was solved by
Seiberg--Witten \cite{Seiberg:1994rs,Seiberg:1994aj}.
The solution was found by using a number of deep physical insights:
electric-magnetic duality, holomorphy, and the existence of
singularities on the moduli space where extra massless BPS particles appear.

In the theory of Seiberg and Witten,
the physical information is encoded in an elegant way: they
introduced a family of algebraic curves (the Seiberg--Witten (SW) curve)
$f(x,y,u, \Lambda)=0$ defined over the Coulomb branch moduli space, and
specified a family of differential forms $\lambda_{SW}$ (the SW differential).
From these geometric objects they were able to extract various interesting
physical properties:
\begin{enumerate}
\item the low energy photon couplings are found from
  periods of the SW differential;
\item the central charges of massive BPS particles are also given by the periods
  of $\lambda_{SW}$;
\item the singularities on the moduli space are identified with the degenerating
  curve in the SW family.
\end{enumerate}

Therefore, the Coulomb branch solution of a 4d $\mathcal{N}=2$ theory is given by SW curve and  SW differential,
and many solutions are henceforth found for simple gauge
theories~\cite{Argyres:1994xh,Klemm:1994qs, Argyres:1995wt, Hanany:1995na} via
this method. Nevertheless, the original method~\cite{Seiberg:1994rs,Seiberg:1994aj} is difficult
to implement in a more general situation.
Instead, a number of solutions were found with the help of other tools. For
example by string theory
constructions~\cite{Kachru:1995fv,Witten:1997sc,Katz:1997eq,Gaiotto:2009we,Xie:2012hs,Xie:2015rpa},
and the relation to integrable system~\cite{Donagi:1995cf, Martinec:1995by}.
The construction of Seiberg and Witten may be generalized since one can try to
define higher dimensional algebraic varieties fibered over the Coulomb branch
moduli space. We call it the associated \textbf{SW geometry}. 

While there are a
number of SW geometries available, it is not straightforward to obtain physical
information form them. Let us explain what are the obstacles.

To begin with, recall that the Coulomb branch vacua are classified by their low
energy behavior: a \emph{generic vacuum} is a vacuum whose the low energy theory
is an abelian gauge theory. If a vacuum is not generic, it is called
\emph{special}.

\medskip\noindent%
\textbf{Generic vacua.}
Given a generic vacuum with Coulomb branch modulus $t$, we
have an \(n\)-dimensional smooth algebraic variety from the SW geometry (we
shall consider the cases when \(n=1\) or \(n=3\)). Its
singular cohomology group $H^n(X_t,\mathbb{C})$%
\footnote{If the SW geometries are curves, the only non-vanishing reduced
  cohomology group is $H^1$; If the SW geometries are three-folds, the only
  non-vanishing reduced cohomology groups are  $H^3$. We can also take
  the coefficient of the cohomology group to be $\mathbb{Z}$ or $\mathbb{Q}$,
  which are useful for the electric-magnetic pairing.}
is the crucial data for us in order to extract physical information.
We face following challenges:
  \begin{enumerate}
  \item The central charges should be given by the periods of the SW
    differential $\lambda_{SW}$. If a theory has global symmetry, then there
    would be periods for \textbf{flavor} central charges. How can we
    distinguish those corresponding to global symmetries from the one that give
    rise the electric-magnetic charges?

  \item The constraint of unitarity requires that there is a family of
    \textbf{abelian varieties} fibered over the Coulomb branch moduli space.
    However, in general, the SW geometries are non-compact algebraic varieties,
    and $H^n(X_t,\mathbb{C})$ does not give an abelian variety.
    How can we find these abelian varieties?
  \item Finally, the Coulomb branch solution carries a
    \textbf{special K\"ahler geometry}. But defining the SKG requires the
    presence of the SW differential \(\lambda_{SW}\). However how to determine
    the SW differential is rather difficult, even if the SW geometry is given.
  \end{enumerate}

\medskip\noindent%
\textbf{Special vacua.}
A special vacuum is a place on the moduli space where the SW geometry
$X_0$ degenerates%
\footnote{A special vacuum does not necessarily means that the algebraic
  variety given by the SW geometry
  is singular, we will describe exactly what this means in section \ref{sec:3}.}.
The low energy theory at these vacua are much more complicated. It could be a
superconformal field theory, an IR free non-abelian gauge theory, or an
interacting theory coupled with an abelian gauge theory. From the SW geometry
we could obtain the cohomology groups $H^n(X_0, \mathbb{C})$,
and a monodromy action on the cohomology groups of fibers near \(X_0\). These
data, however, are not enough to determine the IR theory of the vacuum.

Special vacua are the most interesting and important vacua.
Some of the special vacua of pure $SU(N)$ gauge theory give the
confining vacua of the resulting $\mathcal{N}=1$ theory formed by soft-breaking
of $\mathcal{N}=2$ supersymmetry \cite{Seiberg:1994rs}.

\medskip\noindent%
\textbf{UV theory.}
When we are given just a SW geometry, how do we determine the properties of UV theory from it?

\medskip%
The purpose of this paper is to propose that the \textbf{mixed Hodge structure}
(MHS) can be used to solve above problems.
The reader should probably be familiar with the notion of a Hodge structure,
which is modeled on the Hodge decomposition of a nonsingular projective
variety. By definition, a Hodge structure of weight \(n\) consists of a finite
rank free abelian group
\(H_{\mathbb{Z}}\) together with a decomposition:
\begin{equation*}
H_{\mathbb{C}} := H_{\mathbb{Z}} \otimes \mathbb{C} = \bigoplus_{p+q=n} H^{p,q},
\end{equation*}
such that the complex conjugation of  \(H^{p,q}\) equals \(\overline{H^{q,p}}\).

A Hodge structure of weight \(n\) can be reformulated as a pair
$(H_{\mathbb{Z}}, F^{\bullet})$, where $F^{\bullet}$ is a decreasing
filtration
\begin{equation*}
H_{\mathbb{C}} = F^0\supset F^1\supset \cdots \supset F^{n-1}\supset F^n
\end{equation*}
satisfying the condition $H_{\mathbb{C}}=F^p \oplus \overline{F^{n-p+1}}$ for
every $p$. In this formulation, the \(H^{p,q}\) are recovered by
$H^{p,q}= F^p \cap \overline{F^q}$.

A mixed Hodge structure is a generalization of a HS and was defined by
Deligne~\cite{deligne:hodge2,deligne:hodge3}.
It consists of a triple
$H=(H_{\mathbb{Z}}, F^{\bullet},W_{\bullet})$, where
\begin{itemize}
\item \(H_{\mathbb{Z}}\) is a finite rank free abelian group,
\item $F^{\bullet}$ is an increasing
  filtration on \(H_{\mathbb{C}}\) called Hodge filtration, and
\item  $W_{\bullet}$ is a decreasing filtration on
  \(H_{\mathbb{Q}}=H_{\mathbb{Z}}\otimes \mathbb{Q}\) called the
  weight filtration.
\end{itemize}
They are required to satisfy the following property:
the induced filtration of \(F^{\bullet}\) on the quotient space
$\mathrm{Gr}^{W}_{k}H_{\mathbb{Q}}=W_k/W_{k-1}$ defines a weight $k$ Hodge
structure. In particular
\(\mathrm{Gr}^{W}_{k}H_{\mathbb{Q}} \otimes\mathbb{C}\) has a Hodge
decomposition:
\begin{equation*}
  \mathrm{Gr}^{W}_{k}H_{\mathbb{Q}} \otimes\mathbb{C}
  = \bigoplus H^{p,k-p}.
\end{equation*}

\medskip%
We now explain how the MHS can help us overcome the difficulties listed above.

\medskip\noindent%
\textbf{MHS and generic vacua.}
For a generic vacuum described by a smooth algebraic variety $X_t$, its
cohomology group $H^n(X_t,\mathbb{Q})$ carries a MHS as shown by
Deligne~\cite{deligne:hodge2}.
For the SW geometries studied in the literature, this MH{S} is quite special.
It consists of only weight $n$ and weight $n+1$ part, and the Hodge
decomposition has the following form
\begin{equation*}
  \mathrm{Gr}^{W}_n H \otimes \mathbb{C} =H^{{n+1\over 2},{n-1\over  2}}\oplus H^{{n-1\over2},{n+1\over2}},
  \quad \mathrm{Gr}^{W}_{n+1}H \otimes \mathbb{C}=H^{{n+1\over 2},{n+1\over 2}}.
\end{equation*}
This special MHS can help us solve the previous problems for the generic vacua.
\begin{enumerate}
\item[(a)] The
  weight $n$ part of the cohomology gives the electric-magnetic charges, and
  weight $n+1$ part gives the flavor charge.
\item[(b)] The Hodge structure of weight $n$ part gives an abelian variety. The
  consistent condition is then
  $h^{{n+1\over 2},{n-1\over 2}}=r, h^{{n+1\over 2},{n+1\over 2}}=f$,
  here $r$ is the rank of the theory, and $f$ is the rank of the flavor symmetry
  group.
\item[(c)] We will explain in \S\ref{sec:7} how to find a SW differential from
  the MHS of the SW geometry.
\end{enumerate}

\medskip\noindent%
\textit{Example.}
Let us consider a  geometry  $X: x^3+y^3+1=0$,
which is a generic vacuum of an $(A_2, A_2)$ theory.
The dimension of the cohomology group $H^1(X,\mathbb{Q})$ is 4.
Its MHS has is of the form
\begin{equation*}
  \mathrm{Gr}^{W}_1 H=H^{1,0}\oplus H^{0,1},\quad \mathrm{Gr}_2^{W}H =H^{1,1},
\end{equation*}
The Hodge numbers are $h^{1,0}=h^{0,1}=1$, and $h^{1,1}=2$.
The $(A_2, A_2)$ theory has a rank one Coulomb branch, and a rank two flavor
symmetry group, agreeing the {M}HS computation.

\medskip\noindent%
\textbf{MHS and special vacua.}
Consider a special vaccum corresponding to a variety \(X_0\) given by the SW
geometry. In general the variety \(X_0\) is singular (or has ``hidden
singularities'' as will be explained in \S\ref{sec:4}).
According to Deligne~[loc.~cit.],
its cohomology group $H^n(X_0, \mathbb{Q})$ also carries a MH{S}.
This MH{S} alone is not sufficient to describe the physics at the special
vacuum. In Hodge theory, one needs two extra M{H}S in order to fully
describe the degeneration, they are called
\begin{itemize}
\item the \textbf{nearby cycle}, and
  % It is defined as a limit mixed Hodge structure: the Hodge filtration comes
  % from the limit of the Hodge filtration of the cohomology group of the fibers
  % near \(X_0\), and the weight filtration comes from the monodromy operator.
\item the \textbf{vanishing cycle}.
\end{itemize}
The cohomology of \(X_0\), the nearby cycle, and the vanishing cycle form an
exact sequence of {M}HS. It turns out we can extract the low energy physics from
these M{H}S in a very satisfying manner:
\begin{enumerate}[wide]
\item [(a)] The interacting theory is described by the vanishing cycle, and the
  abelian gauge theory is described by the cohomology group of $X_0$. These two
  parts are coupled together through the exact sequence with the nearby cycle.

\item [(b)] Due to non-trivial monodromy action on vanishing cycle and nearby
  cycle, the M{H}S on the vanishing cycle and nearby cycle have finer
  structures, so that we can get a set of rational numbers called
  \textbf{spectral numbers} from these MH{S}. These numbers can give us
  important information of the Coulomb branch spectrum of the interacting
  theory.
\end{enumerate}

\medskip\noindent%
\textit{Example.} Consider the geometry $X_0: x^3+y^4+y^3=0$, which is a special
vacuum for the $(A_2, A_3)$ theory. $X_0$ has a singularity at $x=y=0$, which is
equivalent to the singularity of that of $x^3+y^3=0$. The MHS for the singular
fiber has the form $H^{1}(X_0,\mathbb{C})=H^{0,0}$ with $h^{0,0}=2$. The MHS for
the vanishing cycle can be computed from the local singularity $x^3+y^3=0$, and
its Hodge numbers are $h^{1,0}=h^{0,1}=1,h^{1,1}=2$.

\begin{figure}[H]
\begin{center}

\tikzset{every picture/.style={line width=0.75pt}} %set default line width to 0.75pt

\begin{tikzpicture}[x=0.55pt,y=0.55pt,yscale=-1,xscale=1]
%uncomment if require: \path (0,479); %set diagram left start at 0, and has height of 479

%Straight Lines [id:da006387412741975851]
\draw    (316,174) -- (341.5,156) ;
%Straight Lines [id:da5779941314917529]
\draw    (316,185) -- (339.5,198) ;
%Shape: Circle [id:dp26478439059933345]
\draw   (340.5,150) .. controls (340.5,142.82) and (346.32,137) .. (353.5,137) .. controls (360.68,137) and (366.5,142.82) .. (366.5,150) .. controls (366.5,157.18) and (360.68,163) .. (353.5,163) .. controls (346.32,163) and (340.5,157.18) .. (340.5,150) -- cycle ;
%Shape: Circle [id:dp8459623487277086]
\draw   (339.5,198) .. controls (339.5,190.89) and (345.26,185.13) .. (352.38,185.13) .. controls (359.49,185.13) and (365.25,190.89) .. (365.25,198) .. controls (365.25,205.11) and (359.49,210.88) .. (352.38,210.88) .. controls (345.26,210.88) and (339.5,205.11) .. (339.5,198) -- cycle ;
%Flowchart: Or [id:dp7452006775594375]
%\draw   (378,174.5) .. controls (378,170.36) and (381.25,167) .. (385.25,167) .. controls (389.25,167) and (392.5,170.36) .. (392.5,174.5) .. controls (392.5,178.64) and (389.25,182) .. (385.25,182) .. controls (381.25,182) and (378,178.64) .. (378,174.5) -- cycle ; \draw   (378,174.5) -- (392.5,174.5) ; \draw   (385.25,167) -- (385.25,182) ;

% Text Node
\draw (302,169) node [anchor=north west][inner sep=0.75pt]   [align=left] {T};
% Text Node
\draw (347,140) node [anchor=north west][inner sep=0.75pt]   [align=left] {1};
% Text Node
\draw (347,188) node [anchor=north west][inner sep=0.75pt]   [align=left] {1};
% Text Node
%\draw (414,163.4) node [anchor=north west][inner sep=0.75pt]    {$U( 1)^{3}$};

\end{tikzpicture}
\end{center}
\caption{The low energy theory for the vacua described by $X_0: x^3+y^4+y^3=0$. Here $T$ is the $(A_2, A_2)$ SCFT.}
\label{intro}
\end{figure}
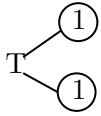

The physical interpretation of the Hodge numbers are: each Hodge number
$h^{1,1}$ of the vanishing cycles gives the mass parameter, and it pairs with
the $h^{0,0}$ part of the singular variety; physically it means that the $U(1)$
flavor groups of the interacting theory is gauged. So the low energy theory is
given in Figure~\ref{intro}.

\medskip\noindent%
\textbf{MHS and UV theory.}
To extract the information of the UV theory, it is natural to take the Coulomb
branch parameter to be $\infty$. An important type of deformation takes the form
$f(x,y, \lambda_i)=t$, here $t$ has the largest scaling dimension, and
$\lambda_i$ are other parameters. The UV limit is recovered by taking $t$
to infinity. In this case, one can also define a limit MHS when $t\to \infty$,
from which one can extract information of the UV theory. The MHS only
depends the UV theory and is independent of the parameters $\lambda_i$ in the
geometry.

\medskip\noindent%
\textit{Example.}
Consider the SW geometry $x^3+y^3+ a x +b  y+ c xy=t$.
This is the generic deformations for the $(A_2, A_2)$ theory.
The MHS for the $t=\infty$  limit (with arbitrary $a,b,c$) is the same as the
MHS for the singularity $x^3+y^3=0$, which gives the correct UV theory.

\medskip\noindent%
\textbf{SW differential.}
The SW differential plays a pivotal role in defining the
special K\"ahler geometry on the Coulomb branch moduli space. It turns out that
one can indeed define a special section on the Coulomb branch (primitive form,
see the discussion in \cite{Li:2018rdd}), and this special section can be identified with
the Seiberg-Witten differential.
The MHS in the $t\rightarrow \infty$ limit will be crucial for finding this
section.

\medskip\noindent%
The great advantage of the approach advocated in this paper is that many
computations are algebraic, and can be carried out explicitly. This is because
for each 4d $\mathcal{N}=2$ theory, one can attach a combinatorial object called
Newton polyhedron and various {M}HSs can be computed from the combinatorial data
of the polyhedron.

This paper is organized as follows.
Section~\ref{sec:2} reviews the basic structure of Coulomb branch solution of 4d
$\mathcal{N}=2$ theories.
Section~\ref{sec:3} describes how to use MH{S}s to study generic vacua.
Section~\ref{sec:4} describes how to find MHS to study special vacua.
Section~\ref{sec:5} describe the MHS associated with UV theory.
While previous section focuses on one particular slice of the Coulomb branch,
Section~\ref{sec:6} describes MHS on the general Coulomb branch space.
Section~\ref{sec:7} describes how to find SW differential from the MHS of the UV
theory. Finally a conclusion is given in section~\ref{sec:8}.

\section{Coulomb branch of 4d $\mathcal{N}=2$ theories}
\label{sec:2}

\subsection{$\mathcal{N}=2$ theory and its Coulomb branch solution}

\medskip\noindent%
\textbf{Basics of $\mathcal{N}=2$ theory.}
We review some basic facts of 4d $\mathcal{N}=2$ supersymmetric field theories,
for more details, see \cite{bagger1990supersymmetry}.

The 4d $\mathcal{N}=2$ superalgebra consists of the generators $M_{\mu\nu}$ for
four dimensional Lorentz group, eight supercharges $Q_{\alpha}^A,$, \(A=1,2\) and
$J_i$ generating a $SU(2)_R$ symmetry. It is also possible to have generators
for other global symmetries $G_F$. The representation theory of the
4d $\mathcal{N}=2$ superalgebra can be found
in~\cite{bagger1990supersymmetry}.

One can construct 4d $\mathcal{N}=2$ supersymmetric field theory using the
superspace method. These models have Lagrangians and admit gauge theory
description. Examples are $\mathcal{N}=2$ SQCD with $SU(N)$ gauge group coupled
with $N_f$ fundamental hypermultiplets (We take $N_f\leq 2N$ so that the theory
is asymptotically free).

4d $\mathcal{N}=2$ theories have two class of half-BPS operators: Coulomb branch
operators $\mathcal{O}$ (singlet under $SU(2)_R$ symmetry), and Higgs branch
operator $\widehat{\mathcal{B}}_R$ which transforms under $SU(2)_R$ symmetries ($R$
denotes the representation type under $SU(2)_R$ symmetry). These operators are
important as they determine the $\mathcal{N}=2$ supersymmetry preserving
deformation of the theory. The deformations of these theories are (we consider
$\mathcal{N}=2$ theories which can be defined for arbitrarily high energy, this
includes the asymptotically free gauge theory and conformal field theory)
classified as follows.
\begin{enumerate}[wide]
\item Relevant or marginal deformation using Coulomb branch operators
  $\mathcal{O}$:
  \begin{equation*}
    \delta S= \lambda \int \mathrm{d}^{4}x\;\mathrm{d}^4\theta\; \mathcal{O}+c.c
  \end{equation*}
  This type of deformation exists only for a subset of Coulomb branch operators%
  \footnote{In the UV limit, we can assign a scaling dimension to $\mathcal{O}$,
    and we require $[\mathcal{O}]\leq 2$ so that the deformation is relevant or
    marginal.} .
\item If our theory has global symmetries, then the theory consists of Higgs
  branch operators $\widehat{\mathcal{B}}_1$, and the following relevant mass
  deformation is also possible:
  \begin{equation*}
    \delta S= m\int \mathrm{d}^4x \; \mathrm{d}^2 \theta \; \widehat{{\mathcal{B}}}_1+c.c
  \end{equation*}
\item We can have Coulomb branch deformation by turning on expectation values of
  Coulomb branch operators $ u_i =\langle \mathcal{O}_i\rangle$. The number of
  Coulomb branch operators is called the \textbf{rank} of the theory. The
  Coulomb branch is not lifted if we turn on deformations of first and second
  kind, but the low energy physics is typically changed. So the Coulomb branch
  is parameterized by $u_i, \lambda, m$.

\item We can have Higgs branch deformation by turning on expectation values of
  Higgs branch operators $\widehat{\mathcal{B}}_R$. These vacua would be lifted
  by turning on deformations of first and second kind.
\item It is also possible to have mixed branch where we turn on expectation
  values of both Coulomb branch and Higgs branch operators.
\end{enumerate}

\begin{figure}[H]
\begin{center}
\includegraphics[width=.9\linewidth]{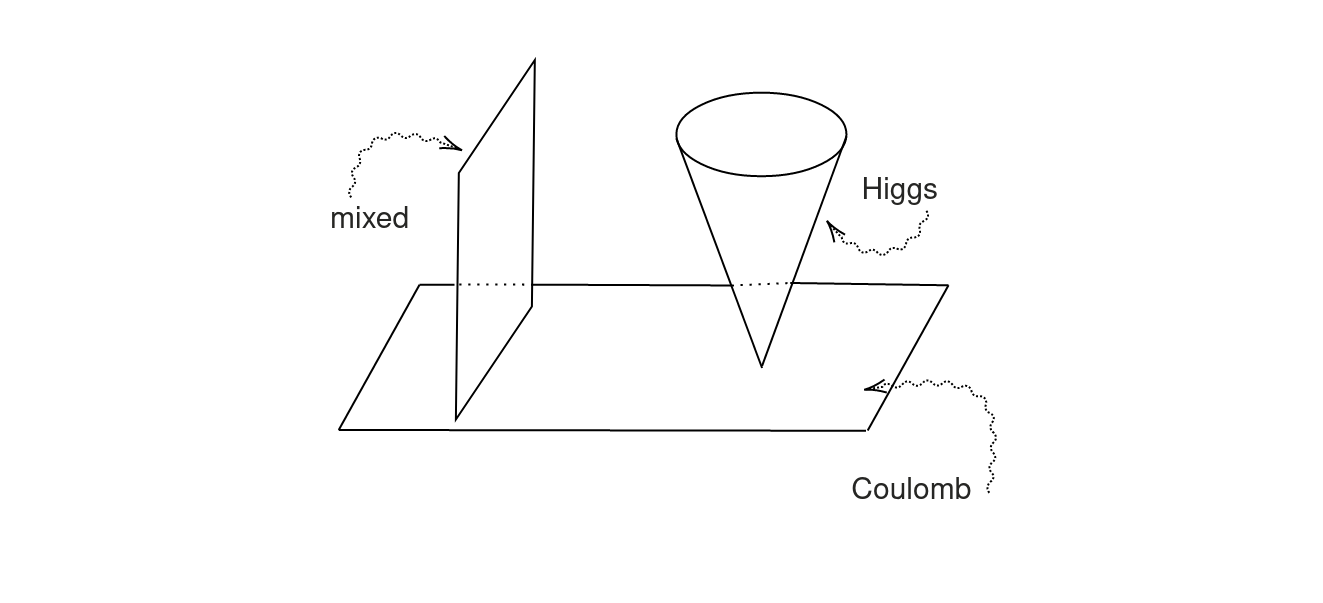}
\end{center}
\caption{The structure of moduli space of 4d $\mathcal{N}=2$ theories. Here Coulomb branch is parameterized by expectation value of Coulomb branch operator ${\cal O}$; the Higgs branch is parameterized by expectation value of Higgs branch
operator $\hat{{\cal B}}_R$; the mixed branch is parameterized by both the Coulomb and Higgs branch operators.}
\label{vacua}
\end{figure}

The deformations listed in 3, 4, 5 are called moduli space of vacua of
$\mathcal{N}=2$ theories, and the typical structure of the moduli space is shown
in Figure~\ref{vacua}.

\medskip\noindent%
\textbf{Superconformal field theory.}
Some $\mathcal{N}=2$ theories have larger symmetry group called $\mathcal{N}=2$
superconformal symmetry. These theories are called superconformal field theories (SCFT),
and play fundamental roles in studying $\mathcal{N}=2$ theories. This is because
general $\mathcal{N}=2$ theories are defined as turning on relevant deformation
of SCFT. SFCTs have an extra $U(1)_R$ symmetry (besides the $SU(2)_R$
symmetry for a general $\mathcal{N}=2$ theory). The bosonic symmetry group of a
general $\mathcal{N}=2$ SCFT is $SO(2,4)\times SU(2)_R \times U(1)_R \times G_F$,
here $SO(2,4)$ is the conformal group, $SU(2)_R\times U(1)_R$ is the $R$
symmetry group which exist for every $\mathcal{N}=2$ SCFT, and $G_F$ are other
global symmetry groups which could be absent for some theories.

The representation theory for 4d $\mathcal{N}=2$ superconformal algebra is
studied in \cite{Dolan:2002zh}. A highest weight representation is labeled as
$|\Delta,R,r, j_1, j_2\rangle$, here $\Delta$ is the scaling dimension, $R$ is
$SU(2)_R$ charge, $r$ is $U(1)_R$ charge, $j_1$ and $j_2$ are left and right
spin. These states could also carry quantum numbers of flavor symmetry group
$G_F$. Representation theory of $\mathcal{N}=2$ superalgebra has been studied in
\cite{Dolan:2002zh}, and its short representations are completely classified in
\cite{Dolan:2002zh}. Three short (Half-BPS) representations that we are
interested in are Coulomb branch operators, Higgs branch operators, and
supercurrent multiplet:
\begin{equation*}
\begin{split}
&\text{Coulomb~branch operators}:~~~~{\cal E}_{r,(0,0)},~~R=0,~~\Delta=r, \nonumber\\
&\text{Higgs~branch operators}:~~~~~~~~\hat{{\cal B}}_R,~~~~~~~r=j_1=j_2=0,~~\Delta=2R, \nonumber\\
&\text{Supercurrent}:~~~~~~~~~~~~~~~~~~~~~\hat{{\cal C}}_{0,(0,0)},~~r=R=0,~~\Delta=2. \nonumber\\
\end{split}
\end{equation*}
$\widehat{\mathcal{B}}_1$ is a multiplet which contains conserved current for
the flavor symmetry group $G_F$, and transforms in adjoint representation of
$G_F$. Here we list the quantum numbers for the bottom component of the
multiplet. ${\cal E}_{r,(0,0)}$ play the same role as the Coulomb branch
operator ${\cal O}$ for the general $\mathcal{N}=2$ theory, here it just carries
an extra $U(1)_R$ quantum number. and $\hat{{\cal B}}_R$ is just the usual Higgs
branch operator we discussed for a general $\mathcal{N}=2$ theory.

For most of theories, Coulomb branch operators form a freely generated ring.
Determining the Coulomb branch chiral ring is equivalent to determining the set
of $U(1)_R$ charges $(r_1,\ldots, r_s)$ (unitarity bound requires these numbers
to be larger than 1). The Higgs branch operators also form a ring, which is in
general quite complicated, see \cite{Xie:2019vzr} for recent development.

An important question for SCFT is to determine its Half-BPS spectrum.
Supercurrent multiplet exists for every $\mathcal{N}=2$ SCFT. There are also a
large class of $\mathcal{N}=2$ SCFTs whose full Coulomb branch spectrum and
Higgs branch spectrum are known
\cite{Gaiotto:2009we,Xie:2012hs,Wang:2015mra,Wang:2018gvb,Xie:2015rpa}.

One can also define central charges $a_{\mathcal{N}=2}$ and $c_{\mathcal{N}=2}$
for a 4d conformal field theories. The central charges $a_{\mathcal{N}=2}$ and
$c_{\mathcal{N}=2}$ are related to the anomalies of $R$ symmetries as follows
\cite{Shapere:2008zf}:
\begin{equation*}
\begin{split}
&\text{Tr}(R_{\mathcal{N}=2}^3)=6(a_{\mathcal{N}=2}-c_{\mathcal{N}=2}),~~\text{Tr}(R_{\mathcal{N}=2})=24(a_{\mathcal{N}=2}-c_{\mathcal{N}=2}), \\
&\text{Tr}(R_{\mathcal{N}=2}R^2_{SU(2)})=(2a_{\mathcal{N}=2}-c_{\mathcal{N}=2}).
\end{split}
\end{equation*}
Here $R_{\mathcal{N}=2}$ is the generator of $U(1)_R$ symmetry, and $R_{SU(2)}$
is the Cartan subalgebra of $SU(2)_R$ algebra. Central charges are computed for
a large class of SCFTs in \cite{Xie:2015rpa}.

Examples of 4d $\mathcal{N}=2$ SCFT are $SU(N)$ gauge theory with $N_f=2N$
fundamental matter, which has conventional Lagrangian description. There are
also non-Lagrangian $\mathcal{N}=2$ SCFTs such as $T_N$ theory
\cite{Gaiotto:2009we}, and Argyres-Douglas theories \cite{Argyres:1995jj}.

\begin{figure}[H]
\begin{center}
\includegraphics[width=.9\linewidth]{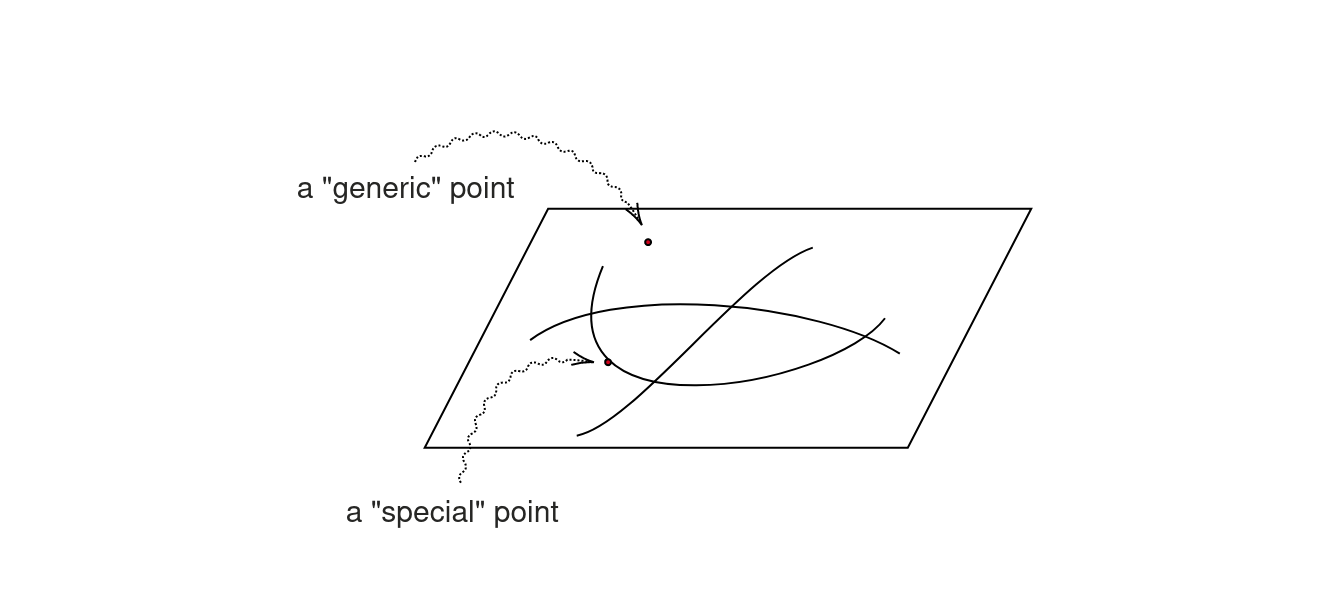}
\end{center}
\caption{The Coulomb branch is separated into generic points and special points: the low energy physics at a generic point is described by a $U(1)^r$ abelian gauge theory; there are extra massless particles at the special point and the low energy physics would be much more complicated.}
\label{coulombbranch}
\end{figure}

\medskip\noindent%
\textbf{Coulomb branch solution.}
In this paper, we focus on the structure of Coulomb branch, which is defined as
the moduli space parameterized by Coulomb branch operators. Since the Coulomb
branch is not lifted by the relevant/marginal and mass deformation, we also
consider turning on these deformations. So we have a family of Coulomb branch
parameterized by the relevant/marginal and mass deformations. One of the basic
question of understanding Coulomb branch is to determine the low energy physics
at every point of the moduli space. The Coulomb branch has following structure
(see Figure~\ref{coulombbranch}):
\begin{enumerate}[wide]
\item The low energy theory at a \textbf{generic} vacuum of Coulomb branch is
  described by an abelian $U(1)^r$gauge theory%
  \footnote{$r$ denotes the number of Coulomb branch operators ${\cal O}_i$, and
    is also called the ``rank'' of the theory.}.
  The important thing is to determine the low energy effective action, which has
  a rigid special Kähler structure (RSK).
  
  At a generic point, there are stable massive BPS particles carrying
  electric-magnetic charge, and flavor symmetry charge. If the rank of the
  theory is $r$ and the rank of the flavor symmetry is $f$, then the dimension
  of the charge lattice of BPS particle is:
  \begin{equation*}
    \mu=2r+f.
  \end{equation*}
  The central charge of a BPS particles with charge $(n_i,n_i^D,n_f)$ is given
  by
  \begin{equation*}
    Z=n_i a^i+n_i^D a_D^i+n_f m_f.
  \end{equation*}
  here $(a^i, a_D^i, m_f)$ are holomorphic functions on the moduli space. An
  important question is to determine those functions so that the central charge
  of BPS particles can be computed, and the mass of the BPS particle is given by
  $M=|Z|$.

\item At special points of Coulomb branch, besides the abelian gauge theory, new
  massless degrees of freedoms appear. The low energy theory would consist of
  several parts: (1) abelian gauge theory; (2) IR free gauge theories: abelian
  or non-abelian gauge theory coupled with enough matter; (3) superconformal
  field theory. In general, these parts are coupled together. The extra massless
  degree of freedoms come from massive BPS particles. The physical question is
  to determine the low energy theory and their physical properties at these
  special points.
\end{enumerate}

It is very difficult to answer above physical questions of Coulomb branch by
using conventional physical tools, as the study often involves complicated
instanton computations in the weakly coupled region or strongly coupled dynamics
of gauge theory. The remarkable discovery of Seiberg and Witten is to encode the
low energy physics into a family of curves fibered over the Coulomb branch, and
such a curve is called the Seiberg-Witten (SW) curve. Seiberg-Witten curve is
denoted as $f(x,y, u,\lambda, m)=0$, and various physical data are read from it
as follows.
\begin{itemize}
\item At generic point of Coulomb branch, the SW curve is smooth and the low
  energy photon couplings are identified with the complex structure of the
  curve. There is a SW differential $\lambda_{SW}$ whose periods of integral over
  cycles of the curve give the central charge of the BPS particles. The
  derivative of these periods along the Coulomb branch coordinate would give the
  complex structure of the curve, and thereby determines the low energy physics.
\item The special point of the Coulomb branch is identified with the
  degenerating curve. The nature of the low energy physics at these special
  points are often complicated and has to be determined carefully, see
  \cite{Seiberg:1994rs,Seiberg:1994aj}.
\end{itemize}

\begin{figure}[H]
\begin{center}
\includegraphics[width=.9\linewidth]{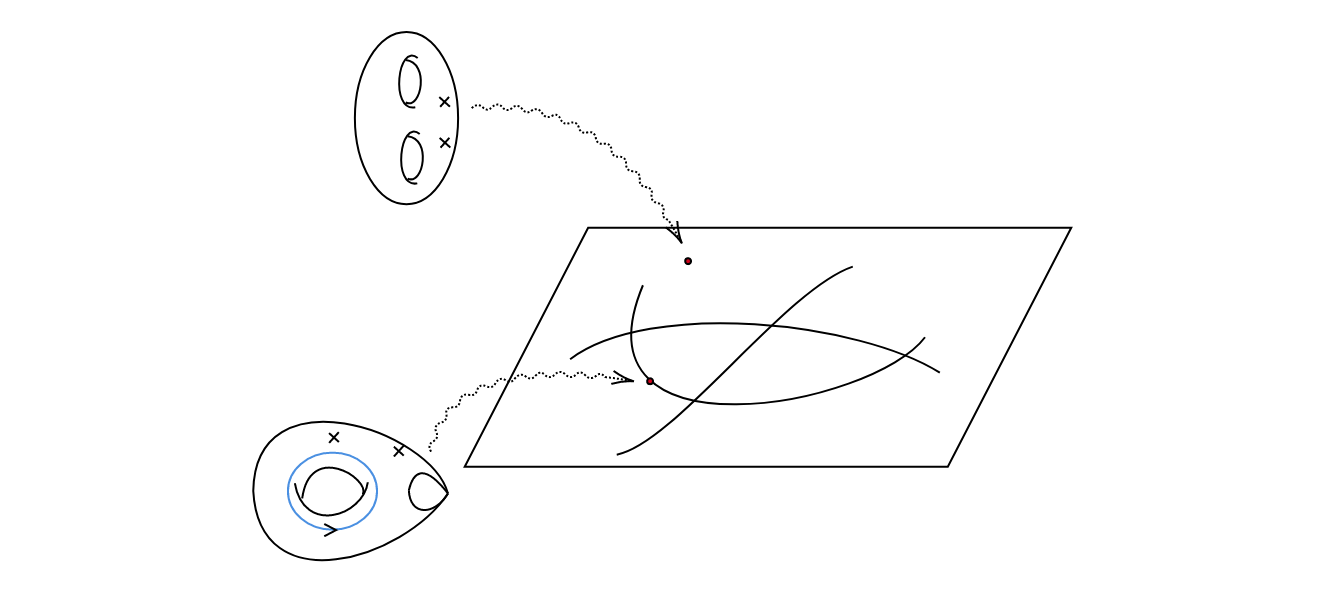}
\end{center}
\caption{There is an auxiliary curve (SW curve) at each point of Coulomb branch. At a generic point, the SW curve is smooth; At a special point, the curve degenerates and often develops singularity.}
\label{SWgeometry}
\end{figure}

See Figure~\ref{SWgeometry} for the illustration of SW curves fibered over the
Coulomb branch. The SW curve is found based on various deep physical inputs such
as holomorphy of superpotential, BPS particles, electric-magnetic duality, etc.

\begin{figure}[H]
  \begin{center}

\tikzset{every picture/.style={line width=0.75pt}} %set default line width to 0.75pt

\begin{tikzpicture}[x=0.45pt,y=0.45pt,yscale=-1,xscale=1]
%uncomment if require: \path (0,300); %set diagram left start at 0, and has height of 300

%Shape: Circle [id:dp035160857567985815]
\draw   (250,133.92) .. controls (250,89.19) and (286.26,52.92) .. (331,52.92) .. controls (375.74,52.92) and (412,89.19) .. (412,133.92) .. controls (412,178.66) and (375.74,214.92) .. (331,214.92) .. controls (286.26,214.92) and (250,178.66) .. (250,133.92) -- cycle ;
%Flowchart: Summing Junction [id:dp3601057476513363]
\draw   (365,139.96) .. controls (365,136.67) and (367.46,134) .. (370.5,134) .. controls (373.54,134) and (376,136.67) .. (376,139.96) .. controls (376,143.25) and (373.54,145.92) .. (370.5,145.92) .. controls (367.46,145.92) and (365,143.25) .. (365,139.96) -- cycle ; \draw   (366.61,135.75) -- (374.39,144.18) ; \draw   (374.39,135.75) -- (366.61,144.18) ;
%Flowchart: Summing Junction [id:dp15543420506038474]
\draw   (295,139.96) .. controls (295,136.67) and (297.46,134) .. (300.5,134) .. controls (303.54,134) and (306,136.67) .. (306,139.96) .. controls (306,143.25) and (303.54,145.92) .. (300.5,145.92) .. controls (297.46,145.92) and (295,143.25) .. (295,139.96) -- cycle ; \draw   (296.61,135.75) -- (304.39,144.18) ; \draw   (304.39,135.75) -- (296.61,144.18) ;
%Flowchart: Summing Junction [id:dp5485018232954428]
\draw   (325,72.96) .. controls (325,69.67) and (327.46,67) .. (330.5,67) .. controls (333.54,67) and (336,69.67) .. (336,72.96) .. controls (336,76.25) and (333.54,78.92) .. (330.5,78.92) .. controls (327.46,78.92) and (325,76.25) .. (325,72.96) -- cycle ; \draw   (326.61,68.75) -- (334.39,77.18) ; \draw   (334.39,68.75) -- (326.61,77.18) ;

% Text Node
\draw (437,133) node [xslant=-0.06] [align=left] {$u$};
% Text Node
\draw (290,159) node   {$\Lambda $};
% Text Node
\draw (380,159) node   {$-\Lambda $};
% Text Node
\draw (352,72) node   {$\infty $};

\end{tikzpicture}
\end{center}
\caption{Coulomb branch moduli space for 4d pure $SU(2)$ SYM theory. Here we use $u$ to parameterize Coulomb branch. At generic point of moduli space, the low energy theory is an abelian $U(1)$ gauge theory.
There are three singularities: $u=\pm \Lambda$ where there is one extra massless hypermultiplet, and $u=\infty$ where the theory is the non-abelian $SU(2)$ gauge theory.}
\label{su2}
\end{figure}
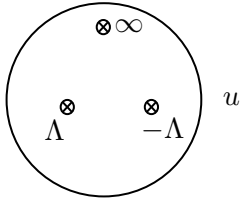

Let's now describe the SW solution of the pure $SU(2)$ gauge theory due to
Seiberg and Witten \cite{Seiberg:1994rs,Seiberg:1994aj}. The solution is given
by the family of curves:
\begin{equation*}
f=x+{\Lambda^2 \over x}+y^2+2u=0;
\end{equation*}
Here $\Lambda$ is the dynamical generated scale, and $u$ is the Coulomb branch
moduli. $x$ is a $\mathbb{C}^*$ variable and $y$ is a $\mathbb{C}$ variable%
\footnote{This is not the original curve given in \cite{Seiberg:1994rs}. It is
  not difficult to prove that the curve given here is equivalent to that of
  \cite{Seiberg:1994rs}.}. Various physical quantities are computed from the
curve as follows:
\begin{itemize}
\item At a point $u$ where the corresponding curve is smooth, the photon
  coupling is identified with the complex structure of the curve. This
  automatically captures the electric-magnetic duality of the low energy $U(1)$
  theory.
\item The curve degenerates at three points $u=\Lambda, -\Lambda, \infty$, see
  Figure~\ref{su2}. The physics at $\infty$ is the non-abelian $SU(2)$ gauge
  theory. At $u=\pm \Lambda$, there is an extra massless hypermultiplet which is
  due to BPS particle becoming massless. The low energy theory is a $U(1)$ gauge
  theory coupled with a hypermultiplet.
\item One has a SW differential $\lambda_{SW}=y {\mathrm{d}x\over x}$, which
  gives us the SKG required by $\mathcal{N}=2$ supersymmetry. Moreover, the
  periods of integral of $\lambda_{SW}$ gives us the central charges for BPS
  particles: the two periods are $a=\int_A \lambda$, \(a_D=\int_B \lambda\)
  (here $A$ and $B$ are two cyles on the curve), and $Z=na+n_D a^D$ for a BPS
  particle with charge $(n,n_D)$.
\end{itemize}

\subsection{Seiberg-Witten geometry and Newton polyhedron}

As we discussed in last subsection, the Coulomb branch solution of a 4d
$\mathcal{N}=2$ theory is encoded by a SW curve fibered over the Coulomb branch%
\footnote{It was found in \cite{Katz:1997eq} that one need to consider higher
  dimensional variety for more general $\mathcal{N}=2$ theories.}, together with
a Seiberg-Witten differential. While this approach of encoding Coulomb branch is
very powerful, there is no systematical way to find SW solution given a 4d
$\mathcal{N}=2$ theory. Usually, one can find SW solution by using the the
connection to integrable systems \cite{Donagi:1995cf}, and string theory
construction \cite{Kachru:1995fv,Witten:1997sc}. The two most powerful
approaches are the Hitchin system
\cite{Nanopoulos:2009uw,Xie:2012hs,Wang:2015mra,Wang:2018gvb} and complex
singularities in one dimension or three dimension
\cite{Chen:2016bzh,Xie:2015rpa,Wang:2016yha, Chen:2017wkw}.

Most of the Seiberg-Witten geometry found in the literature takes the following
form: it is extracted from a function defined on a smooth affine variety
$f:U\rightarrow \mathbb{C}$, $U$ being $\mathbb{C}^2$, $\mathbb{C}^*\times \mathbb{C}$,
$\mathbb{C}^4$, or $\mathbb{C}^*\times \mathbb{C}^3$.%
\footnote{A $\mathbb{C}^*$ variable can be thought of as removing the origin of
  the complex plane. It is an affine variety as its coordinate ring is given by
  $\mathbb{C}[x,y]/(xy-1)$. If we solve $y$ using the defining ideal $xy=1$, the
  coordinate ring of $\mathbb{C}^*$ is then isomorphic to $\mathbb{C}[x,x^{-1}]$, which
  is the ring of Laurent polynomial of variable $x$.}
If the function has a $\mathbb{C}^*$-action with positive  weights, it is
believed that the geometry should describe a SCFT. Here we list several class of
examples (the corresponding 4d theories are SCFTs) that we would like to study
in more detail.

\begin{itemize}[wide]
\item Take a polynomial $f:\mathbb{C}^4\rightarrow \mathbb{C}$ with an
  isolated critical point at origin.
  We require that \(f\) has a $\mathbb{C}^*$ action satisfying
  \begin{equation*}
    f(\lambda^{q_i} x_i)=\lambda f(z_i),\quad \sum q_i>1.
  \end{equation*}
  This type of polynomial has been studied in \cite{Xie:2015rpa}.
  The corresponding 4d theory is an Argyes-Douglas type theory.
  Many properties of those theories have been studied in \cite{Xie:2016evu, Xie:2017vaf}.

\item Take $f: \mathbb{C}^* \times \mathbb{C}^3 \rightarrow \mathbb{C}$
  of following form
  \begin{equation*}
    f=f_{ADE}(x_1, x_2,x_3)+zg(x_1,x_2,x_3, z);
  \end{equation*}
  Here, $f_{ADE}(x_1, x_2,x_3)$ is a polynomial defining a standard two
  dimensional ADE singularity, and $z$ is a $\mathbb{C}^*$ variable; the
  polynomial $g$ is taken so that the polynomial has a $\mathbb{C}^*$ action.
  The corresponding 4d theory can have non-abelian global symmetries. This class
  of theory can be engineered using the 6d $(2,0)$ theory
  \cite{Xie:2012hs,Wang:2015mra,Wang:2018gvb}, and is important to understand
  $S$ duality of Argyres-Douglas theories.

\item Take $f: \mathbb{C}^*\times \mathbb{C} \rightarrow \mathbb{C}$ of the  form
  \begin{equation*}
    f=v^Ng(z)
  \end{equation*}
  Here $z$ is a $\mathbb{C}^*$ variable, and $v$ is a $\mathbb{C}$ variable.
  $g(z)$ is a degree $(n+1)$ polynomial with distinct roots. This type of theory
  can be engineered by using the 6d $(2,0)$ theory on a sphere with regular
  singularities \cite{Gaiotto:2009we}. The physical theories and the above curve
  were found using the M-theory construction \cite{Witten:1997sc}
\end{itemize}

One can also engineer non-conformal theory which is not the deformation of above
quasi-homogeneous polynomials, and some of the examples are

\begin{itemize}[wide]
\item $f=v^{n_1}z^{s+1}+v^{n_2} z^{s}+\ldots+v^{n_{s-1}}z+v^{n_s}$. Here $z$ is
  a $\mathbb{C}^*$ variable, and $v$ is a $\mathbb{C}$ variable. To ensure that
  the 4d theory is UV complete, the constraint on $n_i$ is following
  \begin{equation*}
    2n_i<n_{i-1}+n_{i+1},\quad i=2,\ldots,n_{s-1}
  \end{equation*}
This class of curves  describe the Coulomb branch solution of linear quiver of $SU$ type, and is derived using brane construction \cite{Witten:1997sc}.

\item \(f=f_{ADE}(x_1,x_2,x_3)+z^a+{1\over z^b}\). This curve describes the
  Coulomb branch solution of a single ADE gauge theory coupled with
  Argyres-Douglas matters \cite{Xie:2012hs,Cecotti:2012jx}. Here $z$ is a
  $\mathbb{C}^*$ variable.
\end{itemize}

\medskip\noindent%
\textbf{Newton polyhedron.}
Some properties of most of the SW geometries listed above can be described by
the so-called Newton polyhedron. Here we give a review about how to
associate a Newton polyhedron to a polynomial function.

Let $f: (\mathbb{C}^{*})^n\times \mathbb{C}^m\rightarrow \mathbb{C}$
be given by a Laurent polynomial
\begin{equation*}
  \label{eq:laurent}
  f = \sum \prod_{i=1}^{n+m}x_i^{\omega_i}.
\end{equation*}
For a $\mathbb{C}$ variable in a monomial of above sum, its exponent must be
non-negative. There is no such constraint for $\mathbb{C}^*$ variables. For each monomial
\(\prod_{i}x_i^{\omega_i}\) appearing in $f$, one draws a dot with coordinate
\((\omega_1,\ldots,\omega_{n+m})\) in $\mathbb{R}^{n+m}$; the Newton polyhedron
\(\Delta(f)\) (at infinity) is the convex hull of the origin and these dots.

\begin{figure}[H]
\begin{center}
\includegraphics[width=.9\linewidth]{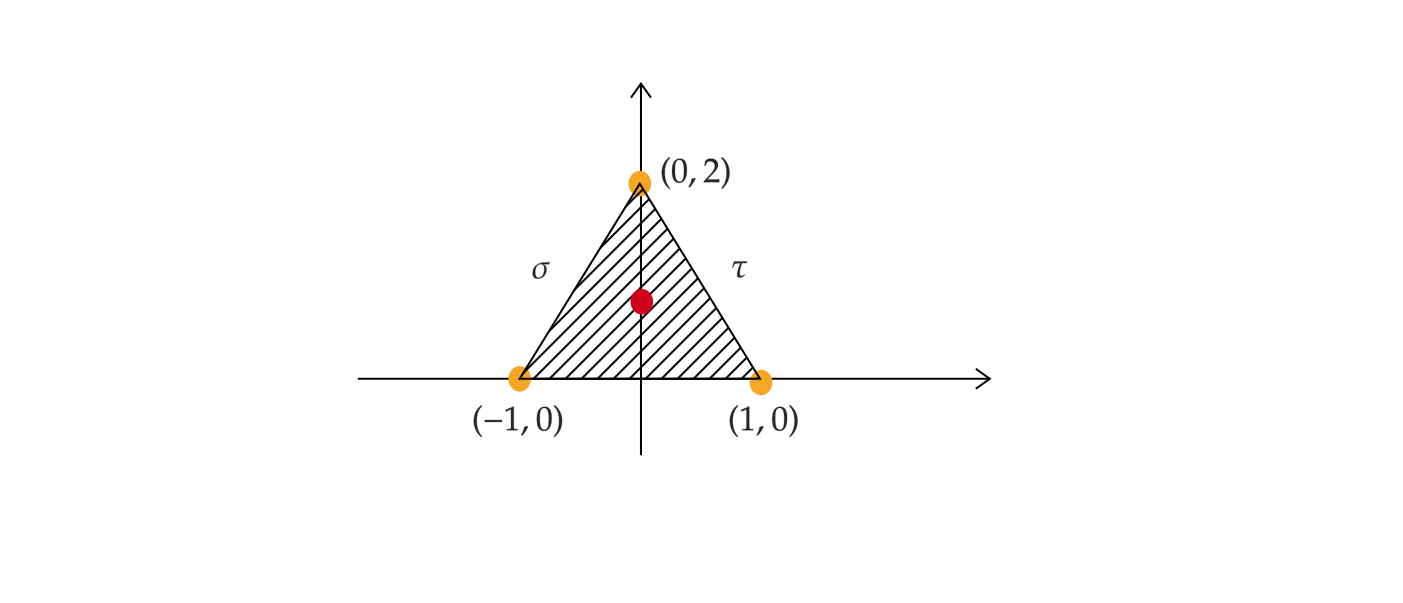}
\end{center}
\caption{The Newton polyhedron for polynomial \(f(x_1,x_2)=x_2^{2}-(x_{1}+x_{1}^{-1})\). Here $x_1$ is a $\mathbb{C}^*$ variable, and $x_2$ is a $\mathbb{C}$ variable.}
\label{newton1}
\end{figure}

For example, the Newton polyhedron of \(f(x_1,x_2)=x_2^{2}-(x_{1}+x_{1}^{-1})\)
is given by Figure~\ref{newton1}.

We say \(f\) is \textbf{non-degenerate} with respect to \(\Delta(f)\) if for every face
\(\sigma\) of \(\Delta(f)\) \textbf{not containing} the origin, the equations
\begin{equation*}
\frac{\partial f_\sigma}{\partial x_1}=\frac{\partial f_\sigma}{\partial x_2}=\ldots=\frac{\partial f_\sigma}{\partial x_{n+m}}=0
\end{equation*}
have no common zeros in \((\mathbb{C}^{\ast})^{n+m}\).
Here
$f_\sigma = \sum_{(\omega_1,\ldots,\omega_{n+m}) \in \sigma}\prod_{i=1}^{n+m}x_i^{\omega_i}$
is the Laurent polynomial defined using the lattice points on the face $\sigma$.

If \(f(x_1,x_2)=x_2^{2}-(x_{1}+x_{1}^{-1})\), there are two faces \(\sigma\) and
\(\tau\) which do not contain the origin, as indicated in Figure~\ref{newton1}.
We have \(f_{\sigma}(x_1,x_2)=x_2^{2} - x_1^{-1}\) and
\(f_{\tau}=x_{2}^{2}-x_{1}\). A simple computation shows that this Laurent
polynomial is nondegenerate with respect to its Newton polyhedron.

\begin{figure}[H]
 \begin{center}
\includegraphics[width=.9\linewidth]{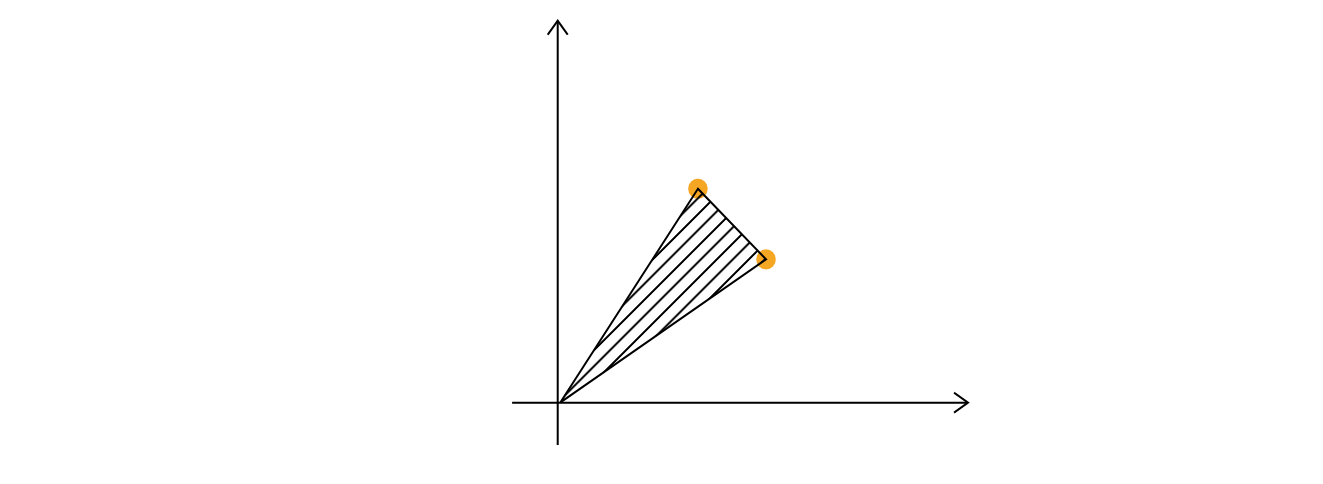}
\end{center}
\caption{The Newton polyhedron of a non-convenient polynomial \(f(x_1,x_2)=1+ x_{1}x_{2}^{2} + x_{1}^{2}x_{2}\).}
\label{convenient}
\end{figure}

Now let us define the concept of a convenient polynomial. If all the variables
in defining $f$ are $\mathbb{C}$ variables, \(f\) is called convenient if and
only if \(f\) contains a term of the form \(ax^{j_i}_i\) for all \(i\), namely
its lattice points intersect with every coordinate axis. For the $\mathbb{C}^*$
variable, the convenient condition is that origin is in the interior (we set
$\mathbb{C}$ variable to be zero). For example, the Laurent polynomial
\(f(x_1,x_2)=x_2^{2}-(x_{1}+x_{1}^{-1})\) is convenient. On the other hand,
\(f(x_1,x_2)=1+ x_{1}x_{2}^{2} + x_{1}^{2}x_{2}\) is not convenient, see
Figure~\ref{convenient}.

\begin{figure}[H]
\begin{center}

 \begin{center}
\includegraphics[width=.9\linewidth]{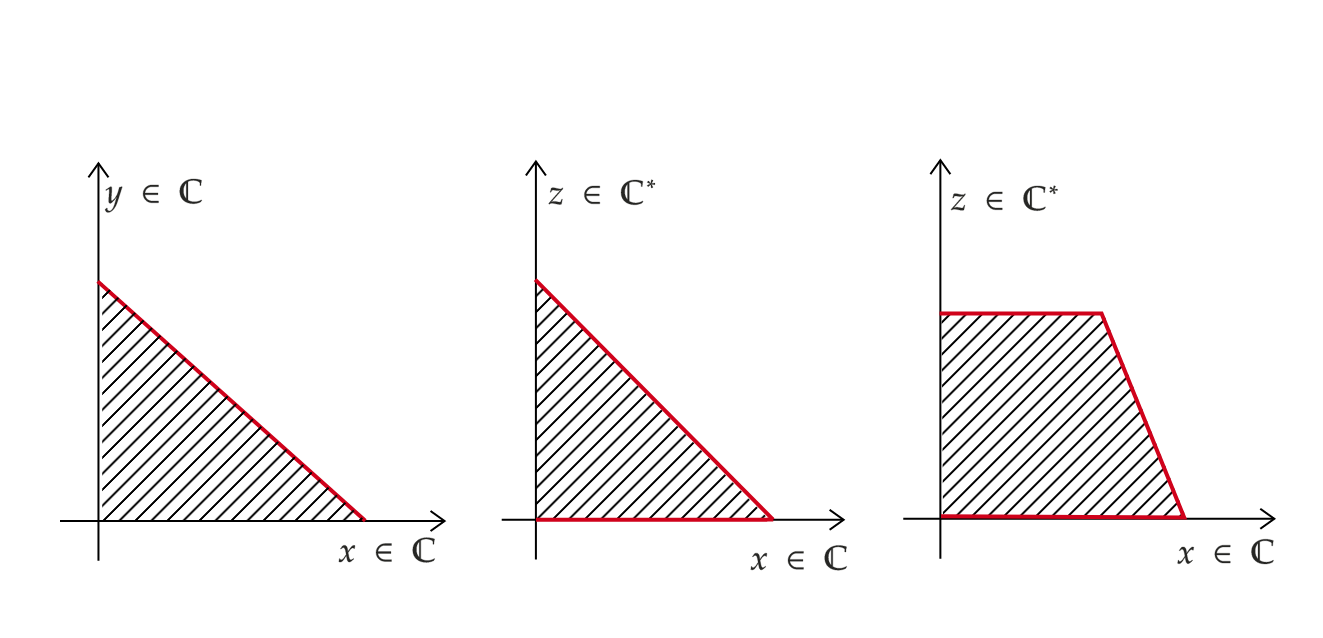}
\end{center}

\caption{The Newton polyhedron for three polynomial functions. Left: Here $f=x^N+y^k$ and both $x$ and $y$ are  $\mathbb{C}$ variable; Middle: Here $f=x^N+z^k$, and $x$ is a $\mathbb{C}$ variable and $z$ is a $\mathbb{C}^*$ variable; Right: Here $f=x^N+x^{n_1} z^b+z^b$, and $x$ is a $\mathbb{C}$ variable and $z$ is a $\mathbb{C}^*$ variable.}
\label{newton2}
\end{center}
\end{figure}

As we will show later, Newton non-degenerate and convenient polynomial have many
nice properties. Some more examples are shown in Figure~\ref{newton2}, and the
properties for the underlying physical theories are:

\begin{enumerate}[wide]
\item The theories described by polynomial $f=x^N+y^k$ are the
  $(A_{N-1}, A_{k-1})$ theories \cite{Cecotti:2010fi}. Here $x$ and $y$ are both
  $\mathbb{C}$ variables. Such theories can be engineered by the 6d $A_{N-1}$ $(2,0)$
  theory on a sphere with one irregular singularity \cite{Xie:2012hs}. Some
  physical data of these theories are: (a) The dimension of charge lattice is
  $\mu=2r+f=(N-1)(k-1)$, here $r$ is the dimension of Coulomb branch and $f$ is
  the number of mass parameters; (b) the number of mass parameters $f$ is given
  by the interior lattice points on the red boundary.

\item The theories described by the polynomial $f=x^N+z^k$, where $x$ is a
  $\mathbb{C}$ variable and $z$ is a $\mathbb{C}^*$ variable, have a $SU(N)$
  flavor symmetry, and can be engineered by the 6d $A_{N-1}$ $(2,0)$ theory on a
  sphere with one irregular singularity and one full type regular singularity
  \cite{Xie:2012hs}. The physical data is summarized as follows: (a) The
  dimension of charge lattice is $\mu=2r+f=(N-1)(k-1)$; (b) the number of mass
  parameters $f$ is given by the interior lattice points on the red boundary.
  In the example described in the figure, we have two red boundaries.

\item The theories described by polynomial $f=x^N+x^{n_1} z^b$, where $x$ is a
  $\mathbb{C}$ variable and $z$ is a $\mathbb{C}^*$ variable, have
  $SU(N)\times U(n_1)$ flavor symmetry, which are found in \cite{Xie:2017vaf}
  using a 6d $(2,0)$ construction. This polynomial is not Newton non-degenerate,
  but the physical spectrum found in \cite{Xie:2017vaf} suggests that we should
  consider Newton polyhedron shown in Figure~\ref{newton2} instead.
\end{enumerate}

\begin{figure}[H]
\begin{center}

\tikzset{every picture/.style={line width=0.75pt}} %set default line width to 0.75pt

\begin{tikzpicture}[x=0.55pt,y=0.55pt,yscale=-1,xscale=1]
%uncomment if require: \path (0,328); %set diagram left start at 0, and has height of 328

%Shape: Axis 2D [id:dp49140336823971564]
\draw  (65,159.27) -- (232,159.27)(80,39.27) -- (80,232.27) (225,154.27) -- (232,159.27) -- (225,164.27) (75,46.27) -- (80,39.27) -- (85,46.27)  ;
%Straight Lines [id:da7564914506205269]
\draw [color={rgb, 255:red, 208; green, 2; blue, 27 }  ,draw opacity=1 ]   (80,60) -- (140,158.27) ;
%Straight Lines [id:da8860231292888918]
\draw [color={rgb, 255:red, 208; green, 2; blue, 27 }  ,draw opacity=1 ]   (140,160) -- (80,220) ;
%Shape: Circle [id:dp24058576584847624]
\draw  [color={rgb, 255:red, 0; green, 0; blue, 0 }  ,draw opacity=1 ][fill={rgb, 255:red, 0; green, 0; blue, 0 }  ,fill opacity=1 ] (77,63) .. controls (77,61.34) and (78.34,60) .. (80,60) .. controls (81.66,60) and (83,61.34) .. (83,63) .. controls (83,64.66) and (81.66,66) .. (80,66) .. controls (78.34,66) and (77,64.66) .. (77,63) -- cycle ;
%Shape: Circle [id:dp7248812947325078]
\draw  [color={rgb, 255:red, 0; green, 0; blue, 0 }  ,draw opacity=1 ][fill={rgb, 255:red, 0; green, 0; blue, 0 }  ,fill opacity=1 ] (137,158.27) .. controls (137,156.61) and (138.34,155.27) .. (140,155.27) .. controls (141.66,155.27) and (143,156.61) .. (143,158.27) .. controls (143,159.92) and (141.66,161.27) .. (140,161.27) .. controls (138.34,161.27) and (137,159.92) .. (137,158.27) -- cycle ;
%Shape: Circle [id:dp7718938591856386]
\draw  [color={rgb, 255:red, 0; green, 0; blue, 0 }  ,draw opacity=1 ][fill={rgb, 255:red, 0; green, 0; blue, 0 }  ,fill opacity=1 ] (77,220) .. controls (77,218.34) and (78.34,217) .. (80,217) .. controls (81.66,217) and (83,218.34) .. (83,220) .. controls (83,221.66) and (81.66,223) .. (80,223) .. controls (78.34,223) and (77,221.66) .. (77,220) -- cycle ;
%Shape: Axis 2D [id:dp5464908397341728]
\draw  (325,160.27) -- (492,160.27)(341,40.27) -- (341,260) (485,155.27) -- (492,160.27) -- (485,165.27) (336,47.27) -- (341,40.27) -- (346,47.27)  ;
%Straight Lines [id:da9660519049799738]
\draw [color={rgb, 255:red, 208; green, 2; blue, 27 }  ,draw opacity=1 ]   (380,100) -- (422,160.27) ;
%Straight Lines [id:da6375982610341171]
\draw [color={rgb, 255:red, 208; green, 2; blue, 27 }  ,draw opacity=1 ]   (422,160.27) -- (340,240) ;
%Shape: Circle [id:dp17363112793408764]
\draw  [color={rgb, 255:red, 0; green, 0; blue, 0 }  ,draw opacity=1 ][fill={rgb, 255:red, 0; green, 0; blue, 0 }  ,fill opacity=1 ] (377,100) .. controls (377,98.34) and (378.34,97) .. (380,97) .. controls (381.66,97) and (383,98.34) .. (383,100) .. controls (383,101.66) and (381.66,103) .. (380,103) .. controls (378.34,103) and (377,101.66) .. (377,100) -- cycle ;
%Shape: Circle [id:dp5718035833766268]
\draw  [color={rgb, 255:red, 0; green, 0; blue, 0 }  ,draw opacity=1 ][fill={rgb, 255:red, 0; green, 0; blue, 0 }  ,fill opacity=1 ] (419,160.27) .. controls (419,158.61) and (420.34,157.27) .. (422,157.27) .. controls (423.66,157.27) and (425,158.61) .. (425,160.27) .. controls (425,161.92) and (423.66,163.27) .. (422,163.27) .. controls (420.34,163.27) and (419,161.92) .. (419,160.27) -- cycle ;
%Shape: Circle [id:dp27739381467419877]
\draw  [color={rgb, 255:red, 0; green, 0; blue, 0 }  ,draw opacity=1 ][fill={rgb, 255:red, 0; green, 0; blue, 0 }  ,fill opacity=1 ] (337,240) .. controls (337,238.34) and (338.34,237) .. (340,237) .. controls (341.66,237) and (343,238.34) .. (343,240) .. controls (343,241.66) and (341.66,243) .. (340,243) .. controls (338.34,243) and (337,241.66) .. (337,240) -- cycle ;
%Straight Lines [id:da9938026908352835]
\draw [color={rgb, 255:red, 208; green, 2; blue, 27 }  ,draw opacity=1 ]   (340,100) -- (380,100) ;
%Shape: Circle [id:dp8318774597955962]
\draw  [color={rgb, 255:red, 0; green, 0; blue, 0 }  ,draw opacity=1 ][fill={rgb, 255:red, 0; green, 0; blue, 0 }  ,fill opacity=1 ] (337,100) .. controls (337,98.34) and (338.34,97) .. (340,97) .. controls (341.66,97) and (343,98.34) .. (343,100) .. controls (343,101.66) and (341.66,103) .. (340,103) .. controls (338.34,103) and (337,101.66) .. (337,100) -- cycle ;

% Text Node
\draw (244,150.4) node [anchor=north west][inner sep=0.75pt]    {$x$};
% Text Node
\draw (58,20.4) node [anchor=north west][inner sep=0.75pt]    {$z$};
% Text Node
\draw (504,151.4) node [anchor=north west][inner sep=0.75pt]    {$x$};
% Text Node
\draw (318,21.4) node [anchor=north west][inner sep=0.75pt]    {$z$};

\end{tikzpicture}

\end{center}
\caption{The Newton polyhedron for the Argyres-Douglas matter. It is convenient and non-degenerate if we turn on generic deformation.}
\label{newpoly}
\end{figure}
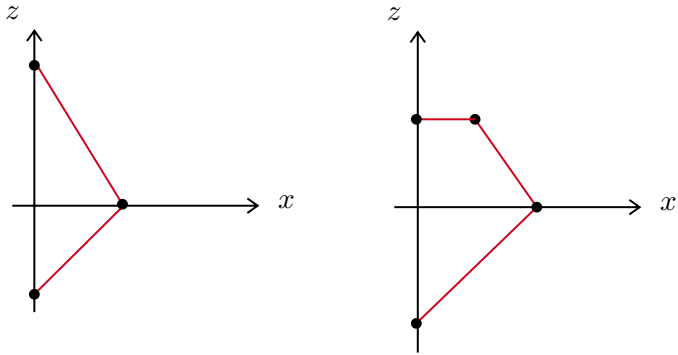

The Newton polyhedrons of the second and third example are not convenient. In
practice, we often perform a coordinate transformation of the form $x=x^{'}z$,
and consider a different Laurent polynomial (take case 2 as an example):
\begin{equation*}
f=x^{'N}+z^{k-N}+{a^N\over z^N}
\end{equation*}
Notice that here we always need to turn on a deformation parameterized by $a$,
so that the polynomial is convenient. The New polyhedron for the case two and
three are shown in figure. \ref{newpoly}.

\medskip\noindent%
\textbf{Constraints on polynomials.}
What kind of polynomials would give us SW geometry of a four dimensional
$\mathcal{N}=2$ theories? If $f$ is quasi-homogeneous, we might use the 4d/2d
correspondence to give a constraint \cite{Shapere:1999xr}: i.e., the
corresponding 2d Landau-Ginzburg model based on $f$ should have a central charge
$\hat{c}\leq 2$. For a $\mathbb{C}^*$ variable, its contribution to the 2d central
charge would be $1$, and a $\mathbb{C}$ variable would contribute to $1-2q$ with
$q$ the weight of the variable.
If our polynomial is given by
$f:(\mathbb{C}^{*})^n\times \mathbb{C}^m\rightarrow \mathbb{C}$,
the central charge of 2d model would be
\begin{equation*}
\hat{c}=n+\sum_{i=1}^m(1-2q_i).
\end{equation*}
If we require $\hat{c}< 2$, we have the following.
\begin{itemize}
\item There is at most one $\mathbb{C}^*$ variable. If there is a $\mathbb{C}^*$
  variable, the contribution to central charge of other $\mathbb{C}$ variables
  should be less than one, which implies that they form a $ADE$ singularity.
\item If all the variables are $\mathbb{C}$ variables ($n=0$), the constraint is
  \begin{equation*}
    \sum_{i}^m (1-2q_i)<2\Longleftrightarrow \sum_{i=1}^m q_i-1>{m-4\over 2}.
  \end{equation*}
  Here we take $m$ to be even, so that the middle homology has a symplectic
  pairing (which would give the Dirac pairing for the BPS particles). For $m=2$,
  there is no constraint; For $m=4$, the constraint is $\sum_{i=1}^m q_i>1$ and
  the list of such polynomials can be found in \cite{Xie:2015rpa}.
  For $m\geq 6$, there are very fewer possibilities, see \cite{DelZotto:2015rca}
  for some interesting examples.
\end{itemize}

\medskip%
We now give a general criterion for whether a polynomial (actually its Newton polyhedron
with respect to infinity) gives a SW geometry for a 4d $\mathcal{N}=2$
theory.
\begin{itemize}
\item If $n+m=2$, the only constraint is $n\leq 1$.
\item If $n+m=4$. There are two constraints. (a) $n\leq 1$;
  (b) there is no interior point in the Newton polyhedron of $\Delta(f)$; (c) if $n=1$, the $\mathbb{C}$ variable part of the polynomial should be of the ADE type.
\end{itemize}
Later on, we will give an explanation of this fact using Hodge structure of the
variety defined by $f$. A generalization to arbitrary $n,m$ will be given there.

\medskip\noindent%
\textbf{Deformations.}
Let us start with a Newton non-degenerate and convenient Laurent polynomial
\(f\) which gives SW geometry for 4d $\mathcal{N}=2$ theory. The deformations of
the polynomial is captured by the Jacobian algebra of $f$:
\begin{equation}
  J_f=
  \frac{\mathbb{C}[x_1,\ldots, x_n,x_{n+1},\ldots, x_{n+m}]}
  {\{x_1{\partial f\over \partial x_1},\ldots,x_n{\partial f\over \partial x_n},{\partial f\over \partial x_{n+1}},\ldots, {\partial f\over \partial x_{n+m}} \}};
\end{equation}
The deformed polynomials are
\begin{equation}
F_\lambda= f+\sum_{i=1}^\mu\lambda_i \phi_i
\end{equation}
where $\mu$ is the dimension of Jacobi algebra, the images of $\phi_i$ form a
basis of $J_f$. When $f$ is quasi-homogeneous, we can assign a scaling dimension
to $\lambda_i$ as follows \cite{Shapere:1999xr}:
\begin{equation}
[\lambda_i]= {2(1-q_i)\over 2-\hat{c}};
\end{equation}
Here $q_i$ are the weights of $\phi_i$ under $\mathbb{C}^*$ action. The parameters
$\lambda_i$ are identified with the parameters of Coulomb branch, and are
classified as follows:
\begin{enumerate}
\item Marginal deformations: if $[\lambda_i]=0$.
\item Relevant deformations: if $0<[\lambda_i]<1$.
\item Mass deformations: if $[\lambda_i]=1$.
\item Coulomb branch moduli: if $[\lambda_i]>1$.
\item Irrelevant deformation: if $[\lambda_i]<0$.
\end{enumerate}
We are particularly interested in the relevant deformations, mass
deformations, and the Coulomb branch moduli, as they would not change the UV theory.
These deformations have a nice geometric descriptions: they are associated with
the so-called under-diagram deformations, namely, the monomials $\phi_\alpha$
are under the boundary of the Newton polyhedron in the region $x_i>0$ (here
$x_i$ is the $\mathbb{C}$ variable). The marginal deformation is associated with
the lattice points on the boundary of Newton polyhedron.

The interesting deformation gives the following SW geometry:
\begin{equation}
F_\lambda= f+\sum_{i=1}^{\mu_{1}}\lambda_{\alpha} \phi_{\alpha};
\end{equation}
Here $\mu_1$ is the total number of relevant/marginal, mass, and Coulomb branch
moduli. The full SW geometry could be regarded as giving a polynomial
\begin{equation}
F: U\times \mathbb{C}^{\mu_1}\to \mathbb{C}^{\mu_1}\times \mathbb{C},
\end{equation}
and the map is given by
\begin{equation}
F: (z_i, \lambda_\alpha)=(\lambda_\alpha, F(z_i, \lambda_\alpha)).
\end{equation}
The above polynomials will be the fundamental objects of this paper.

\medskip\noindent%
\textbf{Polynomial or hypersurface.}
In the usual study of Seiberg-Witten solution, one has a hypersurface fibered
over Coulomb branch. In our study, we find that it is better to attach a
polynomial map $f_\lambda:U\to \mathbb{C}$ for each vacuum, and the usual
hypersurface can be recovered by considering the fiber at point zero
$f_\lambda^{-1}(0)$.

\medskip\noindent%
\textbf{Complete intersection.}
Up to this point, we only consider the SW geometry defined by a polynomial over
an affine space. However, there are lots of $\mathcal{N}=2$ theories whose SW
geometry can not be described by a polynomial: their SW geometries are given by
more general geometry such as complete intersections. Namely, we have map
$f:\mathbb{C}^{*n}\times \mathbb{C}^m \rightarrow \mathbb{C}^k$. The case
$f:\mathbb{C}^5\rightarrow \mathbb{C}^2$ with a $\mathbb{C}^*$ action is studied in
\cite{Chen:2016bzh,Wang:2016yha}. One can similarly attach a Newton polyhedron
for the complete intersection, etc.

\newpage
\section{IR theory: generic vacua}\label{sec:3}

The SW geometries of a large class of $\mathcal{N}=2$ theories can be described
by polynomial maps
$F: U\times \mathbb{C}^{\mu_1}\to \mathbb{C}^{\mu_1}\times \mathbb{C}$.
In most examples, \(U\) is the affine variety
\(\mathbb{C}^{\ast n} \times \mathbb{C}^{m}\). We shall consider this case exclusively.

The low energy theory on the Coulomb branch is then determined by a polynomial
\begin{equation}
F(x;\lambda_{\alpha})=f(x)+\sum_{\alpha=1}^{\mu_1} \lambda_\alpha \phi_\alpha(x);
\end{equation}
where $f(x)$ is a polynomial (which determines the Newton polyhedron),
\(\lambda_{\alpha}\) are  parameters of the relevant deformations, marginal
deformations,  mass deformations, and deformations by Coulomb branch moduli.
The map is then given by
\begin{equation*}
  (x, \lambda_{\alpha}) \mapsto (\lambda_\alpha, F(x,\lambda_{\alpha})).
\end{equation*}

We would like to extract low energy physics from topologies of
these polynomial maps. This is related to the fiber at $t=0$ and its \textbf{nearby} fibers.
This is one of the crucial point of this paper:
\textit{the low energy physics is not completely determined by the hypersurface
  at a particular vacuum but by both the singular fiber and the nearby fibers.}

First, for given physical parameters $\lambda_{\alpha}$, we would like to
determine whether it is a generic vacuum or special vacuum. This question is
related to how the topology of the fibers $F^{-1}(\lambda, t)$ change
with the parameters $\lambda$ and $t$. Vaguely,
if the topology changes at a particular parameters $(\lambda_0, 0)$, it is a
special vacuum, otherwise, it is a generic vacuum. We shall make this precise
in \S\ref{sec:3.1}.

In this section, we would like to discuss the physics associated with generic
vacua, where the low energy physics is given by abelian gauge theories. As we
discussed in the introduction, the cohomology group of the fiber
$F^{-1}(\lambda, t)$ itself is not enough to determine the low energy physics,
one needs to know the mixed Hodge structure defined on the cohomology group. The
MHS is useful in that: (1) It can help us distinguish the cycles relevant for
the flavor central charge from that of the electric-magnetic central charge; (2)
One can define an abelian variety from the cohomology group, which is necessary
from the unitarity requirement of the low energy interaction.

\subsection{Topology of polynomial maps and vacua type}
\label{sec:3.1}

Fix the deformation parameters $\lambda$. We have a polynomial map
\(F: U \to \mathbb{C}\). Physically, this means that we study a particular
deformation of the theory parameterized by the coordinate $t$
(in many cases, this parameter has largest scaling dimension and is the most
important deformation).

\begin{figure}[H]
\begin{center}

 \begin{center}
\includegraphics[width=.9\linewidth]{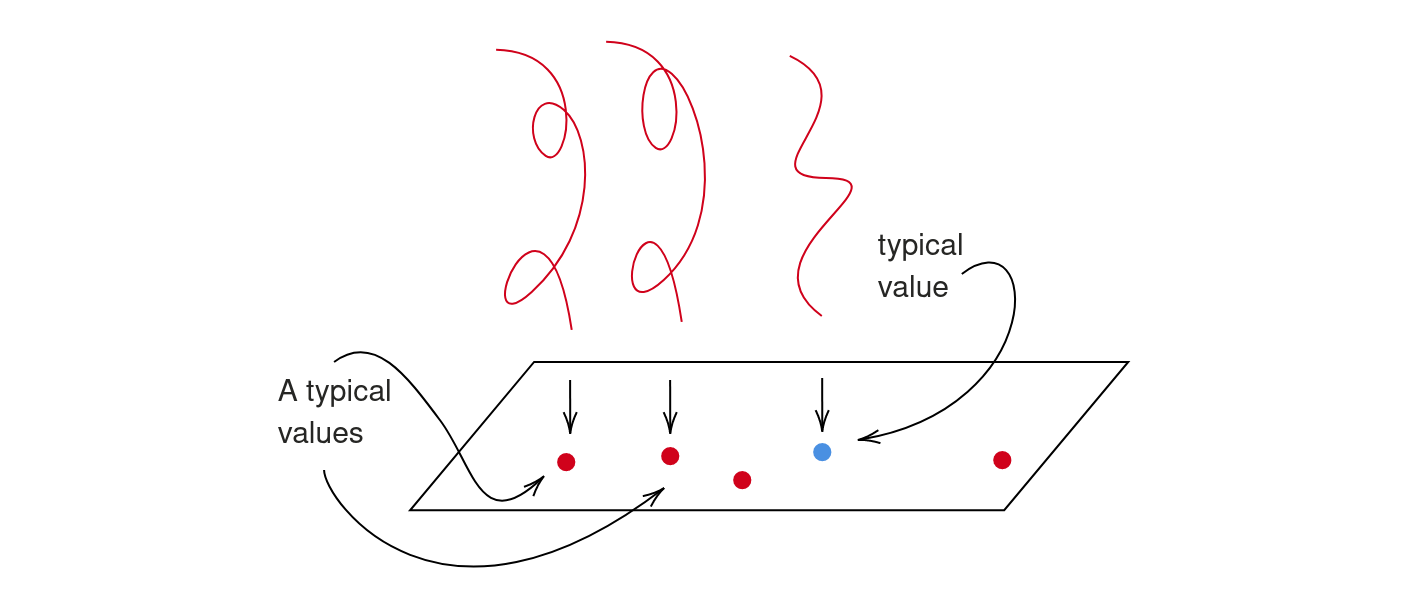}
\end{center}

\caption{A family of hypersurfaces  associated with the polynomial map $F$. Here the typical value $c$ on the $\mathbb{C}$ plane means that locally around $c$, the fibration is  locally trivial.}
\label{atypical}
\end{center}
\end{figure}

Let us review some basic facts about the topology of polynomial maps
\cite{broughton1988milnor}.
Let \(F:\mathbb{C}^{\ast n}\times\mathbb{C}^{m}\to\mathbb{C}\) be a
polynomial map. We shall be concerned with the topology of the fibers of \(F\).
It is known that there is a finite subset \(\alpha_{F}=\{b_1,\ldots,b_s\}\) of
\(\mathbb{C}\) such that
\begin{equation*}
F : F^{-1}(\mathbb{C} - \alpha_{F}) \to \mathbb{C} - \alpha_F
\end{equation*}
is a (topologically) locally trivial fibration. We shall assume that
\(\alpha_{F}\) is the smallest possible such subset, and call it the set of
\emph{atypical values}. A point \(c \in \mathbb{C}-\alpha_{F}\) is thereby called a
\emph{typical value} of the function \(F\). See Figure~\ref{atypical}.

The notion of an atypical value should not be confused with the notion of
a ``critical value''. By definition, a \emph{critical point} of \(F\) is a point
in \(\mathbb{C}^{*n}\times \mathbb{C}^{m}\) where the differential
\(\mathrm{d}F\) equals zero. If \(P\) is a critical point of \(F\), \(c=F(P)\)
is called a \emph{critical value}. A critical value is necessarily an atypical
value. But the converse is not true.

\medskip\noindent%
\textit{Example}~\cite{broughton1988milnor}.
Consider the polynomial \(F(X,Y)=X^{2}Y-X\). Then \(F\) has no critical points at all,
but the topological type of the fiber \(F_{t}=\{(x,y,t): x^{2}y-x-t=0\}\) could
change when \(t\) changes. When \(t=0\), \(F_{t}\) is the union of two curves
\((x=0)\) and \((xy-1=0)\); whereas if \(t\neq 0\),
\(F_{t}\) is isomorphic to \(\mathbb{C}-\{0\}\).

\begin{figure}[H]
\begin{center}
  \includegraphics[width=.9\linewidth]{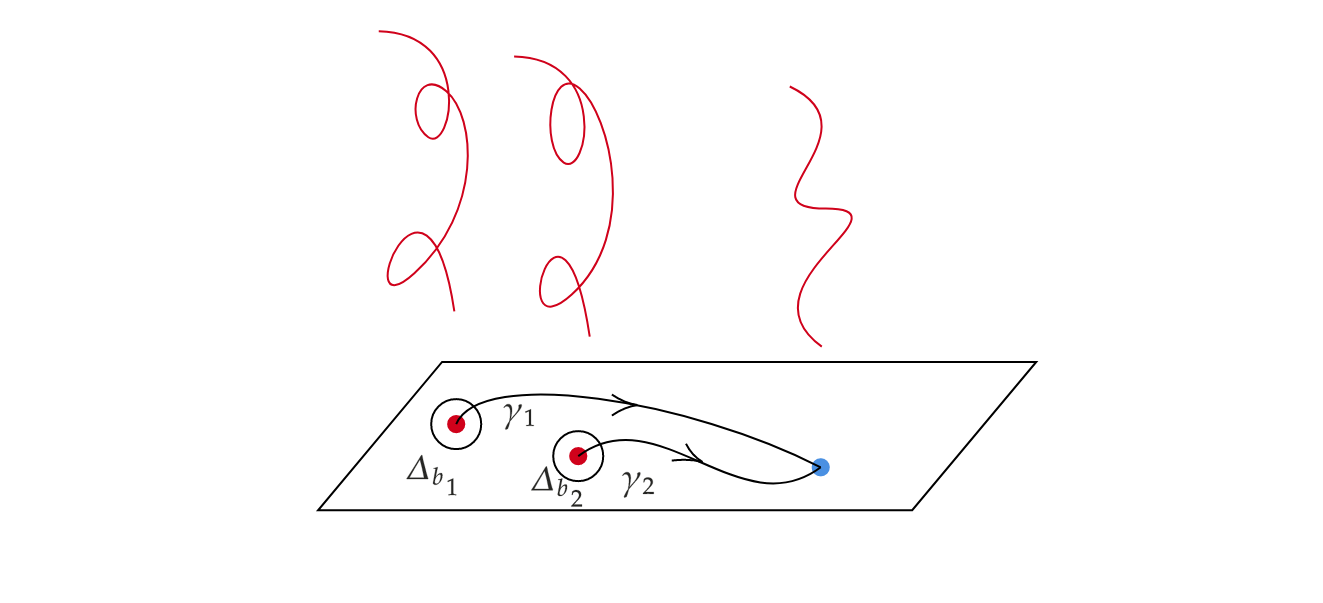}
\end{center}
\caption{The point $b_1$ and $b_2$ are atypical values. For typical value $c$,  one has a path connecting $c$ and $b_1$, and along the path, some of
the cycles on the typical fiber vanishes at $b_1$. The cohomology group of the typical fiber can be computed from the vanishing cohomology along these paths.}
\label{path}
\end{figure}

\medskip
The cohomology of the typical fibers can be computed in terms of the
so-called vanishing cycles. Consider for instance a polynomial map
\(F:\mathbb{C}^{n} \to \mathbb{C}\). We are interested in the cohomology
of \(X_{c}=F^{-1}(c)\) for a typical value \(c\) of \(F\). This will be
described in terms of ``vanishing cohomology groups''. Let \(b\) be an
\emph{atypical value} of \(f\). Then there is a small disk \(\Delta_b\) centered
at \(b\) such that \(\Delta_b-\{b\}\) consists of only typical values. For any
\(\eta \in \Delta_b-\{b\}\), there is a natural ``specialization map''
\begin{equation*}
H^{q}(X_{b}) \to H^{q}(X_{\eta})
\end{equation*}
The cokernel is denoted by \(V^{q}_{b}\), and is called the
\emph{vanishing cohomology} group near \(b\).
Now let \(\alpha_F=\{b_1,\ldots,b_m\}\) be the set of atypical values of \(F\).
Let \(c \in \mathbb{C} \setminus \alpha_F\) be a typical value, see Figure~\ref{path}.
Then we can draw paths
\(\gamma_1,\ldots,\gamma_m\) joining \(c\) and \(b_i\). If we choose
\(\gamma_i\) and \(\Delta_{b_i}\) suitably, we can ensure that the
path \(\gamma_i\) is the only path in the above collection that intersect
\(\Delta_{b_i}\).

Since \(F\) is a locally trivial fibration away from \(\alpha_F\), we
know that parallel transport along \(\gamma_i\) gives rise to an isomorphism
between the cohomology group \(H^{q}(X_{c})\) and
\(H^{q}(X_{\eta_i})\) for any \(\eta_i \in \Delta_{b_i}\).  Thus, we
have a surjective map \(H^{q}(X_{c}) \to V_{b_i}^{q}\). Summing over
all \(i\), we get a surjective map
\begin{equation*}
  H^{q}(X_{c}) \to \bigoplus_{i=1}^{m} V_{b_i}^{q}.
\end{equation*}
which is in fact an isomorphism. In other words, the cohomology of a typical fiber
is the direct sum of the parallel transports of all vanishing cohomology groups.
This fact links the cohomology of a generic fiber with that of
\textbf{vanishing} cohomology.

We assumed the variety \(U\) equals \(\mathbb{C}^{n}\) to fix ideas. In case
there exist \(\mathbb{C}^{\ast}\)-variables, then the formulation should be
changed slightly. For a subvariety \(Z\) of \(U\), the
\textbf{primitive cohomology} of \(Z\) is the cokernel of
\begin{equation*}
  H^q(U) \to H^q(Z).
\end{equation*}
In the case when \(U=\mathbb{C}^{n}\times \mathbb{C}^{\ast m}\),
the correct formulation is then the primitive cohomology of \(Z\) equals the
direct sum of the vanishing cohomologies.

Another useful fact for the cohomology of typical fiber is the relation to the
so-called local Milnor fiber. Assume that $P$ is an \textbf{isolated critical point}
of $F$, then Milnor's Fibration Theorem asserts that there is a small ball $B$
centered around $P$, such that for all sufficiently small discs $\Delta^*$, the
map $F:B\cap F^{-1}(\Delta^*) \rightarrow \Delta^*$ is a locally trivial
fibration. Moreover, the fibers $F^{-1}(c)\cap B$ have the homotopy type of
a bouquet of $\mu_p$ spheres, where $\mu_p$ is called the local Milnor number.
The middle cohomology group has the following structure:

\begin{quote}
  Let $F:\mathbb{C}^{n+1}\to \mathbb{C}$ be a polynomial.
  Let $\bar{\mu}$ be the sum of all the Milnor numbers of $F$
  at isolated critical points of $F$.
  Then for all typical fiber, we have
  \begin{equation}
    H_q(F^{-1}(c))\approx \mathbb{Z}^{\bar{\mu}} \oplus A
  \end{equation}
  where $A$ is some finitely generated abelian group.
\end{quote}
If $F$ has only isolated critical points, then the subgroup $A$ comes from the
contribution of ``vanishing cycles at infinity''. Again, with the presence of
\(\mathbb{C}^{\ast}\) variables, one should use ``primitive homology'', i.e.,
the kernel of \(H_q(F^{-1}(c)) \to H_{q}(U)\), instead of the usual homology.

If the parameter $\lambda$ is varying, we then get the map
$F:\mathbb{C}^{* n}\times \mathbb{C}^{m} \times \mathbb{C}^{\mu_1} \rightarrow\mathbb{C}^{\mu_1}\times \mathbb{C}$,
where the coordinate of
$\mathbb{C}^{\mu_1}$ is given by $\lambda_{\alpha}$, and that of $\mathbb{C}$ is given by
$t$. We are then interested in the fibers $F^{-1}(\lambda, t)$. Similarly, the
points on $\mathbb{C}^{\mu_1}\times \mathbb{C}$ could be classified as typical
and atypical values.
Let
\(F: \mathbb{C}^{* n}\times \mathbb{C}^{m} \times\mathbb{C}^{\mu_1} \to B=\mathbb{C}^{\mu_1}\times\mathbb{C}\)
be a polynomial map. Let \(X_{\lambda,t}\) be the fiber \(F^{-1}(\lambda,t)\).
Then the sheaf \(R^{n+m-1}F_{\ast}\mathbb{Q}\)
(sheaf associated with the presheaf
\(V\mapsto H^{n+m-1}(F^{-1}(V),\mathbb{Q})\))
is an algebraically constructible sheaf. This
means that there is a collection of locally closed subvarieties \(B_{i}\) of
\(B\) such that
\begin{enumerate}
\item \(B = \coprod_{i=0}^{m}B_{i}\),
\item \(B_{i} \subset \overline{B_{j}}\) if \(i > j\)
\end{enumerate}
(when these two conditions hold, we say \(\{B_i\}\) is a \textbf{stratification} of \(B\)),
such that \(R^{n+m-1}F_{\ast}\mathbb{Q}|_{B_j}\) is locally constant.
In particular, \(B_{0}\) is a Zariski open subvariety of \(B\). Points in
\(B_0\) are called \emph{typical values} of the function \(F\). Points in
\(B \setminus B_0\) are called \emph{atypical values} of \(F\).
In particular, atypical values form a proper Zariski closed subset of
\(\mathbb{C}^{\mu_1}\times \mathbb{C}\).

Again, a critical value is necessarily an atypical value, but a regular value
could equally be atypical. A useful fact for us is:
since \(R^{n-1}F_{\ast}\mathbb{Q}|_{B_0}\) is locally constant, its stalk
at a typical value \(c=(\lambda,t) \in B_0\) is equal to the cohomology of the
fiber: \(H^{n+m-1}(X_{\lambda,t})\). The above fact is in general not true for
atypical fibers.

Let us now relate the above discussion of topology of the polynomial map to the
IR physics of the corresponding $\mathcal{N}=2$ theory:

\bigskip

\begin{itemize}
\item \textbf{ The typical fiber at $(\lambda, 0)$ describes the generic vacua
    of $\mathcal{N}=2$ theory, while the atypical fiber at $(\lambda, 0)$
    describes special vacua.}
\end{itemize}

\bigskip

Next, we introduce a special type of polynomial called ``tame'' polynomial, and
for these polynomials the atypical value is the same as the critical value,
which will simplify our computations significantly. This definition is quite
involved \cite{broughton1988milnor}, and in practice not easy to check. But in
our situation, Newton \textbf{non-degenerate} and \textbf{convenient}
polynomials will turn out to be
tame~\cite{broughton1988milnor,sabbah2006hypergeometric}.
As we saw in the examples given in last section, these polynomials give a large
class of SW geometries.

We summarize some of the useful properties of topologies of tame
polynomials.

\begin{enumerate}
\item The \textbf{atypical} value of a tame polynomial is the same as the
  \textbf{critical} value.

\item For tame polynomial, the only non-trivial cohomology of a tame polynomial
  is the middle cohomology group.

\item For the tame polynomials, the stalk of \(R^{n+m-1}F_{\ast}\mathbb{Q}\) at
  an arbitrary point is equal to the cohomology of the fiber.
\end{enumerate}

In Sections~\ref{sec:3}, \ref{sec:4}, \ref{sec:5},
we shall consider exclusively
Seiberg--Witten geometries associated with tame polynomials.

Consider a typical fiber associated with a Newton non-degenerate and convenient
polynomial. We would like to extract the physical information from the variety
defined by $F_{\lambda}(x)=0$. We face the following challenges:
\begin{enumerate}[wide]
\item The central charge of the BPS particles should be given by the periods
  of integral of SW differential. The central charge contains the electric-magnetic
  charge for the $U(1)^r$ gauge groups, and flavor charges. The question is: how
  can we distinguish the electric-magnetic charge and flavor central charge in
  the cohomology group? How can we find a SW differential whose periods would
  give us the central charge?
\item The low energy photon couplings are identified with the complex structure
  of the variety and should have a positivity property. In fact, there is an
  abelian variety fibration over the Coulomb branch moduli space
  \cite{Donagi:1995cf}. In the compact curve case, this abelian variety is given
  by the Jacobian of the curve. How to define
  an abelian variety from the cohomology groups of fibers which are
  non-compact?
\item Finally, as we vary the parameters of the Coulomb branch moduli, we
  should get a special Kahler geometry, a strong physical
  constraint. The key question is to find a Seiberg-Witten
  differential.
\end{enumerate}

In this section, we will answer the first two questions by using the mixed
Hodge structure defined on cohomology groups of the open smooth variety. The
construction of Seiberg-Witten differential will be given in \ref{sec:7}.

\subsection{Mixed Hodge structure}

In this subsection, we recall the concept of a mixed Hodge structure.
The reader with physics background might already be familiar with the Hodge
decomposition of the cohomology group of a compact Kähler manifold, see
\cite{griffiths1978principles}.
\begin{quote}
  On a compact Kähler manifold (e.g., a \emph{projective} algebraic variety)
  \(M\), there exists a canonical decomposition (called the \emph{Hodge
    decomposition}) on the  cohomology group:
  \begin{equation*}
    H^n(M,\mathbb{Q}) \otimes \mathbb{C} \cong \bigoplus_{p+q=n} H^{p,q}(M).
  \end{equation*}
  Under the complex conjugation induced on the second tensor factor, the
  conjugate of the subspace \(H^{p,q}(M)\) is \(H^{q,p}(M)\).
\end{quote}

The Hodge decomposition is already crucial for understanding compact Riemann
surfaces. For a connected compact Riemann surface \(C\), the above theorem
implies that the rank \(b_1\) of $H^1(C,\mathbb{C})$ is necessarily even,
the space $H^1(C,\mathbb{C})$ splits into a direct sum
$H^{1,0}(C) \oplus H^{0,1}(C)$ in a way that the two summands are conjugate to
each other. It can be shown that $H^{1,0}$ is the space of holomorphic one forms.
So we can write $b_1 = 2g$. The number $g$, called the \emph{genus} of $C$,
is the number of linearly independent holomorphic 1-forms on \(C\). A
classical theorem of Torelli says that the abelian group \(H^1(C,\mathbb{Z})\),
the above Hodge decomposition, and the intersection pairing on
\(H^1(C,\mathbb{Z})\) \emph{completely} determines the complex structure of the
Riemann surface \(C\).

\medskip\noindent%
\textbf{Pure Hodge structure.}
A pure Hodge structure is an abstraction of above decomposition of compact
Kähler manifold. By definition, a pure Hodge structure of weight \(n\) consists
of a lattice \(H_{\mathbb{Z}}\) together with a decomposition:
\begin{equation*}
H_{\mathbb{C}} := H_{\mathbb{Z}} \otimes \mathbb{C} = \bigoplus_{p+q=n} H^{p,q},
\end{equation*}
such that the complex conjugation of  \(H^{p,q}\) equals \(\overline{H^{q,p}}\).

Hodge structure can be equivalently given in terms of its
\emph{Hodge filtration}. The \(p\)\textsuperscript{th} Hodge filtration is
\(F^{p}=\bigoplus_{p'>p} H^{p',n-p'}\),
thus $F^0=H, \ldots, F^{n-1}=H^{n,0}\oplus H^{n-1,1},~F^n=H^{n,0}$, and we have
\begin{equation*}
  H_{\mathbb{C}} = F^0 \supset \cdots \supset F^{n},
\end{equation*}
for each \(p\), we have
\begin{equation}
H=F^p\oplus \overline{F^{n-p+1}},
\end{equation}
and finally the Hodge decomposition can be recovered by
\begin{equation}
H^{p,q}=F^p \cap \overline{F^q}.
\end{equation}

\medskip\noindent%
\textit{Example.} A trivial example is the ``Tate structure''
\(\mathbb{Z}(m)\), where \(m\) is an integer. This is a pure Hodge structure of
weight \(-2m\) of one dimensional lattice. The ambient abelian group is
\((2\pi\sqrt{-1})^{m}\mathbb{Z}\), and upon tensoring with \(\mathbb{C}\) all
the \((p,q)\)-components are zero unless \((p,q)=(-m,-m)\). So
\begin{equation}
H=H^{-m,-m}.
\end{equation}
For example, the pure Hodge structure associated with \(H^2(\mathbb{P}^1)\) (one
dimensional vector space) is of weight 2, but it has no \((2,0)\) and \((0,2)\)
parts, and it has only $(1,1)$ part. Thus \(H^2(\mathbb{P}^1)\), as a Hodge
structure, is isomorphic to the Tate structure \(\mathbb{Q}(-1)\).

In general, if \(H\) is a Hodge structure of weight \(n\), we use \(H(m)\) to
denote the tensor product \(H \otimes_{\mathbb{Q}} \mathbb{Q}(m)\). This is a
Hodge structure of weight \(n-2m\). This is called \emph{\(m\)th Tate twist} of
\(H\).

\medskip\noindent%
\textbf{Polarized Hodge structure.}
A polarized Hodge structure of weight $n$ is an HS
$\{H_{\mathbb{Z}}, F^{\bullet}\}$ together with a non-degenerate integral
bilinear form
\begin{equation}
S: H_{\mathbb{Z}}\times H_{\mathbb{Z}} \rightarrow \mathbb{Z}
\end{equation}
This bilinear form is symmetric for $n$ even, and antisymmetric for $n$ odd, and
satisfying following Riemann bilinear relations (first and second):
\begin{align}
& S(H^{p,q}, H^{p', q'})= 0~\text{if $(p',q')\neq (q,p)$)} \nonumber\\
& \sqrt{-1}^{p-q} S(\psi, \bar{\psi})>0~~\text{for}~\psi\in H^{p,q},~\psi\neq 0
\end{align}
In terms of Hodge filtration,  the bilinear relations are:
\begin{align}
& S(F^p,F^{n-p+1})=0,~~\nonumber\\
&  S(C\psi, \bar{\psi})>0~~\text{for}~~\psi\neq 0
\end{align}
Here $C:H\rightarrow H$ is the Weil operator $C|_{H^{p,q}}=i^{p-q}$.

Geometrically, a polarization is induced by an integral K\"ahler form \(\omega\) on
a projective manifold \(M\) and the usual cup product on the cohomology: for
differential $n$ forms \(v\) and \(w\), define
\begin{equation*}
  S(v,w) = (-1)^{n(n-1)/2}\int_{M} \omega^{\dim M-n} \wedge v \wedge w.
\end{equation*}
This induces a polarization on the Hodge structure of the so-called
``primitive cohomology''~\cite[p.122]{griffiths1978principles}.
This is known as the first and second  \emph{Hodge--Riemann bilinear relations}.

For example, on a compact Riemann surface, there is no $(2,0)$- or $(0,2)$-forms,
so the only non-zero cup product exists between a $(1,0)$-form and $(0,1)$-form,
this gives the first bilinear relation. The second bilinear relation is:
\(\sqrt{-1}\int\eta\wedge\overline{\eta}>0\) for a holomorphic 1-form \(\eta\), this is
true because the integral gives the volume of the Riemann surface.

% {\color{red} USELESS PARAGRAPH.
%   The Hodge structure could be used to reflect the complex structure of a
% projective manifold. For instance, if we fix a homology basis of a compact
% Riemann surface of genus \(g\), then how the subspace \(H^{1,0}\) sits
% inside the \(2g\)-dimensional vector space essentially determines the Riemann
% surface up to isomorphism (Torelli's theorem). If \(\lambda_1,\ldots,\lambda_g\)
% is a basis for the holomorphic one forms \(H^{1,0}\), then this position is
% determined by ``periods of integrals''
% \begin{equation*}
% \int_{\gamma_j} \lambda_i, \quad i=1,2,\ldots,g; j=1,2,\ldots,2g.
% \end{equation*}
% Here $\gamma_j$ is the basis of one cycles with standard symplectic intersection
% pairing. The periods of integrals form a $g\times 2g$ matrix $Z$ called period
% matrix. This period matrix determines how $H^{1,0}$ part of $H^{1}$ sits inside
% vector space $H^1$.
% }

\medskip\noindent%
\textbf{Mixed Hodge structure.}
The above story works well for smooth projective varieties. It is natural to ask
what extra structure of the same flavor can be put on a singular or open
variety. In his dissertation ``Th\'eorie de Hodge, II et III'', Pierre Deligne
introduced a new structure called \emph{mixed Hodge structure}, see
\cite{deligne:hodge2}. A mixed Hodge structure (MHS) is a triple
$(H_{\mathbb{Z}}, W_{\bullet}, F^{\bullet})$, where $W_{\bullet}$ is a finite
increasing filtration on $H_{\mathbb{Q}}=H_{\mathbb{Z}}\otimes \mathbb{Q}$, and
$F^{\bullet}$ is a finite decreasing filtration on
$H=H_{\mathbb{Z}}\otimes\mathbb{C}$,
so that the induced filtration on the quotient $\mathrm{Gr}_k^W=W_k/W_{k-1}$
defines a HS of weight $k$. Here $F^p\mathrm{Gr}_k^W H$ is the image of
$F^p\cap W_k H$ in $\mathrm{Gr}_k^W H$:
\begin{equation}
F^p\mathrm{Gr}_k^W H=(F^p\cap W_k+W_{k-1})/W_{k-1}.
\end{equation}
Given a MHS $(H_{\mathbb{Z}}, W_{\bullet},F^{\bullet})$,
$F^\bullet$ is called the Hodge filtration,
and $W^\bullet$ is called weight filtration.
Let $H^{p,q}=\mathrm{Gr}_F^p(\mathrm{Gr}_{p+q}^W H \otimes \mathbb{C})$,
then $\mathrm{Gr}_k^W H \otimes \mathbb{C}= \oplus_{p+q=k} H^{p,q}$. The
number $h^{p,q}=\dim H^{p,q}$ is called Hodge numbers.

If there are two MHS $H$ and $H'$,
then a map $\phi: H\rightarrow H^{'}$ is called morphism of type
$(r,r)$, if $\phi(H_{\mathbb{Z}})\subset H_{\mathbb{Z}}'$,
$\phi(W_k H)\subset W_{k+2r} H'$,
and $\phi(F^pH)\subset F^{p+r} H'$.
The map $\phi$ is called a morphism of mixed Hodge structures if it is of
type $(0,0)$.

\medskip\noindent%
\textit{Example.}
Let \(P\) be a smooth projective variety.
Let \(D \subset P\) be a nonsingular hypersurface.
Set \(U=P-D\). Denote by \(j\) the inclusion map
\begin{equation}
j\colon U=P-D\rightarrow P.
\end{equation}
Then there is a long exact sequence of cohomology groups
\begin{equation*}
\cdots \to H^i(P) \to H^i(U) \to H^{i+1}(P,U) \to \cdots.
\end{equation*}
By excision, the cohomology of the pair \((P,U)\) is the same as the pair
\((N,N-D)\), where \(N\) is the normal bundle of \(D\) inside \(P\).
The ``Thom isomorphism'' in algebraic topology implies that
\begin{equation*}
H^{i+1}(P,U) \cong H^{i+1}(N,N-D) \cong H^{i-1}(D)\otimes \mathbb{Z}(-1)
\end{equation*}
Here we include a Tate twist so that it is an isomorphism of Mixed Hodge Structure.
We get a Gysin sequence
\begin{equation*}
\cdots \to H^{k-1}(U) \xrightarrow{r} H^{k-2}(D) \xrightarrow{i_*} H^{k}(P) \xrightarrow{j^*} H^k(U) \to \cdots.
\label{exact}
\end{equation*}
Here $r$ is the topological residue map; in rational cohomology it is the
transpose of the ``tube over cycle map'' $\tau: H_{k-2}(D;\mathbb{Q})\rightarrow
H_{k-1}(U,\mathbb{Q})$, which associates to the class of a cycle $c$ in $D$ the
class of the boundary in $U$ of a suitable small tubular neighborhood of $c$.
The map $i_*$ is the Gysin map associated to $i$, namely, we have the
restriction map $H^i(P)\xrightarrow{i} H^i(D)$, and the Poincare dual gives the
map $H^{2n-i}(P)\xleftarrow{i_*} H^{2n-2-i}(D)$. Here we used the fact that the
real dimension of $D$ is $2n-2$, and that of $P$ is $2n$.

From the displayed exact sequence we see that \(H^{i}(U)\) has two pieces
of different weights: the piece \(W_{i} = \mathrm{Im}(H^i(P)\to H^i(U))\), which
should inherit the weight of \(H^i(P)\), namely \(i\). Set
\(W_{i+1}=H^{i}(U)\). Then the quotient
\(W_{i+1}/W_i\) is then a subspace of \(H^{i-1}(D)(-1)\), which has weight
\(i+1\). This suggests that in general on a nonsingular algebraic variety \(U\)
there should be more than one weight appearing in its cohomology.

\medskip
In general, Deligne proved that the cohomology group of any open smooth
algebraic variety carries a MHS, and the nontrivial weights on \(H^m(U)\) are in
the range \([m,2m]\).

\medskip\noindent%
\textit{Example.}
Consider \(H^1\) of an open curve $\tilde{C}$ which is
described by a genus \(g\) curve with \(m\) points removed, its MHS is given in
table. \ref{opencurve}.
\begin{table}
\begin{center}
\begin{tabular}{|c|c|c|}
\hline
Weight& \(\mathrm{Gr}_{1}^{W}H\) & \(\mathrm{Gr}_{2}^{W}H\)\\
\hline
HS& \(H^{1,0}\oplus H^{0,1}\) & \(H^{1,1}\)\\ \hline
dimension &\(2g\) dim. & \(m-1\) dim.\\
\hline
\end{tabular}
\caption{The MHS of an open curve derived by removing $m$ points of a genus $g$ compact curve.}
\label{opencurve}
\end{center}
\end{table}

\begin{figure}[H]
\begin{center}
\includegraphics[width=.9\linewidth]{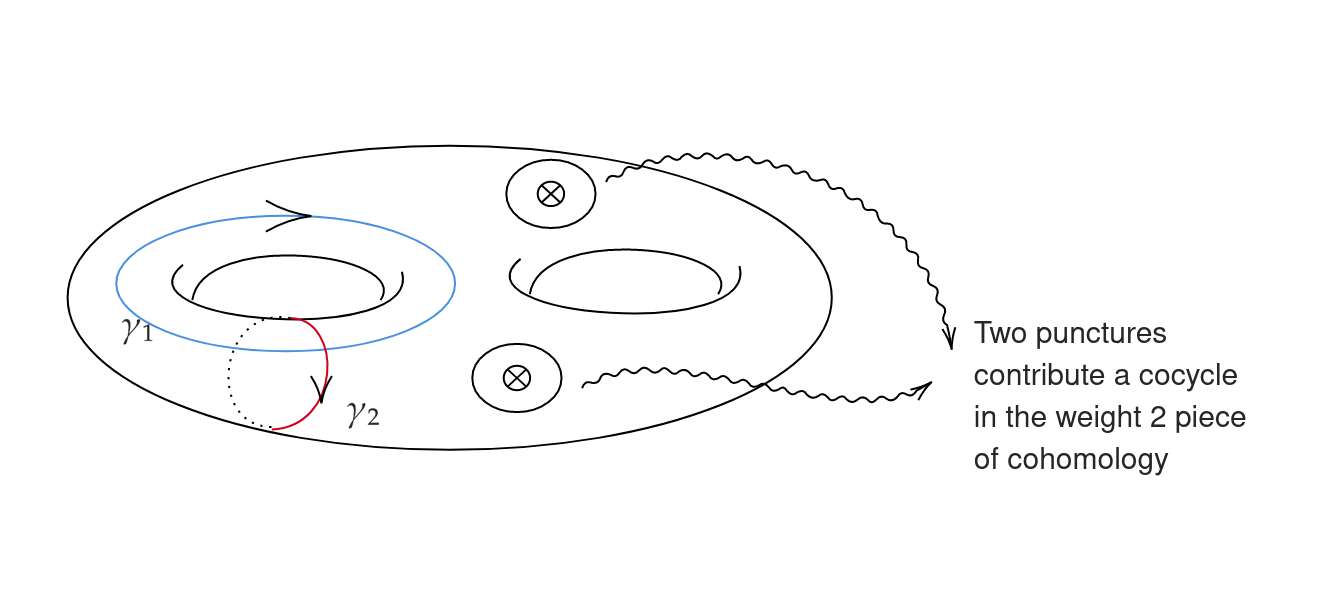}
\end{center}
\caption{An open curve derived by removing $m$ points of genus $g$ compact curve. The weight two part of $H^1$ is associated with the cycles around the removed points, and the weight one
part is associated with the cycles $\gamma_1$ and $\gamma_2$.}
\label{open}
\end{figure}
The above result can be computed using the exact sequence (\ref{exact})
(here we take $P=C$ a compact genus $g$ curve, and $D=m~\text{points}$, so
$U=P-D=\tilde{C}$).
\begin{equation*}
0\to  H^1(C)\to H^{1}(U) \to H^{0}(m~\text{pts})(-1)  \to H^2(C)\to 0
\end{equation*}
Notice that there is a Tate twist for the cohomology group of
$H^{0}(m~\text{points})$,
so it gives a weight 2 piece of $H^1(U)$; the weight $1$ piece of $H^1(U)$ is
then given by $H^1(C)$ of compact curve $C$, see figure.\ref{open}.

\subsection{Computations}\label{sec:3.3}
For a four dimensional $\mathcal{N}=2$ theories, the charge lattice has rank
\begin{equation}
d=2r+f
\end{equation}
Here $r$ is the dimension of Coulomb branch moduli (or the rank of abelian gauge
groups at a generic vacuum) and $f$ is the number of flavor symmetries. This rank
should be equal to the middle (primitive) cohomology of the variety \(X\) in the SW geometry
associated with the vacuum
\begin{equation}
d=H^n(X)
\end{equation}
In this subsection, we shall explain how to determine the Hodge numbers of the
MH{S} on \(H^n(X)\). For curves, we can compute MHS using resolution of
singularity and the result of last subsection. More generally, we can compute
MHS from the combinatorial data of the Newton polyhedron associated with the
polynomial.

\subsubsection*{Direct computation}
One can compute the MHS for an arbitrary smooth curve
$C$ defined by $f(x,y)=0$ using following strategy (here $x$ and $y$ are $\mathbb{C}$ variables):
\begin{enumerate}
\item First we consider a projective completion $\overline{C}$ of $C$ which is
  given by a homogenous polynomial $F(x,y,z)$ ($F(x,y,z)$ is the homogenization
  of $f(x,y)$). This procedure will typically add some extra points called
  points at infinity.
\item The points at infinity might be singular  points of
  \(\overline{C}\),
  and we resolve them (using blow-up, see appendix) to get a smooth curve
  $\tilde{C}$. The Hodge numbers of \(H^1(C)\) are
\begin{equation}
  h^{1,0}(H^{1}(C))=h^{0,1}(H^{1}(C))=g({\tilde{C}}),\quad h^{1,1}(H^{1}(C))=m-1
\end{equation}
\end{enumerate}
Here $m$ are the number of points removed from $\widetilde{C}$. The details of
the computation of these numbers are given in appendix.

\medskip\noindent%
\textit{Example.} Consider the affine curve \(C\) given by $f=x^n+y^n+1$. The
corresponding projective curve \(\overline{C}\) is cut out by
$F(x,y,z)=x^n+y^n+z^n$, and is smooth.
Its genus is $g={(n-1)(n-2)\over 2}$.
The points added at infinity is given by the solution of the equation
$F(x,y,0)=x^n+y^n=0$, and there are $n$ points $(x,y)=(1, y_i),i=1,\ldots,n$
with $y_i^n=-1$. So the Hodge numbers are
\begin{equation}
h^{1,0}=h^{0,1}={(n-1)(n-2)\over2},\quad h^{1,1}=n-1
\end{equation}
The number of Coulomb branch moduli is $r={(n-1)(n-2)\over2}$, and the rank of
flavor symmetries is given by $f=n-1$. This agrees with the physical result
derived in \cite{Xie:2012hs}.

\subsubsection*{Combinatorial formulas}
\label{sec:combina}

The MHS of a smooth hypersurface can be computed using the combinatorial data of
Newton polyhedron, see \cite{danilov1987newton}. We apply their results to
the polynomials relevant to 4d $\mathcal{N}=2$ theories.

We begin by reviewing their formulas. Generalizing the notion of Euler
characteristic, for each pair of integers $(p,q)$, we introduce the following
invariant of algebraic variety $X$:
\begin{equation}
e^{p,q}(X)=\sum_{k} (-1)^k h^{p,q}(H_c^k(X))
\end{equation}
Here $H_c^k(X)$ is the \(k\)\textsuperscript{th} compactly supported cohomolology
(if \(X\) is nonsingular, then as a Hode structure, the compactly supported
cohomology is isomorphic to \(H^{n-k}(X)^{\ast}\otimes \mathbb{Q}(-2n)\)), and
$h^{p,q}$ is the Hodge number defined using MHS. In our case, we only need to
consider middle cohomology group or maximal cohomology group, so $k=n-1$ (the complex dimension of the fiber
is $n-1$). Thus (for $p+q\leq n-1$):
\begin{equation}
  e^{p,q}(X)=(-1)^{n-1}h^{p,q}(H_c^{n-1}(X))
\end{equation}
Hence, knowing the information of $e^{p,q}(X)$ is the same as knowing the Hodge
numbers of the compacted support middle cohomology group, and the information of
the usual middle cohomology group is found using Poincare duality.

Consider a smooth hypersurface $Z$ in $\mathbb{C}^{n}$, and its associated
Newton polyhedron is $\Delta$. We have
\begin{equation}
e^{p,q}(Z)=\sum_{I} e^{p,q}(Z_I).
\label{enumber}
\end{equation}
Here we sum over the index set $I\subset \{1,\ldots, n \}$, and $Z_I$ is the
hypersurface defined by the equation $f_I=0$ in the algebraic torus
$(\mathbb{C}\setminus\{0\})^{|I|}$. So the Hodge number of the hypersurface in
$\mathbb{C}^n$ is known once we know the formula for the hypersurface in
$\mathbb{C}^{*n}$, which is computed in \cite{danilov1987newton}.

% If $Z_a$ is a smooth hypersurface in $\mathbb{C}^{*n}$, one can compute the
% total Euler number from the combinatorial data.The total Euler number
% $E(Z_a)=(-1)^{n-1}dim H^{n-1}(X)+1$, is given by
% \begin{equation}
%   E(Z^a)=\sum_{I}(-1)^{|I|-1}|I|!V_{|I|}(\Delta_I).
%   \label{Euler}
% \end{equation}
% This formula expresses the Euler number in terms of the volume of the Newton
% polyhedrons, and we can easily compute the dimension of $H^{n-1}(X)$.

The generalization to a hypersurface in $\mathbb{C}^{*n}\times \mathbb{C}^m$ is
the following: for the formula \ref{enumber}, the set $I$ is $(1,\ldots, m)$,
namely we consider hypersurfaces $f_I$ in $\mathbb{C}^{*n}\times \mathbb{C}^{*I}$.
instead. In the sum of \ref{enumber}, only $\mathbb{C}$ variable
in the polynomial $f$ can be set to zero.

The Hodge numbers for the middle cohomology $H_c^{n-1}(Z)$ (Here $Z$ is a
hypersurface in $\mathbb{C}^{*n}$) can be computed using the combinatorial data
of the Newton polytope, see table.~\ref{hodgecurve}, \ref{hodgesurface} and
\ref{hodgethree}. These tables can be used to compute MHS for arbitrary
polynomials in dimension less or equal four.

\begin{table}[htp]
\begin{center}
\begin{tabular}{|c|c|} \hline
$h^{0,1}=l^*(\Delta)$ &~  \\  \hline
$h^{0,0}=\Pi-1$&$h^{1,0}=l^*(\Delta)$\\ \hline
\end{tabular}
\end{center}
\caption{Here $\Delta$ is a two dimensional polytope, and $\Pi$ denotes the number of integral points in the 1-skeleton of $\Delta$.}
\label{hodgecurve}
\end{table}

\begin{table}[htp]
\begin{center}
\begin{tabular}{|l|l|l|} \hline
$h^{0,2}=l^*(\Delta)$ & &  \\  \hline
$h^{0,1}=\sum l^*(\Gamma)$ &$h^{1,1}$&\\ \hline
$h^{0,0}=\Pi-1$& $h^{1,0}=\sum l^*(\Gamma)$&$h^{2,0}=l^*(\Delta)$ \\ \hline
\end{tabular}
\end{center}
\caption{Here $\Delta$ is a three dimensional polytope, $\Gamma$ is the two dimensional face of $\Delta$, and $\Pi$ denotes the number of integral points in the 1-skeleton of $\Delta$. Here $h^{1,1}=l^*(2\Delta)-4l^*(\Delta)$.}
\label{hodgesurface}
\end{table}

\begin{table}[H]
\begin{center}
\begin{tabular}{|l|l|l|l|} \hline
\(h^{0,3}=\ell^{\ast}(\Delta)\) &  & & \\ \hline
  \(h^{0,2}=\sum_{\Gamma}\ell^{\ast}(\Gamma)\)& \(h^{1,2}\) & & \\ \hline
\(h^{0,1}=\sum_{F}\ell^{\ast}(F)\) & $h^{1,1}$ & \(h^{2,1}\) & \\ \hline
\(h^{0,0}=\Pi-1 \) &  $h^{1,0}$& $h^{2,0}$  & \(h^{3,0}=l^*(\Delta)\)\\ \hline
\end{tabular}
\caption{Here $\Delta$ is a four dimensional polytope, $\Gamma$ is the three dimensional face of $\Delta$, and $F$ is the two dimensional face, while $\Pi$ is the one dimensional face. We have $h^{2,1}=l^*(2\Delta)-5l^*(\Delta)-\sum l^*(\Gamma)$, and $h^{1,1}=l(2\Delta)-l^*(2\Delta)-5l(\Delta)+5l^*(\Delta)+10+\sum_\Gamma l^*(\Gamma)-\sum_Fl^*(F)$.}
\label{hodgethree}
\end{center}
\end{table}

\medskip\noindent%
\textbf{Curve case.}
Let \(f: \mathbb{C}^2 \to \mathbb{C}\) be a polynomial. Let \(\Delta\) be the
Newton polyhedron of \(f\) at infinity. Then the zero locus \(Z\) of \(f\)
defines a curve in \(\mathbb{C}^{2}\). Assume that \(Z\) is smooth. The MHS for
the cohomology $H^{1}(Z)$ is given in the table \ref{ccurve}.
\begin{table}[H]
\begin{center}
\begin{tabular}{|c|c|} \hline
\(h^{1,0} = \ell^{\ast}(\Delta)\) & \(h^{1,1}=l^*(F)\)\\ \hline
~& \(h^{0,1} = \ell^{\ast}(\Delta)\)\\ \hline
\end{tabular}
\caption{Here $\Delta$ is the two dimensional polytope, and $F$ denotes the one-dimensional boundary where no coordinate axis vanishes identically. Here $\ell^{\ast}(\Delta)$ denotes the interior points of $\Delta$.  }
\label{ccurve}
\end{center}
\end{table}

Let us explain formulas in Table~\ref{ccurve}. According to formula
\ref{enumber}, we have
\begin{equation}
e^{1,0}(Z)=e^{1,0}(Z_{1,2})+e^{1,0}(Z_{1})+e^{1,0}(Z_{0})=e^{1,0}(Z_{1,2})\rightarrow h^{1,0}(X)=h^{1,0}(Z_{1,2})=l^*(\Delta).
\end{equation}
We also have
\begin{equation}
e^{0,0}(Z)=e^{0,0}(Z_{1,2})+e^{0,0}(Z_{1})+e^{0,0}(Z_{2})
\end{equation}
and
\begin{align}
&h^{0,0}(Z)=h^{0,0}(Z_{1,2})-h^{0,0}(Z_{1})-h^{0,0}(Z_{2})=\Pi-1-(\Pi_x-1)-(\Pi_y-1)= \nonumber\\
& \Pi-\Pi_x-\Pi_y+1=l^*(F)
\end{align}
Here $Z_1$ is the hypersurface defined by the equation $f(x_1,0)$ and $x_1\neq
0$, and $Z_2$ is defined by $f(0,x_2)$ with $x_2 \neq 0$. $\Pi_x$ is the number
of lattice points of $\Delta$ at $x$ axis, and similarly for $\Pi_y$. $F$ is the
one dimensional boundary within first quadrant. The above formula gives rise to
the number listed in table.~\ref{ccurve}.

Notice that what the table computes are the Hodge number of the compact support
cohomology, the MHS for ordinary cohomology is found using the Poincare duality.

\medskip\noindent%
\textit{Example.} Consider the polynomial map $f=x^N+y^k$ (here $x,y$ are both
$\mathbb{C}$ variables); We include the deformations so that the zero set of $f$
is smooth, see the Newton polyhedron in Figure~\ref{nkseries}. Assume the
maximal common divisor for $(N,k)$ is $a$. The Hodge numbers are (using table.
\ref{ccurve})
\begin{equation}
h^{1,0}=h^{0,1}={(N-1)(k-1)-a+1\over2},~~h^{1,1}=a-1.
\end{equation}
Here $h^{1,0}$ is the same as rank of the corresponding field theory, and
$h^{1,1}$ is equal to the number of mass parameters.

\begin{figure}[H]
\begin{center}

\tikzset{every picture/.style={line width=0.75pt}} %set default line width to 0.75pt

\begin{tikzpicture}[x=0.55pt,y=0.55pt,yscale=-1,xscale=1]
%uncomment if require: \path (0,479); %set diagram left start at 0, and has height of 479

%Straight Lines [id:da6117605533307109]
\draw    (300,83) -- (300,260) ;
\draw [shift={(300,80)}, rotate = 90] [fill={rgb, 255:red, 0; green, 0; blue, 0 }  ][line width=0.08]  [draw opacity=0] (10.72,-5.15) -- (0,0) -- (10.72,5.15) -- (7.12,0) -- cycle    ;
%Straight Lines [id:da1361861893435674]
\draw    (280,240) -- (437,240) ;
\draw [shift={(440,240)}, rotate = 180] [fill={rgb, 255:red, 0; green, 0; blue, 0 }  ][line width=0.08]  [draw opacity=0] (10.72,-5.15) -- (0,0) -- (10.72,5.15) -- (7.12,0) -- cycle    ;
%Straight Lines [id:da33927932876938804]
\draw    (300,120) -- (380,240) ;
%Shape: Circle [id:dp46682864795174406]
\draw  [fill={rgb, 255:red, 0; green, 0; blue, 0 }  ,fill opacity=1 ] (318,220.75) .. controls (318,219.23) and (319.23,218) .. (320.75,218) .. controls (322.27,218) and (323.5,219.23) .. (323.5,220.75) .. controls (323.5,222.27) and (322.27,223.5) .. (320.75,223.5) .. controls (319.23,223.5) and (318,222.27) .. (318,220.75) -- cycle ;
%Shape: Circle [id:dp4546106152688263]
\draw  [color={rgb, 255:red, 208; green, 2; blue, 27 }  ,draw opacity=1 ][fill={rgb, 255:red, 208; green, 2; blue, 27 }  ,fill opacity=1 ] (337.25,180) .. controls (337.25,178.48) and (338.48,177.25) .. (340,177.25) .. controls (341.52,177.25) and (342.75,178.48) .. (342.75,180) .. controls (342.75,181.52) and (341.52,182.75) .. (340,182.75) .. controls (338.48,182.75) and (337.25,181.52) .. (337.25,180) -- cycle ;
%Shape: Circle [id:dp5111621386838272]
\draw  [fill={rgb, 255:red, 0; green, 0; blue, 0 }  ,fill opacity=1 ] (358.25,220.75) .. controls (358.25,219.23) and (359.48,218) .. (361,218) .. controls (362.52,218) and (363.75,219.23) .. (363.75,220.75) .. controls (363.75,222.27) and (362.52,223.5) .. (361,223.5) .. controls (359.48,223.5) and (358.25,222.27) .. (358.25,220.75) -- cycle ;
%Shape: Circle [id:dp1138896313226283]
\draw  [fill={rgb, 255:red, 0; green, 0; blue, 0 }  ,fill opacity=1 ] (318,160.75) .. controls (318,159.23) and (319.23,158) .. (320.75,158) .. controls (322.27,158) and (323.5,159.23) .. (323.5,160.75) .. controls (323.5,162.27) and (322.27,163.5) .. (320.75,163.5) .. controls (319.23,163.5) and (318,162.27) .. (318,160.75) -- cycle ;
%Shape: Circle [id:dp25975694769849933]
\draw  [fill={rgb, 255:red, 0; green, 0; blue, 0 }  ,fill opacity=1 ] (340,220.75) .. controls (340,219.23) and (341.23,218) .. (342.75,218) .. controls (344.27,218) and (345.5,219.23) .. (345.5,220.75) .. controls (345.5,222.27) and (344.27,223.5) .. (342.75,223.5) .. controls (341.23,223.5) and (340,222.27) .. (340,220.75) -- cycle ;
%Shape: Circle [id:dp17701925816204733]
\draw  [fill={rgb, 255:red, 0; green, 0; blue, 0 }  ,fill opacity=1 ] (318,199.75) .. controls (318,198.23) and (319.23,197) .. (320.75,197) .. controls (322.27,197) and (323.5,198.23) .. (323.5,199.75) .. controls (323.5,201.27) and (322.27,202.5) .. (320.75,202.5) .. controls (319.23,202.5) and (318,201.27) .. (318,199.75) -- cycle ;
%Shape: Circle [id:dp7556860068305304]
\draw  [fill={rgb, 255:red, 0; green, 0; blue, 0 }  ,fill opacity=1 ] (318,181.75) .. controls (318,180.23) and (319.23,179) .. (320.75,179) .. controls (322.27,179) and (323.5,180.23) .. (323.5,181.75) .. controls (323.5,183.27) and (322.27,184.5) .. (320.75,184.5) .. controls (319.23,184.5) and (318,183.27) .. (318,181.75) -- cycle ;
%Shape: Circle [id:dp3833815376281775]
\draw  [fill={rgb, 255:red, 0; green, 0; blue, 0 }  ,fill opacity=1 ] (339,200.75) .. controls (339,199.23) and (340.23,198) .. (341.75,198) .. controls (343.27,198) and (344.5,199.23) .. (344.5,200.75) .. controls (344.5,202.27) and (343.27,203.5) .. (341.75,203.5) .. controls (340.23,203.5) and (339,202.27) .. (339,200.75) -- cycle ;

% Text Node
\draw (452,232.4) node [anchor=north west][inner sep=0.75pt]    {$x$};
% Text Node
\draw (287,57.4) node [anchor=north west][inner sep=0.75pt]    {$y$};

\end{tikzpicture}
\end{center}
\caption{The Newton polyhedron for $f=x^4+y^6+t$ (here $x,y$ are both $\mathbb{C}$ variable and $t$ is chosen so that $f=0$ is smooth). The number of interior points is $7$, and the number of point on the boundary in the first quadrant is $1$.
The Hodge numbers are $h^{1,0}=h^{0,1}=7$, and $h^{1,1}=1$. }
\label{nkseries}
\end{figure}
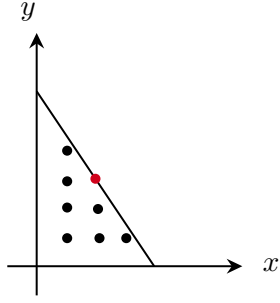

Now we consider the polynomial map
$f:\mathbb{C}^{\ast}\times \mathbb{C}\to \mathbb{C}$.
We only consider the \emph{primitive} cohomology of \(Z = \{f=0\}\).
(in this case, we need to subtract the contribution from the ambient space). The
\textbf{primitive} cohomology carries a MHS, which has a weight one part and a
weight two part. The Hodge numbers are summarized in the table.
\ref{cstarcurve}.
\begin{table}[H]
\begin{center}
\begin{tabular}{|c|c|} \hline
\(h^{1,0} = \ell^{\ast}(\Delta)\) & \(h^{1,1}=l^*(F)+l^*(F_x)\)\\ \hline
 ~& \(h^{0,1} = \ell^{\ast}(\Delta)\)\\ \hline
\end{tabular}
\caption{$\ell^{\ast}(\Delta)$ is the interior points of $\Delta$, $F$ is the one dimensional boundary where no coordinate vanishes identically, and $F_x$ is the boundary on the $x$ axis (here $x$ is a $\mathbb{C}$ variable). }
\label{cstarcurve}
\end{center}
\end{table}
We only need to explain $h^{0,0}$ for the compact support cohomology $H_c^1(Z)$ which is given by (assuming $y$ is the $\mathbb{C}^*$ variable). First, we have (using formula \ref{enumber}):
\begin{equation}
e^{0,0}(Z)=e^{0,0}(Z_{1,2})+e^{0,0}(Z_{2})=-(\Pi-1-(\Pi_y-1))=-(l^*(F)+l^*(F_x)+1)
\end{equation}
Here $Z_2$ is defined by the equation $f(0,x_2)$, and $F$ is the one dimensional boundary inside first quadrant, and $F_x$ is the boundary on the $x$ axis. We find that
\begin{equation}
h^{0,0}(H^1_c(Z))=(l^*(F)+l^*(F_x)+1)
\end{equation}
The primitive part has one less dimension: $h^{0,0}(H^1_c(Z))_{prim}=l^*(F)+l^*(F_x)$, which gives the desired answer.

\medskip\noindent%
\textit{Example.} Consider the polynomial map $f=x^N+z^k+\ldots$, here $x$ is a
$\mathbb{C}$ variable and $z$ is a $\mathbb{C}^{\ast}$ variable; We include the
deformations so that the zero set of $f$ is smooth, see the Newton polyhedron in
Figure~\ref{nkseries}. Assume the maximal common divisor for $(N,k)$ is $a$.
The Hodge numbers are (using formula in table.~\ref{cstarcurve}):
\begin{equation}
  h^{1,0}=h^{1,0}={(N-1)k-(a-1+N-1)\over2},~~h^{1,1}=a-1+N-1.
\end{equation}
Here $h^{1,0}$ is the same as rank of the corresponding field theory, and
$h^{1,1}$ is equal to the number of mass parameters.
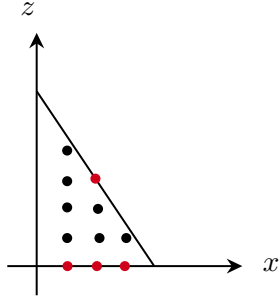
\begin{figure}[H]
\begin{center}

\tikzset{every picture/.style={line width=0.75pt}} %set default line width to 0.75pt

\begin{tikzpicture}[x=0.55pt,y=0.55pt,yscale=-1,xscale=1]
%uncomment if require: \path (0,479); %set diagram left start at 0, and has height of 479

%Straight Lines [id:da6117605533307109]
\draw    (300,83) -- (300,260) ;
\draw [shift={(300,80)}, rotate = 90] [fill={rgb, 255:red, 0; green, 0; blue, 0 }  ][line width=0.08]  [draw opacity=0] (10.72,-5.15) -- (0,0) -- (10.72,5.15) -- (7.12,0) -- cycle    ;
%Straight Lines [id:da1361861893435674]
\draw    (280,240) -- (437,240) ;
\draw [shift={(440,240)}, rotate = 180] [fill={rgb, 255:red, 0; green, 0; blue, 0 }  ][line width=0.08]  [draw opacity=0] (10.72,-5.15) -- (0,0) -- (10.72,5.15) -- (7.12,0) -- cycle    ;
%Straight Lines [id:da33927932876938804]
\draw    (300,120) -- (380,240) ;
%Shape: Circle [id:dp46682864795174406]
\draw  [fill={rgb, 255:red, 0; green, 0; blue, 0 }  ,fill opacity=1 ] (318,220.75) .. controls (318,219.23) and (319.23,218) .. (320.75,218) .. controls (322.27,218) and (323.5,219.23) .. (323.5,220.75) .. controls (323.5,222.27) and (322.27,223.5) .. (320.75,223.5) .. controls (319.23,223.5) and (318,222.27) .. (318,220.75) -- cycle ;
%Shape: Circle [id:dp4546106152688263]
\draw  [color={rgb, 255:red, 208; green, 2; blue, 27 }  ,draw opacity=1 ][fill={rgb, 255:red, 208; green, 2; blue, 27 }  ,fill opacity=1 ] (337.25,180) .. controls (337.25,178.48) and (338.48,177.25) .. (340,177.25) .. controls (341.52,177.25) and (342.75,178.48) .. (342.75,180) .. controls (342.75,181.52) and (341.52,182.75) .. (340,182.75) .. controls (338.48,182.75) and (337.25,181.52) .. (337.25,180) -- cycle ;
%Shape: Circle [id:dp5111621386838272]
\draw  [fill={rgb, 255:red, 0; green, 0; blue, 0 }  ,fill opacity=1 ] (358.25,220.75) .. controls (358.25,219.23) and (359.48,218) .. (361,218) .. controls (362.52,218) and (363.75,219.23) .. (363.75,220.75) .. controls (363.75,222.27) and (362.52,223.5) .. (361,223.5) .. controls (359.48,223.5) and (358.25,222.27) .. (358.25,220.75) -- cycle ;
%Shape: Circle [id:dp1138896313226283]
\draw  [fill={rgb, 255:red, 0; green, 0; blue, 0 }  ,fill opacity=1 ] (318,160.75) .. controls (318,159.23) and (319.23,158) .. (320.75,158) .. controls (322.27,158) and (323.5,159.23) .. (323.5,160.75) .. controls (323.5,162.27) and (322.27,163.5) .. (320.75,163.5) .. controls (319.23,163.5) and (318,162.27) .. (318,160.75) -- cycle ;
%Shape: Circle [id:dp25975694769849933]
\draw  [fill={rgb, 255:red, 0; green, 0; blue, 0 }  ,fill opacity=1 ] (340,220.75) .. controls (340,219.23) and (341.23,218) .. (342.75,218) .. controls (344.27,218) and (345.5,219.23) .. (345.5,220.75) .. controls (345.5,222.27) and (344.27,223.5) .. (342.75,223.5) .. controls (341.23,223.5) and (340,222.27) .. (340,220.75) -- cycle ;
%Shape: Circle [id:dp17701925816204733]
\draw  [fill={rgb, 255:red, 0; green, 0; blue, 0 }  ,fill opacity=1 ] (318,199.75) .. controls (318,198.23) and (319.23,197) .. (320.75,197) .. controls (322.27,197) and (323.5,198.23) .. (323.5,199.75) .. controls (323.5,201.27) and (322.27,202.5) .. (320.75,202.5) .. controls (319.23,202.5) and (318,201.27) .. (318,199.75) -- cycle ;
%Shape: Circle [id:dp7556860068305304]
\draw  [fill={rgb, 255:red, 0; green, 0; blue, 0 }  ,fill opacity=1 ] (318,181.75) .. controls (318,180.23) and (319.23,179) .. (320.75,179) .. controls (322.27,179) and (323.5,180.23) .. (323.5,181.75) .. controls (323.5,183.27) and (322.27,184.5) .. (320.75,184.5) .. controls (319.23,184.5) and (318,183.27) .. (318,181.75) -- cycle ;
%Shape: Circle [id:dp3833815376281775]
\draw  [fill={rgb, 255:red, 0; green, 0; blue, 0 }  ,fill opacity=1 ] (339,200.75) .. controls (339,199.23) and (340.23,198) .. (341.75,198) .. controls (343.27,198) and (344.5,199.23) .. (344.5,200.75) .. controls (344.5,202.27) and (343.27,203.5) .. (341.75,203.5) .. controls (340.23,203.5) and (339,202.27) .. (339,200.75) -- cycle ;
%Shape: Circle [id:dp817254628931132]
\draw  [color={rgb, 255:red, 208; green, 2; blue, 27 }  ,draw opacity=1 ][fill={rgb, 255:red, 208; green, 2; blue, 27 }  ,fill opacity=1 ] (338.25,240) .. controls (338.25,238.48) and (339.48,237.25) .. (341,237.25) .. controls (342.52,237.25) and (343.75,238.48) .. (343.75,240) .. controls (343.75,241.52) and (342.52,242.75) .. (341,242.75) .. controls (339.48,242.75) and (338.25,241.52) .. (338.25,240) -- cycle ;
%Shape: Circle [id:dp374998046172454]
\draw  [color={rgb, 255:red, 208; green, 2; blue, 27 }  ,draw opacity=1 ][fill={rgb, 255:red, 208; green, 2; blue, 27 }  ,fill opacity=1 ] (357.25,240) .. controls (357.25,238.48) and (358.48,237.25) .. (360,237.25) .. controls (361.52,237.25) and (362.75,238.48) .. (362.75,240) .. controls (362.75,241.52) and (361.52,242.75) .. (360,242.75) .. controls (358.48,242.75) and (357.25,241.52) .. (357.25,240) -- cycle ;
%Shape: Circle [id:dp04414192479712398]
\draw  [color={rgb, 255:red, 208; green, 2; blue, 27 }  ,draw opacity=1 ][fill={rgb, 255:red, 208; green, 2; blue, 27 }  ,fill opacity=1 ] (318.25,240) .. controls (318.25,238.48) and (319.48,237.25) .. (321,237.25) .. controls (322.52,237.25) and (323.75,238.48) .. (323.75,240) .. controls (323.75,241.52) and (322.52,242.75) .. (321,242.75) .. controls (319.48,242.75) and (318.25,241.52) .. (318.25,240) -- cycle ;

% Text Node
\draw (452,232.4) node [anchor=north west][inner sep=0.75pt]    {$x$};
% Text Node
\draw (287,57.4) node [anchor=north west][inner sep=0.75pt]    {$z$};

\end{tikzpicture}
\end{center}
\caption{The Newton polyhedron for $f=x^4+z^6+t$ (here $t$ is chosen so that $f=0$ is smooth. The number of interior lattice points is $7$, and the number of interior lattice points on two boundaries are $4$ (not including the boundary of $z$ axis).  }
\label{nkseries}
\end{figure}

\medskip\noindent%
\textbf{Threefold case.}
\label{sec:org260fe2e}
Let \(f: \mathbb{C}^{4} \to \mathbb{C}\) be a polynomial map. Let \(\Delta\) be
the Newton polyhedron of \(f\) at infinity which is Newton non-degenerate. Then
the zero locus \(Z\) of \(f\) defines a 3-fold in \(\mathbb{C}^{4}\). Assume
that \(Z\) is smooth. Then the cohomology \(H^{3}(Z)\) is a mixed Hodge
structure which possibly has weights \(3,4,5,6\). These numbers are summarized
in the table. \ref{threefold}.
\begin{table}[H]
\begin{center}
\begin{tabular}{|c|c|c|c|} \hline
\(h^{0,3}=\ell^{\ast}(\Delta)\) &  \(h^{1,3}=\sum_{\Gamma^{'}}\ell^{\ast}(\Gamma^{'})\) & \(h^{2,3}=\sum_{F^{'}}\ell^{\ast}(F^{'})\) & \(h^{3,3}=0 \)\\ \hline
 & \(h^{1,2}\) & \(h^{2,2}\) & \(h^{3,2}\)\\ \hline
 &  & \(h^{2,1}\) & \(h^{3,1}\)\\ \hline
 &  &  & \(h^{3,0}=l^*(\Delta)\)\\ \hline
\end{tabular}
\caption{The Hodge numbers for $H^3(Z)$ of a smooth hypersurface $Z$ in $\mathbb{C}^4$. Here $l^*(\sigma)$ denotes the internal lattice points of a face $\sigma$. Here $\Gamma^{'}$ denotes the three dimensional face where no coordinates vanish identically, similarly $F^{'}$ denotes
two dimensional face where no coordinates vanish identically. }
\label{threefold}
\end{center}
\end{table}

Let us explain the entries in Table~\ref{threefold}. We take
$h^{1,0}(H_{c}^3(Z))$%
\footnote{This is equal to $h^{2,3}(H^3(Z))$ due to
  Poincaré duality.}
as an example. Using formula.~\ref{enumber}, we have
\begin{align}
&h^{1,0}(Z)=h^{1,0}(Z_{1234})-h^{1,0}(Z_{123})-h^{1,0}(Z_{124})-h^{1,0}(Z_{134})-h^{1,0}(Z_{234}) \nonumber\\
&+h^{1,0}(Z_{12})+h^{1,0}(Z_{13})+h^{1,0}(Z_{14})+h^{1,0}(Z_{23})+h^{1,0}(Z_{24})+h^{1,0}(Z_{34})=\nonumber\\
&\sum_{2d~F~\in \Delta} l^*(F)-\sum_{2d~F \in \Delta_{ijk}} l^*(\Gamma)+\sum_{2d~F~\in~\Delta_{ij}} l^*(F)=\sum l^*(F^{'})
\label{honezero}
\end{align}
In the last row, the first term consists of sum of the face of polytope
$\Delta$, and the second term consists of sum of the face of the three
dimensional polytope $\Delta_{ijk}$ (defined by setting one of the coordinate in
$f$ to be zero), and finally the last term consists of sum of the two
dimensional polytope $\Delta_{ij}$. Finally $F^{'}$ is the two dimensional face
of three dimensional face $\Gamma^{'}$ where no coordinate vanishes identically.
One can similarly get the result for $h^{2,0}(Z)$ (for the compactly supported
cohomology).

Let us now impose the condition that $f$ should describe the SW geometry of a 4d
$\mathcal{N}=2$ theory. Then there is no internal points in $\Delta$ ($l^*(\Delta)=0$), and also
no internal points on $\Gamma^{'}$ and $F^{'}$ (This is true for
quasi-homogeneous case with $\sum q_i>1$ and we have to impose this for the general case); So
$h^{3,0}=h^{0,3}=0$, $h^{1,3}=h^{3,1}=0$, and $h^{2,3}=h^{3,2}=0$.
The only non-trivial Hodge numbers are:
\begin{align}
&h^{1,2}=h^{2,1}=l^*(2\Delta)-\sum_{\Gamma} l^*(\Gamma),~~\nonumber \\
&h^{2,2}=-E+1-2h^{1,2}=\sum_{I}(-1)^{|I|-1}|I|!V_{|I|}(\Delta_I)-2h^{1,2}+1
\label{3fold}
\end{align}

So we need to know the lattice points of polytope in order to compute $h^{1,2}$.
Let us review some basic facts about lattice points of polytope. For a \(d\)
dimensional polytope ${\cal P}$, we have following polynomials
\begin{align}
&i({\cal P}, n)=\# \{n{ \cal P} \cap \mathbb{Z}^d \} \nonumber \\
& \bar{i}({\cal P},n)=\# \{n{ \cal P}^o \cap \mathbb{Z}^d \}
\end{align}
Here $i({\cal P},n)$ counts the lattice points of polytope $n {\cal P}$, and
$\bar{i}({\cal P},n)$ counts the interior lattice points of $n {\cal P}$.
$i({\cal P}, n)$ is a polynomial in $n$ and is called Ehrhart polynomial.
Moreover, we have
 \begin{equation}
 \bar{i}({\cal P},n)=(-1)^d i({\cal P}, -n)
 \end{equation}
The Ehrhart polynomial takes the following form
\begin{equation}
i({\cal P}, n)=\mathrm{Vol}(P)n^d+\ldots
\end{equation}
In practice, once we know the polytope ${\cal P}$ (One can use Macaulay2
\cite{mac}
to compute it), we can compute $h^{1,2}$ and $h^{2,2}$ explicitly.

\medskip\noindent%
\textit{Example.}
Let us consider polynomial map given by \(f=x^3+y^3+z^3+w^3\).
The corresponding Ehrhart polynomial is given by
\begin{equation}
i({\cal P}, n)={27\over8}n^4+{45\over4}n^3+{105\over8}n^2+{25\over4}n+1.
\end{equation}
and so
\begin{equation}
\bar{i}({\cal P}, n)={27\over8}n^4-{45\over4}n^3+{105\over8}n^2-{25\over4}n+1.
\end{equation}
Now we we use Ehrhart polynomial and formula (\ref{3fold}) to compute the Hodge
numbers: We have $h^{2,1}=h^{1,2}=\bar{i}({\cal P}, 2)=5$ (There is no
contribution from the interior points of five faces of the polytope). To compute
$h^{2,2}$, we need to compute the volume of various polytopes, and we have
\begin{equation}
E=-4!{27\over 8}+4\times 3! {9\over2}-6\times 2! {9\over2}+4\times{3}=-15
\end{equation}
So we find $h^{2,2}=6$. The Coulomb branch spectrum of this theory can be found
in \cite{Xie:2015rpa}, and the rank of Coulomb branch is 5, the number of mass
parameter is $6$. This gives the identification $r=h^{1,2}$ and $f=h^{2,2}$.

Now, let us consider a polynomial map
\(f:\mathbb{C}^{\ast}\times\mathbb{C}^{3}\to\mathbb{C}\) which is nondegenerate
with respect to its Newton polyhedron. If we impose the condition that it
describes the SW geometry of a 4d $\mathcal{N}=2$ theory, then the non-trivial
Hodge numbers are again $h^{1,2}=h^{2,1}$ and $h^{2,2}$, see formula
(\ref{3fold}). The formula for $h^{1,1}$ and $h^{1,2}$ is the same. One thing
that we need to be careful is the formula for $h^{2,2}$
\begin{align}
&h^{1,2}=h^{2,1}=l^*(2\Delta)-\sum_{\Gamma} l^*(\Gamma),~~\nonumber \\
&h^{2,2}=-E-2h^{1,2}=\sum_{I}(-1)^{|I|-1}|I|!V_{|I|}(\Delta_I)-2h^{1,2}
\label{3fold}
\end{align}
We used the fact that $h^3(X)=-E$ for the current case, and the sum in the second equation includes the set of $\mathbb{C}$ variable. So we see that the
formula for $h^{1,2}$ is not changed, but the formula for $h^{2,2}$ is changed,
as we only consider the set $I\subset \{1,2,3\}$ with $x_1, x_2, x_3$
$\mathbb{C}$ variables, and there is a difference one because we only consider primitive cohomology.

\medskip\noindent%
\textit{Example.} Consider the polynomial map given by
\(f=x_1^2+x_2^2+x_3^3+z^9\), where $z$ is a $\mathbb{C}^*$ variable. The
polytope is generated by four lattice points
\begin{equation}
(0,0,0,0),(2,0,0,0),(0,2,0,0),(0,0,3,0),(0,0,0,9)
\end{equation}
and the Ehrhart polynomial is given by
\begin{equation}
i({\cal P}, n)={9\over2}n^4+{27\over2}n^3+{14}n^2+{6}n+1.
\end{equation}
So $l^*(2\Delta)=\bar{i}(P,2)=i(P,-2)=9$ (no contribution), and the following two faces would also contribute to the computation:
\begin{align}
& \Gamma_1:    (0,0,0,0),(2,0,0,0),(0,0,3,0),(0,0,0,9) \nonumber \\
&\Gamma_2: (0,0,0,0),(0,2,0,0),(0,0,3,0),(0,0,0,9)
\end{align}
The corresponding Ehrhart polynomial is $i(\Gamma, n)=9n^3+{27\over2}n^2+{13\over2}n+1$, so we find $l^*(\Gamma_1)=l^*(\Gamma_2)=i(P,-1)=1$.  This gives that $h^{1,2}=9-2=7$.
The Euler number of the hypersurface is
\begin{equation}
E=-4!{9\over2}+3!( 6+9\times 2)-1\times 2! \times {27\over2}-2\times 2!\times 9+9=-18
\end{equation}
so we find $h^{2,2}=4$. This equals the number of mass parameters of the
corresponding four dimensional $\mathcal{N}=2$ model, see \cite{Xie:2012hs}.

\medskip\noindent%
\textbf{Summary.}
From explicit computation, we see that the weight filtration of middle
cohomology $H_{prim}^{n-1}(X)$ have two weights $n-1$ and $n$ ($n=2$ or $n=4$)
\footnote{Here the embedding dimension of hypersurface $X$ is $n$.}. The weight
$n-1$ part has the Hodge decomposition
$\mathrm{Gr}^W_{n-1} H^{n-1}(X)=H^{{n\over 2},{n\over 2}-1}\oplus H^{{n\over2}-1,{n\over2}}$,
while the weight $n$ piece has the decomposition
$\mathrm{Gr}^W_{n}H^{n-1}(X)=H^{{n\over 2},{n\over 2}}$. Based on this fact, we would make the
following statement:

\begin{statement}
  The weight $n-1$ piece of the primitive middle cohomology $H^{n-1}_{prim}(X)$
  gives the electric-magnetic charge, and the weight $n$ part gives the flavor
  charge.
\end{statement}

This result implies that dimension of weight $n$ part of MHS is equal to $2r$
where $r$ is the rank of the theory, and dimension of weight $n+1$ part of Mixed
Hodge Structure is equal to $f$ which is the number of mass parameters. So the
MHS would solve the problem regarding separating the electric-magnetic and
flavor charges.

\medskip\noindent%
\textbf{Constraints on the Newton polyhedra.}
If we require a hypersurface to have above type of MHS, then the Newton
polyhedron for the polynomial $f:U\to \mathbb{C}$ is constrained. The
constraints are as follows.

\begin{itemize}
\item \textit{If the hypersurface is one dimensional, there is no constraint on $U$ and
  $f$.} $U$ can be either $\mathbb{C}\times \mathbb{C}$,
  $\mathbb{C}\times \mathbb{C}^*$, or $\mathbb{C}^*\times \mathbb{C}^*$.
  To describe the Coulomb branch of 4d $\mathcal{N}=2$ theories, we require $U$
  to be $\mathbb{C}\times \mathbb{C}$ or $\mathbb{C}\times \mathbb{C}^*$. The
  case $\mathbb{C}^*\times \mathbb{C}^*$ is relevant for the study of circle
  compactification of 5d $\mathcal{N}=1$ SCFT.
\item If the hypersurface is three dimensional, there are strong constraints on
  $U$ and $f$.
\begin{enumerate}[wide]
\item If $U=\mathbb{C}^4$, the polytope $\Delta$ has to satisfy following
  condition: $l^*(\Delta)=0, l^*(F')=0,l^*(\Gamma')=0$. Here $\Gamma'$
  and $F'$ are the three dimensional face and two dimensional face, where no
  coordinate vanishes identically. In the quasi-homogeneous case, this implies
  that $\sum q_i>1$, where $q_i$ are the weights of the coordinate $x_i$.
\item Now let us consider $U$ consists of one $\mathbb{C}^*$ variable. To ensure
  $h^{3,0}$ to be zero, the polytope $\Delta$ should have no interior point. We
  now also have new constraint from vanishing of $h^{1,0}$ and $h^{2,0}$. Now
  $h^{1,0}$ and $h^{2,0}$ (of compactly supported cohomology) takes following form
  (see formula \ref{honezero}):
  \begin{equation}
    h^{1,0}(Z)=\sum l^*(F^{'})+ \sum_{F\in \Delta_{123}} l^{*}(F),
    \quad h^{2,0}=\sum l^{*}(\Gamma^{'})+l^*(\Delta_{123})
  \end{equation}
  Here $\Delta_{123}$ is the polytope formed from $f$ by setting the
  $\mathbb{C}^*$ variable to be zero. The vanishing of $h^{1,0}(Z)$ then implies
  $l^{*}(F)=0,~F\in \Delta_{123}$, which is a new constraint. Similarly, the vanishing of $h^{2,0}(Z)$
  implies $l^*(\Delta_{123})=0$. For a quasi-homogeneous case, this implies that
  the polynomial $f_{123}$
  (formed from $f$ by setting $\mathbb{C}^*$ variable to be zero):
\begin{equation}
  \sum_{i=1}^3 q_i > 1
\end{equation}
If the sum of the weights are bigger than one, the polynomial $f_{123}$ is
constrained to be ADE singularity.

We get above constraint by requiring the middle cohomology has only weight 3,
and weight 4 pieces. We might further relax the constraints by requiring that
each weight has only one type of Hodge numbers, then we have:
\begin{align}
& h^{3,0}=0,\quad \rightarrow~l^*(\Delta)=0 \nonumber \\
& h^{2,0}=0, \quad \rightarrow~l^*(\Delta_{123})=0
\end{align}
and for the quasi-homogeneous case, this implies that the weights of polynomials
$f_{123}$ should satisfy the following condition
\begin{equation}
\sum_{i=1}^3 q_i\geq 1.
\end{equation}
The case $\sum_{i=1}^3 q_i=1$ is new and they are useful to study the
torus compactification of 6d $(1,0)$ theory. 

\item $U$ consists of maximal two $\mathbb{C}^*$ variables, and the polynomial
  takes the form $f(z_1,z_2)+x^2+y^2$, here $x,y$ are $\mathbb{C}$ variables,
  and $z_1, z_2$ are $\mathbb{C}^*$ variables. This case does not add anything new.
\end{enumerate}
\end{itemize}

\subsection{Abelian variety and mixed Hodge structure }

As we discussed in section~\ref{sec:2}, the low energy photon couplings should
be identified with the complex structure of the abelian variety, and so we
should get an abelian variety from the SW geometry \cite{Donagi:1995cf}. In the
compact curve case, the abelian variety is found from the Jacobian variety of the
curve. Now that we have the open variety, it is not easy to get an abelian
variety from the cohomology of the variety. We now show that we did get an
abelian variety from MHS for the polynomials used to describe a $\mathcal{N}=2$
theory.

Let us briefly review basics about abelian variety. An $n$ dimensional complex
torus is defined as $\mathbb{C}^n/\Lambda$, Here $\Lambda$ is a $2n$ dimensional
lattice. The lattice is generated by $2n$ linearly independent vectors in the vector
space $\mathbb{R}^{2n}$. Using the standard complex structure, we can use a $n\times 2n$
complex matrix $A$ to encode the lattice. By choosing a suitable basis of the
lattice, the matrix for the lattice can be taken to be of form $[I_n|B]$. If the
imaginary part of $B$ is positive definite, the corresponding complex torus is
called an abelian variety.

Abelian varieties arise naturally from the polarized Hodge structure of weight
\(1\). Let \(H=(H_{\mathbb{Z}},H^{1,0},H^{0,1})\) be a Hodge structure of weight \(1\).
Let \(2g\) be the rank of \(H_{\mathbb{Z}}\). Then
\(A=H^{1,0}/H_{\mathbb{Z}}\cong \mathbb{C}^{g}/\mathbb{Z}^{2g}\) is a a complex
torus.

A complex torus \(A\) is an abelian variety if and only
if it can be holomorphically embedded into a complex projective space. In
general, the complex torus \(A\) constructed from a plain Hodge structure of
weight \(1\) is not an abelian variety. A sufficient and necessary condition for
\(A\) to be an abelian variety is that \(H\) is polarizable. If \(S\) is a
polarization on \(H\), it gives rise to a line bundle on \(A\), and the
Hodge--Riemann bilinear relation translates to the positivity of this line
bundle.

The weight one part of the MHS of an open curve is polarizable (the weight one
part comes from compact Riemann surface which carries a polarizable weight one
Hodge structure), we can obtain an abelian variety from it.

Now consider a Hodge structure
\(H=(H_{\mathbb{Z}}, H^{3,0}, H^{2,1}, H^{1,2}, H^{0,3})\) of weight \(3\).
Let again \(2g\) be the rank of \(H_{\mathbb{Z}}\).
Then we can mimic the previous construction and define a complex torus
\(\mathrm{IJ}=(H^{3,0}\oplus H^{2,1})/H_{\mathbb{Z}}\) of dimension \(g\).

This time, even if \(H\) admits a polarization, the Hodge--Riemann bilinear
relation does not ensure that \(\mathrm{IJ}\) is an abelian variety. Because it
only says that \(\sqrt{-1}S(v,\overline{v})\) is positive definite on
\(H^{2,1}\), but is negative definite on \(H^{3,0}\). So the line bundle defined
by \(S\) is not positive.

Fortunately, for the open variety we considered, $H^{3,0}=0$. See the remark at
the end of~\S\ref{sec:3.3}
The positivity provided by the Hodge--Riemann bilinear relation tells us that
\(\mathrm{IJ}=H^{2,1}/H_{\mathbb{Z}}\) is indeed an abelian variety.

\newpage
\section{IR theory: special vacua}\label{sec:4}

In last section, we studied the MHS defined on the cohomology of a typical
fiber of the SW geometry. The typical fiber describes the generic vacuum, and the
MHS is crucial for defining an abelian variety, and useful in separating the
electric-magnetic and flavor charges. In this section, we study special
vacua, which are described by atypical fibers.

Related to the cohomology of atypical fibers are three mixed Hodge structures:
\begin{enumerate}
\item In ``Th\'eorie de Hodge III''~\cite{deligne:hodge3},
  Deligne proves that the cohomology groups of the singular fiber
  $F_\lambda=0$ carries a functorial mixed Hodge structure.
\item As we discussed in last section, the topology of the atypical fiber is
  related to the behavior of the nearby fibers. We have shown that the every
  nearby fiber carries a MHS. When we vary the parameters $\lambda$, this MH{S}
  changes accordingly and we get a ``variations of MHS''. For this variation,
  there is a \textbf{limit} MHS as the parameters approach to atypical fiber,
  due to Schmid~\cite{schmid:variation-hodge-structure} and
  Steenbrink--Zucker~\cite{steenbrink-zucker:vmhs-1}.
  The Hodge filtration is given by the limit Hodge filtration of the smooth
  fibers, and the weight filtration now is given by the nilpotent part of the
  monodromy matrix.
\item When we pushing to the atypical fiber, some elements of the cohomology
  group become zero. These vanishing cycles also carry a limit MHS, due to
  Steenbrink~\cite{steenbrink:mixed-hodge-on-vanishing-cohomology}.

\end{enumerate}
These three MHS are not independent, they form an exact sequence.
We refer the reader to~\cite{kulikov1998mixed} for a comprehensive review.
We shall explain how can one determine the low energy physics at a special vacuum
using these three MHS.

\subsection{MHS for the singular variety}

The middle cohomology group of an open smooth variety carries a MHS as we
reviewed in last section. The weights of the MHS on middle cohomology group
takes the range $n\leq w \leq 2n$ (Here $n$ is the complex dimension of the
variety). For an open singular variety, it is still possible to define a MHS on
it \cite{deligne:hodge3}. This time, the range of weights \(w\) on the middle
cohomology satisfy $0\leq w\leq 2n$. We do not attempt to give a general
description of MHS of singular variety, but content to giving an example to show
how the computation can be reduced to the that of smooth variety reviewed in
last section.

Consider a singular variety \(X\), and assume that the singular
locus \(S\) of \(X\) is smooth. Then we can find a resolution of singularity:
\begin{equation*}
  \begin{tikzcd}
    E \ar[r] \ar[d] & \widetilde{X} \ar[d] \\
    S \ar[r] & X
  \end{tikzcd}
\end{equation*}
where \(E\) is a normal crossing divisor in \(\widetilde{X}\).
Then the cohomology of \(X\) is determined by those of \(\widetilde{X}\), \(S\),
and \(E\), via an exact sequence
\begin{equation*}
\cdots \to H^i(X) \to H^{i}(\widetilde{X}) \oplus H^{i}(S) \to H^{i}(E) \to H^{i+1}(X)\to \cdots
\label{singularexact}
\end{equation*}
As we have explained, the cohomology of a smooth variety always has a MHS,
therefore the mixed Hodge structure of \(H^i(X)\) can be computed by above sequences.

In general, the singular locus \(S\) of \(X\) may again be singular.
Nevertheless, we know the singular locus has lower dimension. Thus we could
inductively determine the MHS of \(S\) using the procedure above, and then
proceed to determine that of \(X\).

\medskip\noindent%
\textit{Example 1.}
Let \(C\) be a possibly singular Riemann surface with \(n\) ordinary double
points. Let \(\widetilde{C}\) be its normalization (resolution). Assume that
\(\widetilde{C}\) is obtained from a \emph{compact} Riemann surface of genus \(g\)
with \(m\) points removed. Then the mixed Hodge structure of \(C\) is summarized
as follows (\(H=H^{1}(C)\)):
\begin{table}
\begin{center}
\begin{tabular}{|c|c|c|}
\hline
\(\mathrm{Gr}_{0}^{W}H\) & \(\mathrm{Gr}_{1}^{W}H\) & \(\mathrm{Gr}_{2}^{W}H\)\\
\hline
\(H^{0,0}\) & \(H^{1,0}\oplus H^{0,1}\) & \(H^{1,1}\)\\ \hline
\(n\) dim. & \(2g\) dim. & \(m-1\) dim.\\
\hline
\end{tabular}
\caption{The MHS of a genus $g$ curve with $n$ double points, and $m$  points removed.}
\label{}
\end{center}
\end{table}
The weight \(2\) terms of the mixed Hodge structure are from the two punctures,
whereas the weight zero part of the cohomology are from the nodal points.
\begin{figure}[H]
\begin{center}
\includegraphics[width=.9\linewidth]{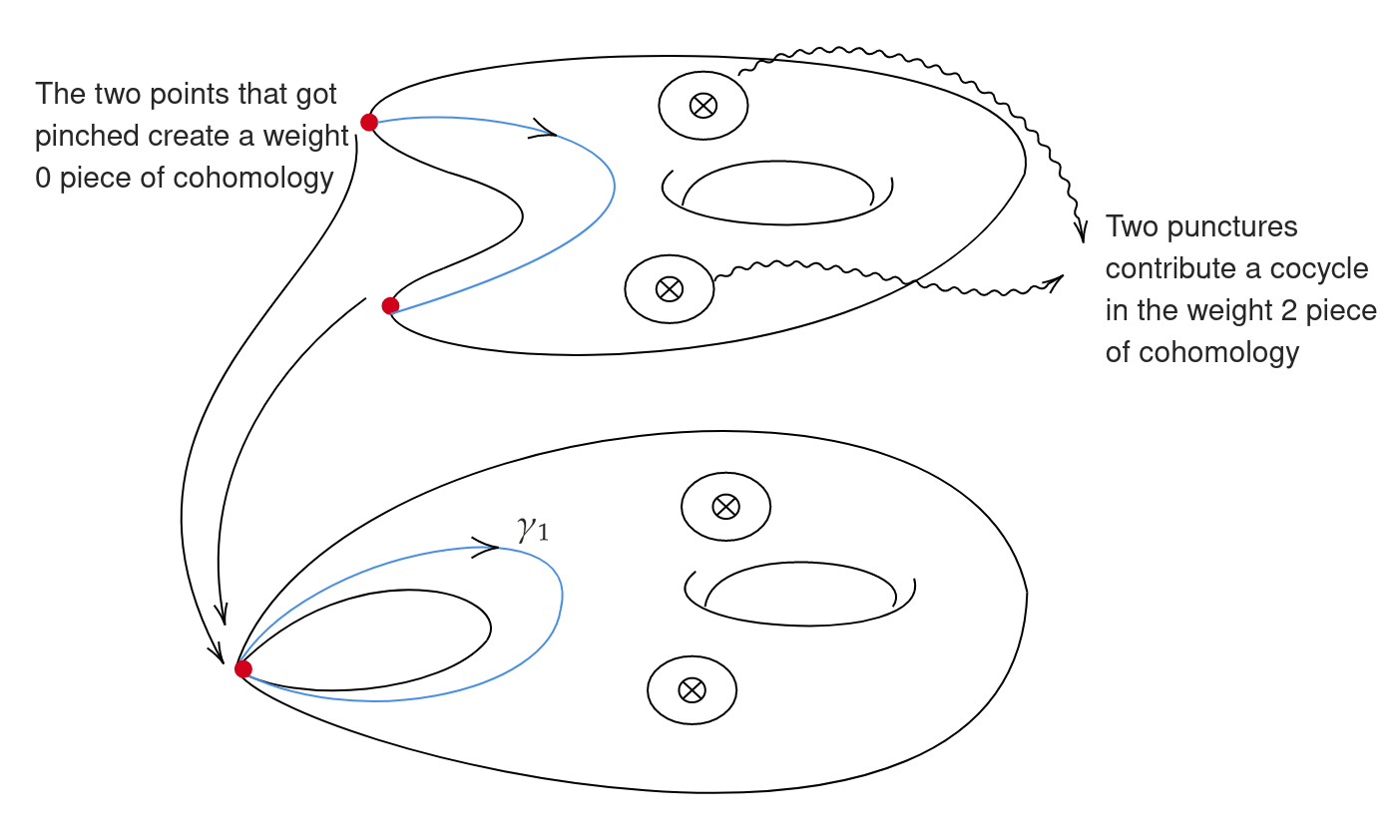}
\end{center}
\caption{The MHS on a $H^1(C)$, here $C$ is a curve with derived from removing $m$ points of a genus $g$ compact Riemann surface with a double point. The weight one part of $H^1(C)$ comes from
the cycles of the original compact Riemann surface, and weight two comes from the cycles around the removed points. Weight zero part $(\gamma_1^*)$ comes from the cycles around the double point.}
\label{singular}
\end{figure}

\medskip\noindent%
\textit{Example 2.}
Let $C$ be a singular curve defined by the affine equation
$x^3+y^3+y^4=0$. This curve has a singular point at the origin $S=(0,0)$. The
singularity of \(S\) is equivalent to that of $x^3+y^3=0$ at the origin.
The singularity can be resolved by blowing up the \((x,y)\)-plane at the point
\(S\). Then  \(E\) consists of $3$ points. The MHS of $\tilde{C}$ can be computed
using the method given in last section (one first compute the genus of
$\tilde{C}$ using the Pluker formua, the weight two part is found by doing
blow-up of the singular point at infinity), and we have $g=0$. We have following
exact sequence (see formula \ref{singularexact}):
\begin{align*}
&0\to H^0(C) \to H^{0}(\widetilde{C}) \oplus H^{0}(S) \to H^{0}(E) \to \boxed{H^{1}(C)}\to H^1(\tilde{C})\oplus H^1(S) \to H^1(E) \nonumber\\
  &\to H^2(C)\to H^2(\tilde{C})\oplus H^2(S)\to 0
\end{align*}
From which we find that $h^{1,0}=h^{0,1}=0$, and $h^{0,0}=2$.

\medskip\noindent%
\textbf{Physical interpretation.} Now we explain how to extract  the low energy
physics from the MHS associated with the singular variety. Let us
assume the SW geometries are curves, and the general cases are similar.
The MHS for $H^1(C)$ has weight zero, weight one, and weight two parts.
The weight one part gives an abelian gauge theory; the weight two part gives the
flavor central charge. The weight zero part is new: they are interpreted as
gauging the $U(1)$ flavor symmetry for the interacting theory associated with
the singular points. In next subsection, we will discuss how to find the
interacting theory from the so-called limit MHS.

\subsection{Limit Mixed Hodge Structure}

We have discussed the MHS on the cohomology of the singular fiber. Its weight
one piece was interpreted as the abelian gauge theory part of the special
vacuum. However, the singular points of the variety should be able to describe a
more complicated part of the theory, i.e., a SCFT or an IR free non-abelian
gauge theory. We argue that the limit mixed Hodge structure will allow us to
extract this more intricate piece of information.

\medskip\noindent%
\textbf{Variation of (mixed) Hodge structure}:
Let us first describe briefly variation of (mixed) Hodge structures.
Let \(f: U \to B\) be a map of algebraic varieties which is a locally
topologically trivial fibration. This is the case for example when \(f\) is
proper and is a submersion of manifolds (Ehresmann's lemma). In this case, the
cohomology \(H^m(U_b)\) of the fibers of \(f\) form a local system
\(\mathbb{V}\) on \(B\). Let
\(\mathcal{V}=\mathbb{V}\otimes_{\mathbb{Q}}\mathcal{O}_B\) be the associated
vector bundle on \(B\). Thus \(\mathcal{V}\) is equipped with a flat connection
\(\nabla\) known as the Gauss--Manin connection.
The following are known.
\begin{itemize}
\item The fiber-by-fiber weight filtrations on \(\mathbb{V}\) are horizontal,
  i.e., they form a sub-local system \(W_{\bullet}\) of \(\mathbb{V}\).
\item The Hodge filtration \(F^{\bullet}\mathcal{V}\) are holomorphic subbundles
  of \(\mathcal{V}\), but they are not horizontal (i.e., \(\nabla\) does not
  respect \(F^{\bullet}\mathcal{V}\) in general).
\item Instead, the so-called ``infinitesimal period relation'' holds: for any
  tangent vector field \(v\) on \(B\)
  \begin{equation*}
    \nabla_{v} F^{i}\mathcal{V} \subset F^{i-1}\mathcal{V}.
  \end{equation*}
\end{itemize}
It is therefore not difficult to make an abstract version of the above and
define a variation of MHS. A \textbf{variation} of mixed Hodge
structure on a complex manifold \(B\) then consists of
\begin{itemize}
\item a local system \(\mathbb{V}\) of \(\mathbb{Q}\)-vector spaces, whose
  associated vector bundle with connection is denoted by \((\mathcal{V},\nabla)\);
\item sub-local systems \(\mathbb{W}_{\bullet}\)
  such that \(\mathbb{W}_i \subset \mathbb{W}_{i+1}\); and
\item holomorphic subbundles \(F^{\bullet}\mathcal{V} \subset \mathcal{V}\).
\end{itemize}
These data satisfy the following requirements:
\begin{itemize}
\item the infinitesimal period relation holds:
  \(\nabla_{v}F^{i}\mathcal{V}\subset F^{i-1}\mathcal{V}\);
\item for each \(b \in B\), the triple
  \((\mathbb{V}_b,\mathbb{W}_{k,b},F_{k,b})\) is a mixed Hodge structure.
\end{itemize}
If in a variation of mixed Hodge structure the fiberwise mixed Hodge structures
are pure, we then say this is a variation of pure Hodge structure, or simply
variation of Hodge structure. A \emph{polarization} of a variation of Hodge
structure \((\mathbb{V},F^{\bullet})\) on a complex manifold \(B\) is a
horizontal map
\begin{equation*}
S: \mathbb{V} \otimes \mathbb{V} \to \mathbb{C}_B
\end{equation*}
that induces a polarization on each fiber. If \((\mathbb{V},\mathbb{W},F)\) is a
variation of mixed Hodge structure, such that \((\mathrm{Gr}^{W}_k, F)\) admits
a polarization, then we say \((\mathbb{V},\mathbb{W},F)\) is
\emph{graded polarizable}.

For example, if \(f: X \to B\) is a smooth projective map of algebraic varieties
the cohomology bundle \(R^{m}f_{\ast}\mathbb{Q}\) then underlies a variation of
\emph{pure} Hodge structure which admits a polarization.

If \(f:X \to S\) is a morphism of complex algebraic varieties, then there exists
a Zariski open subset \(U \subset S\) such that \(R^{m}f_{\ast}\mathbb{Q}\)
underlies a variation of mixed Hodge structure. If \(f\) is quasi-projective,
then \(R^{m}f_{\ast}\mathbb{Q}\) underlies a graded polarizable variation of
mixed Hodge structure.

If we have a variation of polarized Hodge structure on a punctured disk $\Delta^*$,
then a theorem of Schmid~\cite{schmid1973variation}
says that the Hodge filtration (holomorphic subbundles)
can be extended to the origin. This limit filtration, together with a monodromy
weight filtration (see below), form a MH{S} called the limit MHS. The limit Hodge filtration
comes from the Hodge structure of the nearby fibers, while the limit weight
filtration comes from the monodromy operator. This result is extended to a nice
class of variations of mixed Hodge structure by
Kashiwara~\cite{kashiwara:a-study-vmhs} and
Steenbrink--Zucker~\cite{steenbrink-zucker:vmhs-1}.
In the situation that interests us, we should really use the theory of
Kashiwara--Steenbrink--Zucker. However in this paper we only need to perform
computations about Hodge numbers, so we confine ourselves to review the result
of Schmid only.

\medskip\noindent%
\textbf{Monodromy theorem.} In order to define the monodromy weight filtration
of the limit MHS, we need some knowledge about the monodromy action.
Suppose we have a continuous map of
topological spaces
\begin{equation*}
\pi: X \to \Delta.
\end{equation*}
Denote by \(X_{t}\) the fiber \(\pi^{-1}(t)\).
Assume that \(\pi|_{X - X_0}: X \setminus X_0 \to \Delta \setminus \{0\}\) is a locally
topologically trivial fibration.
The cohomology groups \(H^i(X_t)\), with varying \(t\), patch together and form
a ``cohomology bundle'' \(\mathbf{V}\) on \(\Delta\setminus\{0\}\) (thus the fiber of
\(\mathbf{V}\) at \(t\) is \(H^i(X_t)\)).

For a generic element \(\eta\in\Delta\setminus\{0\}\), we can draw a circle \(\xi\)
based on \(\eta\) and contains \(0\in \Delta\) in its interior. Then parallel
transport along the path \(\eta\) defines an automorphism
\begin{equation*}
h: H^i(X_{\eta}) \to H^i(X_{\eta})
\end{equation*}
called the \emph{monodromy transform} or \emph{Picard--Lefschtz transform}.
Thus, to describe the cohomology bundle we only need two pieces of information:
a vector space \(V\) (which is the vector space that underlies the various
\(H^i(X_t)\)) and a linear operator \(h\) (which prescribes the monodromy
operator).

A typical example is demonstrated in the following diagram. In this diagram, we
have a family of tori fibered over the punctured disk. When approaching to zero, a cycle
\(\delta\) is shrinking to zero. Going around the circle, the cycle \(\gamma\)
is transformed via the monodromy transform to \(\gamma+\delta\). This picture is
known in topology as the ``Dehn twist''.
\begin{center}
  \includegraphics[width=.9\linewidth]{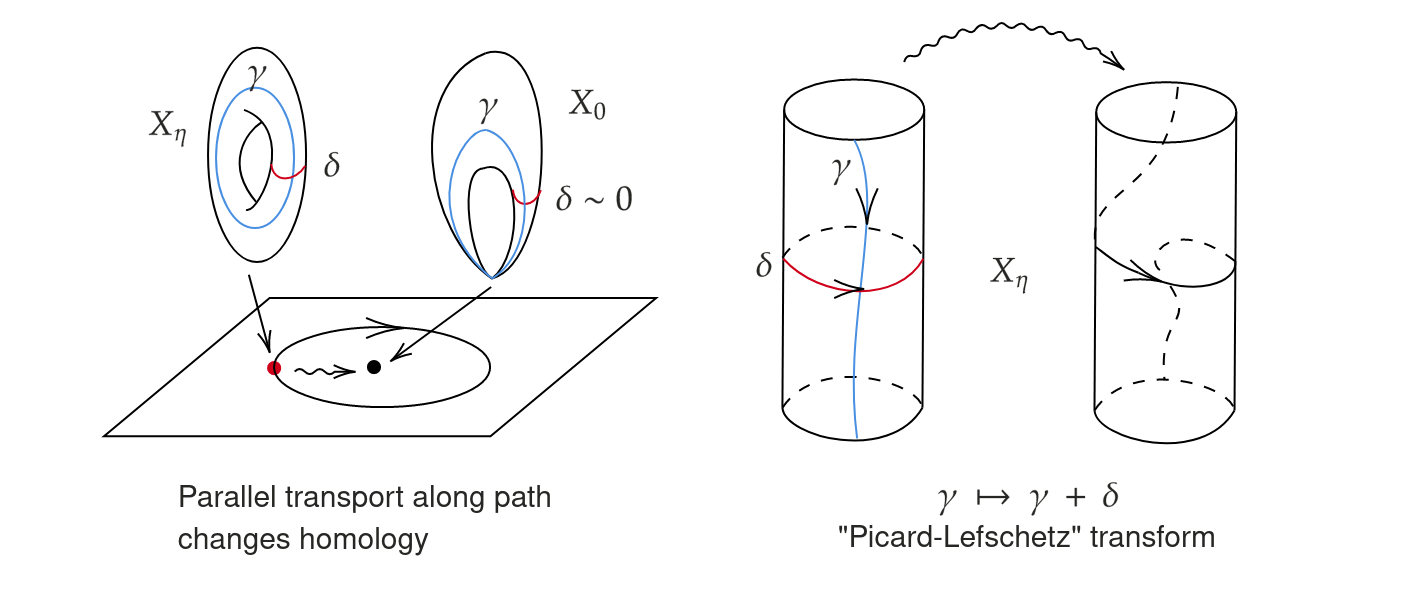}
\end{center}
In this case, the monodromy group action is expressed in terms of famous
Picard-Lefschtz formula:
\begin{equation}
h(\sigma)=\sigma+(\delta,\sigma)\delta;
\end{equation}
Here $\sigma$ is an arbitrary cycle, $\delta$ is the vanishing cycle, and
$(\sigma, \delta)$ is the intersection number of these two cycles.

If \(\pi\colon X \to \Delta\) is a degenerating family of algebraic varieties,
then we have the famous monodromy theorem: the monodromy operator $h$ is
quasi-unipotent:
\begin{equation}
(h^q-1)^{n+1}=0
\end{equation}
Here $n$ is the dimension of the variety. The monodromy theorem implies that
\begin{enumerate}
\item the eigenvalues of the monodromy operator are roots of unity,
\item the size of the Jordan block of $T$ is less than $n+1$.
\end{enumerate}
In the above setting, we discussed the monodromy action $h$ acting on the
homology group. If we look at the cohomology group, the monodromy group is given
by $h^*=(h^T)^{-1}$.

\subsection*{The limit Hodge filtration} %

Now assume that we have a variation of polarized Hodge structure (with
weight $\omega$) over a punctured disk $\Delta^*$ with coordinate $t$. So we
have a local system $\mathbf{V}$ together with a monodromy operator $h$
(satisfying monodromy theorem), holomorphic sub-bundles $F^{\bullet}(t)$, and
polarization $S(t)$. Denote by
\begin{align*}
  p: \mathfrak{h} &= \{z \in \mathbb{C} : \mathrm{Im}(z) > 0\}
                    \to \Delta^*, \\
  z &\mapsto t = \exp(2\pi\mathrm{i}z)
\end{align*}
the universal covering of $\Delta^*$ by the upper half plane. Since
\(\mathfrak{h}\) is simply connected, the pullback
bundle $p^{-1}\mathbf{V}$ is a trivial local system on $\mathfrak{h}$. Fix
a trivialization of $p^{-1}\mathbf{V}$ on $\mathfrak{h}$ and denote by $V$
the complex vector space represents a typical fiber of $\mathbf{V}$. The space
$V$ has the following structures.

\begin{enumerate}[wide]
\item
  \label{point:polarization-sesquilinear}
  \textbf{Polarization.}
  The polarization $S$ induces a sesquilinear form on the complex vector
  space $V$, which we still denote by $S$.
\item \textbf{Monodromy}.
  Since the space $\mathfrak{h}$ endows an automorphism $z \mapsto z + 1$
  which generates the group of deck transformations of $p$, the monodromy action
  defines an endomorphism $T$, called the \emph{monodromy operator}, on $V$.
  The operator $T$ preserves the polarization.
\item  \textbf{Hodge Filtrations}.
  The Hodge filtration $F^\bullet$ on $\mathcal{V}$ induces a holomorphically
  varying family of flags on $V$ which we shall denote by $F^\bullet(z)$. Then by
  the definition of $T$ we have $hF^\bullet(z) = F^\bullet(z+1)$.
\end{enumerate}

The monodromy theorem implies that $h$ can be written as
\begin{equation}
  h = T_s \cdot \exp(2\pi\mathrm{i}N)
\end{equation}
of a semisimple matrix whose entries are all of length $1$, and the exponential
of a nilpotent matrix, in a way that $[T_s,N] = 0$. We can decompose the space
$V$ into the sum of the eigenspaces of $T_s$:
\begin{equation}
  \label{eq:1}
  V = \bigoplus_{|\lambda|=1} E_\lambda(T_s)
\end{equation}
where the nilpotent operator $N$ preserves this decomposition. In particular,
the monodromy invariant part (invariant vector) of $V$ equals $\mathrm{Ker}(N|_{E_1(T_s)})$, and
the monodromy coinvariant part of $V$ equals $\mathrm{Coker}(N|_{E_1(T_s)})$.
It can be shown that the decomposition~\eqref{eq:1}
is orthogonal with respect to the polarization form $S$.

{Schmid} studied the behavior of the the filtrations $F^\bullet(z)$
when $\mathrm{Im}(z)$ turns to $\infty$. He proved the the limit of a suitable
twist (see~\eqref{eq:nilpotent-orbit} below) of $F^\bullet(z)$ exists. We write
\begin{equation}
\label{eq:nilpotent-orbit}
F^\bullet = \lim_{\mathrm{Im}(z) \to \infty} \exp(-2\pi\mathrm{i}zN) \cdot F^\bullet(z).
\end{equation}
for the limit of the Hodge filtrations. This is called the
\textit{limit Hodge filtration}. Note that one can easily verify, using the fact
$hF^\bullet(z)=F^\bullet(z+1)$, that we have
\begin{equation}
\label{eq:descend-flag}
\exp(2\pi\mathrm{i}(z+1)N)F^\bullet(z+1) = T_s \exp(2\pi\mathrm{i}zN)F^\bullet(z).
\end{equation}
Hence the limit Hodge filtration is thus preserved by the semisimple part $T_s$
of the monodromy operator:
\begin{equation}
h\cdot F = F,
\end{equation}
hence the limit Hodge filtration is compatible with the
decomposition~\eqref{eq:1}.

The vector space $V$ now has been endowed with

(a) a polarization $S$,

(b) a monodromy operator $T = T_s \exp(2\pi\mathrm{i}N)$,

(c) an $S$-orthogonal decomposition~\eqref{eq:1}, and

(d) a limit Hodge filtration $F^\bullet$.

Deligne observed that the presence of the nilpotent operator $N$ also puts
another increasing filtration $M_\bullet$, the
\emph{monodromy weight filtration}, on the space $V$. Assume $N^{k+1} = 0$, then
this filtration satisfies the following properties:

\begin{enumerate}
\item $N(M_\ell) \subset M_{\ell+2}$,
\item $M_\ell = 0$ if $|\ell|> k$,
\item $N^\ell : \mathrm{gr}^M_{\ell}(V) \to \mathrm{gr}^M_{-\ell}(V)$ is an
  isomorphism.
\end{enumerate}

One defines the \emph{primitive part} of $\mathrm{gr}^M_\ell(V)$ to be the
kernel of $N^{\ell+1}$. Then the usual $\mathfrak{sl}_2$-argument decomposes the
spaces $\mathrm{gr}^M_\ell(V)$ into the sum of primitive parts.

Deligne conjectured that the filtration $M$, plus the limit Hodge filtration
$F$, defines a mixed Hodge structure on the vector space $V$. This is proved by
Schmid \cite{schmid1973variation}. Precisely, we have

\medskip\noindent%
\textbf{Schmid's theorem.}
\textit{The data $(V,M_\bullet,F^\bullet,N,S)$ is a graded polarized mixed Hodge
structure on $V$, in the sense that
\begin{equation*}
(\mathrm{gr}^M_\ell(V){}_{\textup{prim}},F,S(\cdot,N^\ell\cdot))
\end{equation*}
is a polarized Hodge structure of weight $w+\ell$.}

\medskip%
This mixed Hodge structure is called the \emph{limit mixed Hodge structure} of
$\mathcal{V}$. A very useful interpretation is following: Deligne defines a
canonical extension of the holomorphic vector bundle on $\Delta^*$ to the whole
disk $\Delta$, and the limit MHS can be thought of as a natural mixed Hodge structure on the
fiber of canonical extension at the origin of the disk.

\medskip\noindent%
\textit{Example.}
Let us give an example of limit MHS coming from degeneration
of Riemann surfaces of genus \(2\). For instance, such a degeneration is given
by the projective completion of
\begin{equation*}
X_{t} : y^2  = x(x-1)(x-2)(x-3)(x-t)
\end{equation*}
where \(t\) is a number sufficiently close to \(0\). In this example, the
monodromy operation on \(X_t\) is already unipotent. To determine the Hodge
filtration on cohomology of nearby cycles, we first fix a homology basis
\(\gamma_1,\gamma_2,\gamma_3,\gamma_4\) of the curve \(X_t\). Then the Hodge
filtration on \(F_{t}\) is determined by the period matrix \(Z(t)\) (i.e., the
matrix whose column vectors are the integration of the two linearly independent
abelian differentials on \(\gamma_i\)). The multivalued matrix function \(Z(t)\)
should be thought of as a single valued function from \(\mathfrak{h}\) into
period domain. A standard trick of ``untwisting'' gives rise to a different
matrix valued function:
\begin{equation*}
\widehat{Z}(t) = \exp(-2\pi\sqrt{-1}tN) Z(t)
\end{equation*}
which does not land in the period domain, but is single valued. At any rate,
Schmid's nilpotent orbit theorem states that \(\widehat{Z}(0)\) exists in a
suitable flag space containing the period domain. This gives the limit Hodge
filtration.

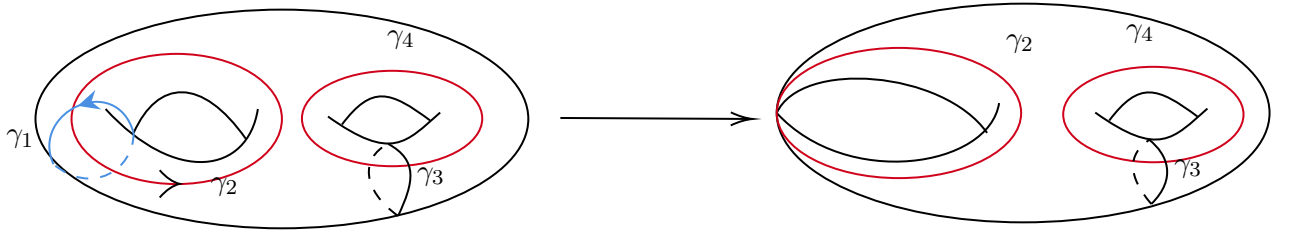
\begin{figure}
\begin{center}

\tikzset{every picture/.style={line width=0.75pt}} %set default line width to 0.75pt

\begin{tikzpicture}[x=0.75pt,y=0.75pt,yscale=-1,xscale=1]
%uncomment if require: \path (0,328); %set diagram left start at 0, and has height of 328

%Shape: Ellipse [id:dp8712561559344043]
\draw   (35,112.5) .. controls (35,82.12) and (90.07,57.5) .. (158,57.5) .. controls (225.93,57.5) and (281,82.12) .. (281,112.5) .. controls (281,142.88) and (225.93,167.5) .. (158,167.5) .. controls (90.07,167.5) and (35,142.88) .. (35,112.5) -- cycle ;
%Straight Lines [id:da4312036780177493]
\draw    (297,112) -- (391,112.49) ;
\draw [shift={(393,112.5)}, rotate = 180.3] [color={rgb, 255:red, 0; green, 0; blue, 0 }  ][line width=0.75]    (10.93,-3.29) .. controls (6.95,-1.4) and (3.31,-0.3) .. (0,0) .. controls (3.31,0.3) and (6.95,1.4) .. (10.93,3.29)   ;
%Curve Lines [id:da538027067237288]
\draw    (188,115.27) .. controls (203,95.27) and (212,98.27) .. (232,113.27) ;
%Curve Lines [id:da9510525038646791]
\draw    (181,111.27) .. controls (213,134.27) and (220,124.27) .. (237,109.27) ;
%Curve Lines [id:da02491283581481385]
\draw    (84,119.5) .. controls (96,92.5) and (118,91) .. (140,122.5) ;
%Curve Lines [id:da5013348713125849]
\draw    (70,107.5) .. controls (105,146) and (141,140) .. (146,107.5) ;
%Shape: Ellipse [id:dp7285749862475819]
\draw  [color={rgb, 255:red, 208; green, 2; blue, 27 }  ,draw opacity=1 ] (53,112.5) .. controls (53,94.41) and (76.51,79.75) .. (105.5,79.75) .. controls (134.49,79.75) and (158,94.41) .. (158,112.5) .. controls (158,130.59) and (134.49,145.25) .. (105.5,145.25) .. controls (76.51,145.25) and (53,130.59) .. (53,112.5) -- cycle ;
%Curve Lines [id:da292999058446604]
\draw [color={rgb, 255:red, 74; green, 144; blue, 226 }  ,draw opacity=1 ]   (84,119.5) .. controls (75,88.27) and (32,109.27) .. (44,135.27) ;
\draw [shift={(56.12,105.92)}, rotate = 343.16999999999996] [fill={rgb, 255:red, 74; green, 144; blue, 226 }  ,fill opacity=1 ][line width=0.08]  [draw opacity=0] (10.72,-5.15) -- (0,0) -- (10.72,5.15) -- (7.12,0) -- cycle    ;
%Curve Lines [id:da36356803675456084]
\draw [color={rgb, 255:red, 74; green, 144; blue, 226 }  ,draw opacity=1 ] [dash pattern={on 4.5pt off 4.5pt}]  (84,119.5) .. controls (84,138.27) and (63,151.27) .. (44,135.27) ;
%Shape: Ellipse [id:dp1367210859024104]
\draw   (405,109.5) .. controls (405,79.12) and (460.07,54.5) .. (528,54.5) .. controls (595.93,54.5) and (651,79.12) .. (651,109.5) .. controls (651,139.88) and (595.93,164.5) .. (528,164.5) .. controls (460.07,164.5) and (405,139.88) .. (405,109.5) -- cycle ;
%Curve Lines [id:da47796388304266113]
\draw    (405,109.5) .. controls (417,82.5) and (488,88) .. (510,119.5) ;
%Curve Lines [id:da0006614588739335403]
\draw    (405,109.5) .. controls (440,148) and (511,137) .. (516,104.5) ;
%Shape: Ellipse [id:dp7293873730988203]
\draw  [color={rgb, 255:red, 208; green, 2; blue, 27 }  ,draw opacity=1 ] (405,109.5) .. controls (405,91.41) and (432.31,76.75) .. (466,76.75) .. controls (499.69,76.75) and (527,91.41) .. (527,109.5) .. controls (527,127.59) and (499.69,142.25) .. (466,142.25) .. controls (432.31,142.25) and (405,127.59) .. (405,109.5) -- cycle ;
\draw   (97,138.27) .. controls (100.33,142.08) and (103.67,144.37) .. (107,145.13) .. controls (103.67,145.9) and (100.33,148.18) .. (97,152) ;
%Curve Lines [id:da33914405710544204]
\draw    (571,113.27) .. controls (586,93.27) and (595,96.27) .. (615,111.27) ;
%Curve Lines [id:da9820730972962666]
\draw    (564,109.27) .. controls (596,132.27) and (603,122.27) .. (620,107.27) ;
%Shape: Ellipse [id:dp052731388059593876]
\draw  [color={rgb, 255:red, 208; green, 2; blue, 27 }  ,draw opacity=1 ] (168,112.27) .. controls (168,99.01) and (188.15,88.27) .. (213,88.27) .. controls (237.85,88.27) and (258,99.01) .. (258,112.27) .. controls (258,125.52) and (237.85,136.27) .. (213,136.27) .. controls (188.15,136.27) and (168,125.52) .. (168,112.27) -- cycle ;
%Curve Lines [id:da1680934672738459]
\draw    (211,125) .. controls (226,133.27) and (224,144.27) .. (216,161.27) ;
%Curve Lines [id:da9189185095603905]
\draw  [dash pattern={on 4.5pt off 4.5pt}]  (216,161.27) .. controls (198,148.27) and (197,132.27) .. (211,125) ;
%Shape: Ellipse [id:dp7437307766513432]
\draw  [color={rgb, 255:red, 208; green, 2; blue, 27 }  ,draw opacity=1 ] (548,110.27) .. controls (548,97.01) and (568.15,86.27) .. (593,86.27) .. controls (617.85,86.27) and (638,97.01) .. (638,110.27) .. controls (638,123.52) and (617.85,134.27) .. (593,134.27) .. controls (568.15,134.27) and (548,123.52) .. (548,110.27) -- cycle ;
%Curve Lines [id:da42626551311787253]
\draw    (591,123) .. controls (600,130.27) and (605,142.27) .. (592,155.27) ;
%Curve Lines [id:da8426469586797853]
\draw  [dash pattern={on 4.5pt off 4.5pt}]  (592,155.27) .. controls (580,141.27) and (580,130.27) .. (591,123) ;

% Text Node
\draw (121,140.4) node [anchor=north west][inner sep=0.75pt]    {$\gamma _{2}$};
% Text Node
\draw (19,116.4) node [anchor=north west][inner sep=0.75pt]    {$\gamma _{1}$};
% Text Node
\draw (518,68.4) node [anchor=north west][inner sep=0.75pt]    {$\gamma _{2}$};
% Text Node
\draw (224,135.4) node [anchor=north west][inner sep=0.75pt]    {$\gamma _{3}$};
% Text Node
\draw (209,66.4) node [anchor=north west][inner sep=0.75pt]    {$\gamma _{4}$};
% Text Node
\draw (602,131.4) node [anchor=north west][inner sep=0.75pt]    {$\gamma _{3}$};
% Text Node
\draw (578,63.4) node [anchor=north west][inner sep=0.75pt]    {$\gamma _{4}$};

\end{tikzpicture}

\end{center}
\caption{The degeneration of a genus two Riemann surface. Here $\gamma_1$ is a vanishing cycle. }
\label{limitmhs}
\end{figure}

The monodromy weight filtration is a purely topological quantity. In the present
case, the degeneration is caused by the vanishing cycle $\gamma_1$, see figure.
\ref{limitmhs} (the intersection number is given as $(\gamma_1, \gamma_2)=-1$).
And the monodromy action on the \textbf{cohomology} group is given by the
Picard--Lefschetz formula:
\begin{equation}
T=\left[\begin{array}{cccc}
1&0&0&0\\
1&1&0&0\\
0&0&1&0\\
0&0&0&1\\
\end{array}\right]
\end{equation}
Write  \(T=(\exp(2\pi\sqrt{-1}N)\), and $N$ takes the following form
\begin{equation}
N=\left[\begin{array}{cccc}
0&0&0&0\\
1&0&0&0\\
0&0&0&0\\
0&0&0&0\\
\end{array}\right]
\end{equation}
The weight filtration is then given by (here $\gamma_i^*$ is the dual basis in the cohomology groups):
\begin{equation*}
W_0 =\{ \gamma_2^* \}, \quad W_1 = \mathrm{Ker}(N)=\{ \gamma_2^*, \gamma_3^*, \gamma_4^* \}, \quad W_2=\{\gamma_1^*, \gamma_2^*, \gamma_3^*, \gamma_4^* \}
\end{equation*}
The monodromy invariant part of the cohomology is generated by the basis $\{\gamma_2^*, \gamma_3^*, \gamma_4^*\}$, and is the same as the cohomology of the singular
variety.

\subsection{Limit MHS on vanishing cohomology}

We have recalled two MHS associated with singular fiber: the MHS associated with
singular variety itself, and the limit MHS using the variation of Hodge
structures of nearby fibers. Here we will introduce a third MHS associated with
singular fiber: the MHS on vanishing cohomology. We refer the reader
to~\cite{kulikov1998mixed} for a more detailed introduction.

We assume that the atypical fiber has at worst isolated singularities.
A feature of the vanishing cohomology is that it is concerns only the local behavior of the
polynomial map near its critical points.
Thus we only need to consider the function germ modeling the local behavior.
Let \(f: (\mathbb{C}^{n+1},0) \to (\mathbb{C},0)\) be a germ of a holomorphic
function. Let
\(\partial_{i}f=\partial f/\partial x_i\). Define the Jacobian algebra to be
\begin{equation*}
  J_{f}=\frac{\mathbb{C}[\![x_0,\ldots,x_n]\!]}{(\partial_0 f, \cdots,\partial_n f)}
  =\frac{\mathbb{C}\{x_0,\ldots,x_n\}}{(\partial_0 f, \cdots,\partial_n f)}.
\end{equation*}
We say \(0\) is an isolated critical point of \(f\) if \(J_{f}\) is a finite
dimensional \(\mathbb{C}\)-vector space.

It is known \cite{milnor:singular-points-complex-hypersurfaces}
that there is a neighborhood \(X\) of \(0\) in
\(\mathbb{C}^{n}\) and a small neighborhood
\(S=\{t\in\mathbb{C}:|t|<\epsilon\}\) of \(0 \in \mathbb{C}\) such that \(f\)
defines a holomorphic map \(f:X \to S\) such that
\begin{itemize}
\item \(X - f^{-1}(0) \to S-\{0\}\) is a locally topologically trivial
  fibration,
\item \(f^{-1}(0)\) is a strong deformation retract of \(X\), and
\item \(X\) is homeomorphic to a cone over \(f^{-1}(t)\)  .
\end{itemize}
We shall refer to \(f:X \to S\) as a \emph{good representative} of \(f\).
Moreover, when \(f\) has an isolated critical point at \(0\), Milnor proved that
for \(t \in S -\{0\}\), \(f^{-1}(t)\) is homotopy equivalent to a bouquet of
\(n\)-dimensional spheres. Therefore,
\begin{equation*}
H^{n}(f^{-1}(t),\mathbb{Z}) \cong \mathbb{Z}^{\mu}, \quad \forall t \in S-\{0\},
\end{equation*}
and we shall refer to this cohomology group as the \emph{vanishing cohomology}
of the germ \(f\). It is a theorem of Milnor that \(\mu=\dim_{\mathbb{C}}J_{f}\).

One can define a MHS on the vanishing
cohomology of a germ of a holomorphic function
with isolated critical point, see \cite{steenbrink1976mixed}.
In the following, we will give an algebraic way of computing MHS for the
vanishing cohomology, due to Varchenko, Morihiko Saito, Scherk and
Steenbrink~\cite{scherk1985mixed}.

\medskip\noindent%
\textbf{Gauss--Manin system.} First, we would like to describe cohomological
bundles. Given a good representative \(f: X \to S\), the sheaf
\(R^{n}f_{\ast}\mathbb{C}\) capturing the cohomology of fibers of \(f\) admits
an equivalent description in terms of D-modules. D-modules refer to the (sheaves
of) modules of the ring of differential operators defined on $S$. Denote the local
coordinate on $S$ as $t$, and the derivative as $\partial_t$, the ring of
differential operators is generated by $t$ and $\partial_t$ with relation
$[\partial_t, t]=1$.

The vanishing cohomology groups $H^{n}(f^{-1}(t),\mathbb{C})$ can be described
by a D-module, known as the ``Gauss--Manin system'' of \(f\), denoted
by \(M\). It is defined as follows. First we consider the ``twisted'' de~Rham
complex
\begin{equation*}
  (\Omega^{\bullet}_X[\tau], \mathrm{d} + \tau\mathrm{d}f\wedge),
\end{equation*}
where $\tau$ is a formal parameter, and $\Omega^{m}_X$ is the sheaf of
holomorphic \(m\)-forms on $X$.
We define an action of \(\partial_{t}\) on  complex $\Omega^{\bullet}_X[\tau]$ by
\begin{equation*}
  \partial_{t}(\tau^{m} \omega) = \tau^{m+1}\omega,
\end{equation*}
and define an action of \(\mathcal{O}_{S}\) by
\begin{equation*}
t\cdot ( \tau^{m}\omega ) = \tau^{m}f\omega -m\tau^{m-1}\omega.
\end{equation*}
One checks that the actions of \(t\) and \(\partial_{t}\) commutes with the
differential \(d+\tau \mathrm{d}f\wedge\). Let \(M\) be the
\((n+1)\)\textsuperscript{th} cohomology of
this complex. Then \(M\) has an action of \(\partial_{t}\) and \(t\), thereby
obtaining a structure of D-module on \(S\). We call \(M\) the
\emph{Gauss--Manin system} of the \(f\). An important property of $\partial_t$
action in the current case is

\medskip\noindent%
\textbf{Theorem (Brieskorn~\cite{brieskorn:monodromy}).}
\textit{The map \(\partial_{t}: M \to M\) is an isomorphism (as a map between sheaves of \textbf{abelian groups}).}

\medskip%
This implies that $\partial_t$ is invertible on $M$, and the inverse is
denoted as $\partial_t^{-1}$. The upshot of this construction is that the
D-module \(M\) gives an ``analytic'' way to visualize the cohomology bundle. For
instance, the information of the monodromy action can be decoded from \(M\) via
the actions of \(t\) and \(\partial_{t}\).
Malgrange~\cite{malgrange:asymptotic-integrals-and-monodromy}
proves that \(M\) has \(0 \in S\) as a regular singular point.
More or less equivalent to this statement
is that \(M\) admits a ``root decomposition''
\begin{equation*}
  M = \mathop{\widehat{\bigoplus}}_{\alpha \in \mathbb{Q}} M^{\alpha}
\end{equation*}
where
\begin{equation*}
  \boxed{
    M^{\alpha} = \bigcup_{i=1}^{\infty} \mathrm{Ker}(t\partial_{t} - \alpha)^{i}
  }
\end{equation*}
and the hat above the direct sum sign means that \(M\) is contained in
\(\prod_{\alpha\in\mathbb{Q}} M^{\alpha}\) is generated by the direct sum
\(\bigoplus M^{\alpha}\) as a \(\mathbb{C}\{t\}\)-submodule. Moreover, for each
\(\alpha \in \mathbb{Q}\), the root space \(M^{\alpha}\) is isomorphic to the
generalized eigenspace of \(T\) on the Milnor cohomology with eigenvalue
\(\exp(-2\pi\sqrt{-1}\alpha)\).
Thus, that this decomposition is indexed by rational numbers is a consequence of
the monodromy theorem.

The root decomposition defines a (decreasing) \emph{Kashiwara--Malgrange}
filtration on \(M\) defined by
\begin{equation*}
  V^{\alpha}M = \mathop{\widehat{\bigoplus}}_{\beta \geq \alpha}M^{\beta},\quad
  V^{>\alpha}M = \mathop{\widehat{\bigoplus}}_{\beta > \alpha}M^{\beta}
\end{equation*}
and it satisfies the following conditions:
\begin{itemize}
\item \(V^{\alpha}M\) are finite \(\mathbb{C}\{t\}\)-submodules of \(M\) and
\(M=\bigcup_{\alpha}M^{\alpha}\).
\item \(tV^{\alpha}M \subset V^{\alpha+1}M\),
\(\partial_{t}V^{\alpha}M \subset V^{\alpha-1}M\), and for \(\alpha>0\),
\(tV^{\alpha}M = V^{\alpha+1}M\).
\item \(t\partial_{t}-\alpha\) is a nilpotent operator on
\(\boxed{\mathrm{Gr}_{V}^{\alpha}M = V^{\alpha}M/V^{>\alpha}M}\).
\end{itemize}

The idea of the Kashiwara--Malgrange filtration comes from
``asymptotic expansions''. On the punctured disk \(S^{\ast} = S \setminus
\{0\}\) the \(\mathcal{D}\)-module \(M\) determines a local system
\(\mathbf{H}\) (i.e., the cohomology bundle of the Milnor fibration) and we can
form its dual local system \(\mathbf{H}^{\vee}\) on \(S^{\ast}\) (the homology
bundle). For an element \(\omega \in M\) defines a holomorphic section (i.e., a
family of differential forms on the fibers of the Milnor fibration) of
\(\mathbf{H}\otimes_{\mathbb{C}}\mathcal{O}_{S^{\ast}}\) (upon shrinking \(S\),
if necessary). Thus (as \(M\) is a \(\mathcal{D}\)-module with regular
singularity at \(0\)) for each multivalued horizontal section \(\gamma\) of
\(\mathbf{H}^{\vee}\), the pairing
\(s[\omega,\gamma]=\langle\gamma,\omega\rangle\) (i.e., period integrals
\(\int_{\gamma}\omega\)) is a multivalued holomorphic function on \(S^{\ast}\)
with logarithmic singularity at \(0\). Whence we have an asymptotic expansion
\begin{equation*}
  s[\omega,\gamma] = \sum_{\alpha \in \mathbb{Q}} \sum_{m=0}^{\infty} a_{\alpha,m,\gamma}t^{\alpha-1} (\log t)^{m}
\end{equation*}
with \(\inf\{\alpha:a_{\alpha,m,\gamma} \neq 0\} > -\infty\), where the
infimum runs through all multivalued sections \(\gamma\) of
\(\mathbf{H}^{\vee}\), all integers \(m\), and all rational numbers \(\alpha\).
This number is then called the \emph{order} of \(\omega\).

\medskip\noindent%
\textbf{Theorem.} \textit{Let \(\omega\) be a section of \(M\). Then \(\omega
  \in V^{\alpha}M / V^{>\alpha}M\) if and only if the order of \(\omega\) is
  \(\alpha\).}

\medskip
The \emph{canonical lattice} is the \(\mathbb{C}\{t\}\)-module
\begin{equation*}
\mathcal{L} = V^{>-1}M.
\end{equation*}
The \emph{vanishing cohomology of the canonical Milnor fiber} is
\begin{equation*}
  \boxed{
    H^{n}(X_{\infty}) = \mathcal{L}/t\mathcal{L} = \bigoplus_{-1< \alpha\leq 0} M^{\alpha}.
  }
\end{equation*}
On the vector space \(H^{n}(X_{\infty})\) one has a ``monodromy
action'' \(T:H^{n}(X_{\infty}) \to H^{n}(X_{\infty})\) defined by
\begin{equation*}
  T|_{M^{\alpha}} = \exp\left\{-2\pi\sqrt{-1}\alpha\right\}\mathrm{Id} \cdot
  \exp\left\{2\pi\sqrt{-1}(t\partial_{t} -\alpha)\right\}.
\end{equation*}
In other words, \(M^{\alpha}\) is the root space of the operator \(T\)
associated with the eigenvalue
\(\lambda=\exp\left\{-2\pi\sqrt{-1}\alpha\right\}\).  We would like to define a MHS on vector space $H^{n}(X_{\infty})$, which could be thought of the zero fiber of the Deligne extension of
the cohomological bundle.

\medskip\noindent%
\textbf{The Brieskorn lattice.} To define the Hodge filtration on vanishing
cohomology, we will need a filtration on the Gauss-Manin system. The
Gauss--Manin system \(M\) comes with an exhaustive filtration \(F^{\bullet}M\),
with \(F^{n-i}=\partial_{t}F^{i}\), and
\begin{equation*}
  \boxed{
    F^{n}M=M_{0}=\frac{\Omega^{n+1}_X}{\mathrm{d}f\wedge\mathrm{d}\Omega^{n-1}_X}
  }
\end{equation*}
called the \emph{Brieskorn lattice}. The filtration looks like
\begin{equation}
F^nM=M_0,~~F^{n-1}M=\partial_t M_0,~~\ldots,~~~, F^{n-k}M=\partial_t^k M_0,~\ldots,
\end{equation}
By a theorem of Sebastiani, \(M_{0}\) is a finite free
\(\mathbb{C}\{t\}\)-module. This lattice plays an important role in defining Mixed Hodge Structure
on vanishing cohomology.

The Brieskorn lattice is not stable under the action of \(\partial_{t}\), rather
it is \textbf{stable} under \(\partial_{t}^{-1}\), indeed, if \(\omega\) is an
\((n+1)\)-form representing an element in the Brieskorn lattice, then
\begin{equation*}
  \boxed{
    \partial_{t}^{-1} \omega = \mathrm{d}f \wedge \eta
  }, \quad \text{ where }
  \omega = \mathrm{d}\eta.
\end{equation*}
In particular,
\begin{equation*}
  \mathop{\mathrm{Im}}\partial_{t}^{-1} = \frac{\mathrm{d}f \wedge \Omega_X^{n}}{\mathrm{d}f\wedge\mathrm{d}\Omega^{n-1}_X}.
\end{equation*}
It follows that \(M_{0}/\partial^{-1}_{t}M_{0} = \Omega_{f}\), where
\begin{equation*}
  \boxed{
    \Omega_{f} = \frac{\Omega^{n+1}_{X}}{\mathrm{d}f\wedge \Omega^{n}_X}
  }.
\end{equation*}
An useful property of Brieskorn lattice is that it is contained in the canonical
lattice
$M_0\subset V^{>-1}M$, and in general $M_0$ is smaller than $V^{>-1}M$.

\medskip\noindent%
\textbf{Hodge filtration after Saito--Scherk--Steenbrink.}
We now explain how to define the Hodge filtration on vanishing cohomology
according to Morihiko Saito and Scherk--Steenbrink
\cite{saito1982gauss, scherk1985mixed}.

\begin{figure}[H]
\begin{center}

\tikzset{every picture/.style={line width=0.75pt}} %set default line width to 0.75pt

\begin{tikzpicture}[x=0.55pt,y=0.55pt,yscale=-1,xscale=1]
%uncomment if require: \path (0,479); %set diagram left start at 0, and has height of 479

%Straight Lines [id:da3737584424905769]
\draw    (63.5,241) -- (825.5,239) ;
%Shape: Circle [id:dp8534482489094473]
\draw  [fill={rgb, 255:red, 0; green, 0; blue, 0 }  ,fill opacity=1 ] (197,240.75) .. controls (197,239.23) and (198.23,238) .. (199.75,238) .. controls (201.27,238) and (202.5,239.23) .. (202.5,240.75) .. controls (202.5,242.27) and (201.27,243.5) .. (199.75,243.5) .. controls (198.23,243.5) and (197,242.27) .. (197,240.75) -- cycle ;
%Shape: Circle [id:dp9702619042732747]
\draw  [fill={rgb, 255:red, 0; green, 0; blue, 0 }  ,fill opacity=1 ] (298.5,240.75) .. controls (298.5,239.23) and (299.73,238) .. (301.25,238) .. controls (302.77,238) and (304,239.23) .. (304,240.75) .. controls (304,242.27) and (302.77,243.5) .. (301.25,243.5) .. controls (299.73,243.5) and (298.5,242.27) .. (298.5,240.75) -- cycle ;
%Flowchart: Process [id:dp22662057734085717]
\draw   (241,79) -- (258.5,79) -- (258.5,240) -- (241,240) -- cycle ;
%Straight Lines [id:da32379569273981157]
\draw    (240,200) -- (260,200) ;
%Flowchart: Process [id:dp4157711554567227]
\draw   (342.5,80) -- (360,80) -- (360,241) -- (342.5,241) -- cycle ;
%Straight Lines [id:da26922786230296136]
\draw    (341,200) -- (361,200) ;
%Straight Lines [id:da3834725942723325]
\draw    (340,180) -- (360,180) ;
%Straight Lines [id:da14926919690170593]
\draw    (246,200) .. controls (248.34,199.74) and (249.64,200.78) .. (249.9,203.12) .. controls (250.16,205.47) and (251.46,206.51) .. (253.81,206.25) .. controls (256.15,205.99) and (257.45,207.03) .. (257.71,209.37) -- (258.5,210) -- (258.5,210) ;
%Straight Lines [id:da8592016318737954]
\draw    (240.5,213) .. controls (242.79,212.45) and (244.22,213.32) .. (244.77,215.61) .. controls (245.32,217.9) and (246.74,218.76) .. (249.03,218.21) .. controls (251.32,217.66) and (252.75,218.53) .. (253.3,220.82) .. controls (253.85,223.11) and (255.28,223.98) .. (257.57,223.43) -- (258.5,224) -- (258.5,224) ;
%Straight Lines [id:da0350871974192446]
\draw    (240.5,229) .. controls (242.79,228.45) and (244.22,229.32) .. (244.77,231.61) .. controls (245.32,233.9) and (246.74,234.76) .. (249.03,234.21) .. controls (251.32,233.66) and (252.75,234.53) .. (253.3,236.82) .. controls (253.85,239.11) and (255.28,239.98) .. (257.57,239.43) -- (258.5,240) -- (258.5,240) ;
%Shape: Circle [id:dp9709054912373418]
\draw  [fill={rgb, 255:red, 0; green, 0; blue, 0 }  ,fill opacity=1 ] (397.5,239.75) .. controls (397.5,238.23) and (398.73,237) .. (400.25,237) .. controls (401.77,237) and (403,238.23) .. (403,239.75) .. controls (403,241.27) and (401.77,242.5) .. (400.25,242.5) .. controls (398.73,242.5) and (397.5,241.27) .. (397.5,239.75) -- cycle ;
%Straight Lines [id:da3262857858225925]
\draw    (343,189) .. controls (345.29,188.45) and (346.72,189.32) .. (347.27,191.61) .. controls (347.82,193.9) and (349.24,194.76) .. (351.53,194.21) .. controls (353.82,193.66) and (355.25,194.53) .. (355.8,196.82) .. controls (356.35,199.11) and (357.78,199.98) .. (360.07,199.43) -- (361,200) -- (361,200) ;
%Straight Lines [id:da2534814380215371]
\draw    (350,180) .. controls (352.34,179.74) and (353.64,180.78) .. (353.9,183.12) .. controls (354.16,185.47) and (355.46,186.51) .. (357.81,186.25) .. controls (360.15,185.99) and (361.45,187.03) .. (361.71,189.37) -- (362.5,190) -- (362.5,190) ;
%Straight Lines [id:da6120548911098533]
\draw    (215.5,201) -- (232.5,201) ;
%Straight Lines [id:da7784864282541728]
\draw    (222.96,206) -- (222.54,237) ;
\draw [shift={(222.5,240)}, rotate = 270.77] [fill={rgb, 255:red, 0; green, 0; blue, 0 }  ][line width=0.08]  [draw opacity=0] (10.72,-5.15) -- (0,0) -- (10.72,5.15) -- (7.12,0) -- cycle    ;
\draw [shift={(223,203)}, rotate = 90.77] [fill={rgb, 255:red, 0; green, 0; blue, 0 }  ][line width=0.08]  [draw opacity=0] (8.93,-4.29) -- (0,0) -- (8.93,4.29) -- cycle    ;
%Straight Lines [id:da17288979688455797]
\draw  [dash pattern={on 0.84pt off 2.51pt}]  (180.5,181) -- (333.5,180) ;
%Straight Lines [id:da3733184210531688]
\draw    (180.47,184) -- (180.03,237) ;
\draw [shift={(180,240)}, rotate = 270.49] [fill={rgb, 255:red, 0; green, 0; blue, 0 }  ][line width=0.08]  [draw opacity=0] (10.72,-5.15) -- (0,0) -- (10.72,5.15) -- (7.12,0) -- cycle    ;
\draw [shift={(180.5,181)}, rotate = 90.49] [fill={rgb, 255:red, 0; green, 0; blue, 0 }  ][line width=0.08]  [draw opacity=0] (8.93,-4.29) -- (0,0) -- (8.93,4.29) -- cycle    ;
%Flowchart: Process [id:dp10020268865961279]
\draw   (439.5,79) -- (457,79) -- (457,240) -- (439.5,240) -- cycle ;
%Straight Lines [id:da30461368158498625]
\draw    (439,180) -- (459,180) ;
%Straight Lines [id:da7183360250761888]
\draw    (439,140) -- (459,140) ;
%Straight Lines [id:da42254553942724904]
\draw    (440.5,148) .. controls (442.79,147.45) and (444.22,148.32) .. (444.77,150.61) .. controls (445.32,152.9) and (446.74,153.76) .. (449.03,153.21) .. controls (451.32,152.66) and (452.75,153.53) .. (453.3,155.82) .. controls (453.85,158.11) and (455.28,158.98) .. (457.57,158.43) -- (458.5,159) -- (458.5,159) ;
%Straight Lines [id:da8172867032630649]
\draw    (440.5,163) .. controls (442.79,162.45) and (444.22,163.32) .. (444.77,165.61) .. controls (445.32,167.9) and (446.74,168.76) .. (449.03,168.21) .. controls (451.32,167.66) and (452.75,168.53) .. (453.3,170.82) .. controls (453.85,173.11) and (455.28,173.98) .. (457.57,173.43) -- (458.5,174) -- (458.5,174) ;
%Straight Lines [id:da5065182630996401]
\draw    (446,139) .. controls (448.34,138.74) and (449.64,139.78) .. (449.9,142.12) .. controls (450.16,144.47) and (451.46,145.51) .. (453.81,145.25) .. controls (456.15,144.99) and (457.45,146.03) .. (457.71,148.37) -- (458.5,149) -- (458.5,149) ;
%Flowchart: Process [id:dp7277585251232173]
\draw  [fill={rgb, 255:red, 0; green, 0; blue, 0 }  ,fill opacity=1 ] (341.5,200) -- (361,200) -- (361,239) -- (341.5,239) -- cycle ;
%Flowchart: Process [id:dp7380047340294213]
\draw  [fill={rgb, 255:red, 0; green, 0; blue, 0 }  ,fill opacity=1 ] (440,179) -- (457,179) -- (457,240) -- (440,240) -- cycle ;
%Straight Lines [id:da7770639193608879]
\draw  [dash pattern={on 0.84pt off 2.51pt}]  (137.5,140) -- (439,140) ;
%Straight Lines [id:da27695050116200703]
\draw    (137.5,143) -- (137.5,241) ;
\draw [shift={(137.5,244)}, rotate = 270] [fill={rgb, 255:red, 0; green, 0; blue, 0 }  ][line width=0.08]  [draw opacity=0] (10.72,-5.15) -- (0,0) -- (10.72,5.15) -- (7.12,0) -- cycle    ;
\draw [shift={(137.5,140)}, rotate = 90] [fill={rgb, 255:red, 0; green, 0; blue, 0 }  ][line width=0.08]  [draw opacity=0] (8.93,-4.29) -- (0,0) -- (8.93,4.29) -- cycle    ;

% Text Node
\draw (243,243.4) node [anchor=north west][inner sep=0.75pt]   [font=\tiny]  {$\alpha $};
% Text Node
\draw (341,248.4) node [anchor=north west][inner sep=0.75pt]  [font=\tiny]  {$\alpha +1$};
% Text Node
\draw (209,183.4) node [anchor=north west][inner sep=0.75pt]  [font=\tiny]  {$F^{n}$};
% Text Node
\draw (168,161.4) node [anchor=north west][inner sep=0.75pt]  [font=\tiny]  {$F^{n-1}$};
% Text Node
\draw (436.5,248.4) node [anchor=north west][inner sep=0.75pt]  [font=\tiny]  {$\alpha +2$};
% Text Node
\draw (295,252) node [anchor=north west][inner sep=0.75pt]  [font=\scriptsize] [align=left] {0};
% Text Node
\draw (188,251) node [anchor=north west][inner sep=0.75pt]  [font=\scriptsize] [align=left] {\mbox{-}1};
% Text Node
\draw (108,125.4) node [anchor=north west][inner sep=0.75pt]  [font=\tiny]  {$F^{n-2}$};

\end{tikzpicture}

\end{center}
\caption{The black block with order $\alpha+p$ are elements of $M_0$ which come from the action of $\partial_t^{-1}$ on elements of $M_0$ with order $\alpha+p-1$.}
\label{17}
\end{figure}
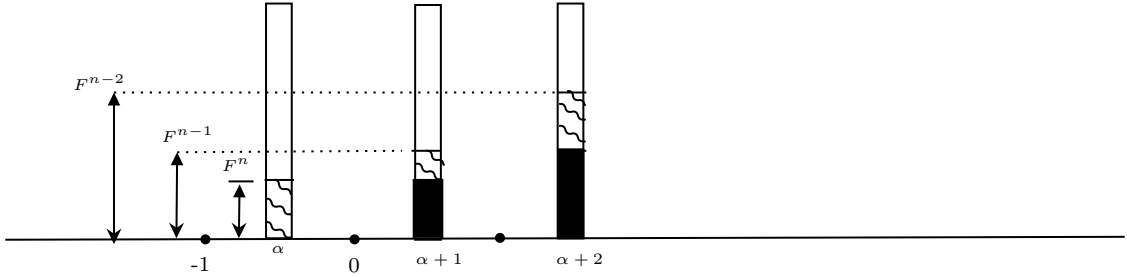

We start with a germ of a holomorphic function with an isolated critical point,
and we then consider its Gauss-Mannin system and Brieskorn lattice $(M,M_0)$.
The vanishing cohomology can be seen from the \(V\)-filtration by
\(\phi_tM=\bigoplus_{-1<\alpha \leq 0}M^{\alpha}\), where \(M^{\alpha}\) is the
root space in the Kashiwara--Malgrange filtration. To put a Hodge
filtration on each individual root space \(M^{\alpha}\) in $\phi_tM$,
Morihiko Saito and Scherk--Steenbrink use the formula
\begin{equation*}
  \boxed{
    F^{p} M^\alpha = \mathrm{Gr}_{V}^{\alpha}F^{p}M = \frac{V^{\alpha}M \cap F^{p}M + V^{>\alpha}M}{V^{>\alpha}M}
  },
  \quad -1 < \alpha \leq  0.
\end{equation*}
The formula for the Hodge filtration then reduces to
\begin{align}
  F^pM^{\alpha}
  &= \frac{V^{\alpha} M \cap \partial_{t}^{n-p}M_{0}+V^{>\alpha} M}{V^{>\alpha} M}
  \label{crucial}
\end{align}
Notice that $\partial_t$ action does not preserve $M_0$.
The above formula shows that one can compute the Hodge filtration on vanishing cohomology by using
Brieskorn lattice $M_0$:
\begin{enumerate}
\item $F^nM^\alpha$ consists of elements of $M_0$ whose order is exactly $\alpha$.
\item $F^{n-1} M^\alpha$ consists of elements of $M_0$ whose order is
  $\alpha+1$. These can be separated into two kinds: one kind takes the form
  $\partial_t^{-1} \omega$ with $\omega \in F^n M^\alpha$, and others are new.
  See Figure~\ref{17}.
\end{enumerate}

\medskip\noindent%
\textbf{Weight filtration.} The weight filtration comes from the nilpotent part
of the monodromy operator $h$ (this is the monodromy operator acting on
vanishing cohomology associated with the singularity). It is given by
$N=\partial_t t -\alpha$ on $M^{\alpha}$. Given a nilpotent operator $N$ acting
on a space $H$, one can define a filtration as follows. Choose a Jordan
basis $u_1,\ldots, u_q$, so that $N$ takes the form
\begin{equation}
N=\left(\begin{array}{cccc}
0&~&~& ~\\
1&0&~&~\\
~&1&0&~\\
~&~&1&0\\ \end{array}\right),~~Nu_1=u_2,\ldots,Nu_{q-1}=u_q,~~Nu_q=0
\end{equation}
We renumber the indices of the basis so that the action of $N$ is
$N(u_i)=u_{i-2}$, so we have
\begin{enumerate}
\item If $q=2k+1$ is odd, the basis is numbered as $u_{2k},\ldots,u_2,u_0,u_{-2},\ldots, u_{-2k}$.
\item if $q=2k$ is even, the basis is numbered as $u_{2k-1},\ldots,u_2,u_0,u_{-2},\ldots, u_{-(2k-1)}$.
\end{enumerate}
In this way, each basis has a weight corresponding its index, and the nilpotent
operator induces a filtration on $H$: $W_k H$ consists of the subspace generated by
the vectors of weight $\leq k$.

\medskip\noindent%
\textbf{Definition.}
The weight filtration of a nilpotent operator $N$ with center $n$ is a filtration
${}^nW$ which is defined as ${}^nW_k=W_{-n+k}$.

\medskip%
Let's go back to the root space $M^\alpha$, if $\alpha \neq 0$, the weight
filtration is given by ${}^{-n}W$, and if $\alpha=0$, the weight filtration is
given by ${}^{-n-1}W$.

\medskip\noindent%
\textbf{Spectral pairs.} We have explained how to define a Hodge and weight
filtration on each root space $M^\alpha$ of vanishing cohomology. The discrete
invariant of MHS on vanishing cohomology is recorded by the so-called spectral
pairs. Let \(f\) be a function germ with isolated singularity
$f:\mathbb{C}^{n+1}\to \mathbb{C}$, and its vanishing cohomology admits a limit
mixed Hodge structure. Thus the vanishing cohomology admits a decomposition
\begin{equation*}
\bigoplus H_\lambda
\end{equation*}
according to the eigenvalues of the monodromy operator.
Each \(H_{\lambda}\) has two filtrations: the Hodge filtration
\(F^{\bullet}H_{\lambda}\) and the weight filtration \(W_{\bullet}H_{\lambda}\).

Thus for each integer \(w\), the associated graded vector space
\begin{equation*}
H_{\lambda,w}=\mathrm{Gr}_{w}^{W}H_{\lambda}
\end{equation*}
is a pure Hodge structure of weight \(w\), the Hodge filtration being the one
induced by the limit Hodge filtration which we still denote by \(F^{\bullet}\).
For an element in  $F^p\mathrm{Gr}_w^W H_\lambda$,
we define the spectral pairs $(\alpha, l)$ as follows:
\begin{equation}
\alpha=-{1\over 2\pi i} \log \lambda,\quad n-p-1<\alpha \leq n-p
\end{equation}
and
\begin{equation}
l=\begin{cases}
			w & \text{if $\lambda \neq 1$ }\\
          w-1, & \text{$\lambda=1$}
		 \end{cases}
\end{equation}

We say a pair of numbers \((\alpha,l)\) is a \emph{spectral pair of the vanishing
cohomology of multiplicity \(m=n_{\alpha,l}\)} if
\begin{equation*}
m = \dim F^{p}H_{\lambda,w}/F^{p+1}H_{\lambda,w},
\end{equation*}

These spectral numbers play a very important role in determining the physical theory associated with the local singularity, see  \cite{Xie:2015rpa}. The spectral numbers have following important properties:
\begin{enumerate}
\item The spectral number is paired: $\alpha_i+\alpha_j=n-1$, namely, the spectral number is symmetric with respect to ${n-1\over2}$. The corresponding weight number is also paired $w_1+w_2=2(n+1)$.
\item If we have two singularities $f$ and $g$ whose spectral numbers are $\alpha_i, \beta_j$ respectively, then the singularity $f+g$ has spectra $\alpha_i+\beta_j$.
\item The multiplicity of the minimal spectra $\alpha_1$ is one.
\end{enumerate}

\medskip\noindent%
\textit{Example.} Tthe spectral pairs with nonzero multiplicities of the function germ
\(f(x,y)=x^5+x^{2}y^2+y^5\)
are given by the following table
\begin{center}
  \begin{tabular}{|c|c|}
    \hline
  spectral pair & multiplicity\\
  \hline
\((-\frac{1}{2},2)\) & 1\\
\((-\frac{3}{10},1)\) & 2\\
\((-\frac{1}{10},1)\) & 2\\
\((0,1)\) & 1\\
\((\frac{1}{10},1)\) & 2\\
\((\frac{3}{10},1)\) & 2\\
    \((\frac{1}{2},0)\) & 1\\
    \hline
\end{tabular}
\end{center}

One can extract the physical information from the spectral pairs as follows (We
shall only discuss 3-fold case, so we consider $f:\mathbb{C}^4\to \mathbb{C}$.
This is because the curve singularity case can be lifted to a
three-fold one by adding two extra quadratic terms):
\begin{enumerate}
\item The spectral number $1$ is identified with the mass parameters.
\item For a three dimensional quasi-homogeneous singularity, assume the spectra
  is ordered as $(\alpha_1,\ldots, \alpha_\mu)$ with
  $\alpha_1\leq \ldots \leq \alpha_\mu$, the scaling dimension of Coulomb branch
  spectrum associated with a spectral number $\alpha_i$ is
  \begin{equation}
    \Delta_i={1\over \alpha_1}+{\alpha_1-\alpha_i\over \alpha_1}
  \end{equation}
\end{enumerate}

\medskip\noindent%
\textbf{Summary.}
Let \(f: (\mathbb{C}^{n+1},0) \to (S,0)\) be a germ of a holomorphic
function, and $f=0$ has an isolated singularity at the origin. We associate a
Gauss-Manin system $M$ which is a D-module over $S$. The MHS on vanishing
cohomology can be found as follows:

\begin{enumerate}
\item There is a Kashiwara-Magrange filtration on $M$ so that $M =
  \mathop{\widehat{\bigoplus}}_{\alpha \in \mathbb{Q}} M^{\alpha}$. Here
  \(\exp(-2\pi\sqrt{-1}\alpha)\) is the eigenvalue of the monodromy operator.
\item The vanishing cohomology is given by
  \(\bigoplus_{-1<\alpha\leq 0}M^{\alpha}\).
\item In defining Hodge filtration on vanishing cohomology, we can use Brieskorn
  module $M_0$ which induces a Hodge filtration $F^\bullet$ on $M$.
\item The weight filtration is given by the operator $t\partial_t -\alpha$ on $M^\alpha$.
\end{enumerate}

\medskip\noindent%
\textit{Example 1.}
Let's consider an isolated quasi-homogeneous singularity $f$, and the weights of
the variables are $q_1,\ldots, q_n$ so that the weight of $f$ is one. The
Bireskorn lattice is generated by the differential forms $\omega_m=x^m dx^m$,
here $x^m$ denotes the monomial basis of the Jacobi algebra $J_f$. The
$\partial_t$ action is \cite{kulikov1998mixed}:
\begin{equation}
t\partial_t \omega_m=(a(m)-1) \omega_m
\end{equation}
here $a(m)=\sum m_i q_i+\sum q_i$, here $m_i$ is the exponents of $x^m dx^m=x_1^{m_1}\ldots x_n^{m_n}dx_1\wedge \ldots \wedge dx_n$. The above equation means that the order of $\omega_m$ is
\begin{equation}
a(m)-1
\end{equation}
Let's denote the space $A_p$ as the set of elements $\omega_m$ such that $ n-p-1 <  a(m)-1\leq n-p$. Then the Hodge filtration is given by
\begin{equation}
F^p=A_p\cup A_{p+1} \ldots
\end{equation}
Since the monodromy is semi-simple, the weight filtration is simple, with
\begin{equation}
0\subset W_{n-1}\subset W_n \subset W_{n+1},~~~W_{n}=\bigoplus_{-1<\alpha<0} M^\alpha,~~W_{n+1}/W_n=M^0.
\end{equation}

\medskip\noindent%
\textit{Example 2.}
One can actually compute the spectral pairs using the software ``Singular''
\cite{DGPS}. The following is a computer code:
\begin{verbatim}
LIB "gmssing.lib";
ring R=0,(x,y),ds;
poly t=x5+x2y2+y5;
sppprint(sppairs(t));
==> ((-1/2,2),1),((-3/10,1),2),((-1/10,1),2),((0,1),1),((1/10,1),2),((3/10,1)\
   ,2),((1/2,0),1)
\end{verbatim}
The output consists of triples $((\alpha, l), m)$ which are spectra number, weight number, and multiplicity.

\medskip\noindent%
\textbf{Physical interpretation.}
Associated with a local singularity, one has an interacting 4d
$\mathcal{N}=2$ theories. If the local singularity is quasi-homogenenous, the 4d
theory is superconformal and a lot of studies for them were performed in
\cite{Xie:2015rpa,Xie:2016evu,Xie:2019vzr}, in particular the spectral numbers
play a crucial role there. The study of 4d theories associated with general local
singularity will be given in \cite{xie:2021abc}.

\subsection{Nearby and vanishing cycles}

For the atypical fiber associated with a special vacuum, we have defined limit
MHS on the cohomology space of a nearby fiber, the MH{S} on the vanishing
cohomology for its critical points (in general, there are more than one local
singularities on the singular fiber, and the vanishing cohomology is the direct
sum of them). In this subsection, we will give a unified description using the
language of Gauss-Manin system associated with the polynomial map, in terms of
the language of \textbf{nearby cycle and vanishing cycle functors}.

Let \(f\) be a tame polynomial
$f:\mathbb{C}^n\times \mathbb{C}^{*m}\rightarrow \mathbb{C}$. Let \(c\) be an
atypical value of \(f\). Without loss of generality assume \(c=0\), and $t$
denotes the local coordinate around $c$. Then in a small neighborhood of \(0\),
the Gauss--Manin system \(M=f_+{\cal O}_U\)
($U=\mathbb{C}^n\times \mathbb{C}^{*m}$) allows the following root decomposition
(which is in parallel with the local case studied in last subsection):
\begin{equation*}
  M = \mathop{\widehat{\bigoplus}}_{\alpha \in \mathbb{Q}} M^{\alpha}
\end{equation*} where
\begin{equation*}
  \boxed{
    M^{\alpha} = \bigcup_{i=1}^{\infty} \mathrm{Ker}(t\partial_{t} - \alpha)^{i}
  }
\end{equation*}
The root decomposition defines a (decreasing) \emph{Kashiwara--Malgrange}
filtration on \(M\) defined by
\begin{equation*}
  V^{\alpha}M = \mathop{\widehat{\bigoplus}}_{\beta \geq \alpha}M,\quad
  V^{>\alpha}M = \mathop{\widehat{\bigoplus}}_{\beta > \alpha}M
\end{equation*}
which satisfies the same properties as in previous section.
Using this filtration,
we can define two vector spaces: nearby cycle and vanishing cycle
\begin{equation}
\text{Nearby cycle}:~\psi_tM=\bigoplus_{-1<\alpha\leq0}M^{\alpha}
\end{equation}
and
\begin{equation}
\text{Vanishing cycle}:~\phi_tM=\bigoplus_{-1\leq\alpha<0}M^{\alpha}
\end{equation}
There is a limit MHS on these two spaces too: the Hodge filtration is induced
from the Hodge filtration of the Gauss-Manin system, and the weight filtration
is given by the monodromy. The computations are quite similar to the isolated
singularity case. Here we just want to make one important remark: in current
case, we usually need to consider variation of mixed Hodge structure, and in
general there have two weights of a general fiber (which correspond to
electric-magnetic charge and flavor charge), so in considering the limit MHS, we
need to consider variation of Hodge structure of each weight separately. In
particular, in defining the weight filtration of limit MHS, the shift would be
different depending the weights of varying Hodge structure. For example, consider
the curve case so that the nearby fiber has weight one and weight two, for the
weight one part, the limit MHS has a weight filtration ${}^{-1}N$, and for the
weight one part, the limit MHS has a weight filtration ${}^{-2}N$.

Therefore we have three MHS associated with the atypical fiber, and there is a
distinguished triangle for them which will induce a long exact sequence for the
cohomology groups \cite{dimca2004sheaves}. Since we consider only tame
polynomials in this section, the vanishing cycles can be computed from the local singularity of the
singular fiber (the Milnor fiber). For a tame polynomial, the only non-trivial
part of the exact sequences of the cohomology groups are:
\begin{equation}\label{eq:okk}
 	0 \to
	H^{n}(X_{0}) \xrightarrow{\mathrm{sp}}
	H^{n}(\psi_tM) \xrightarrow{\mathrm{can}}
	H^{n}(\phi_tM) \to 0.
\end{equation}
All items in this sequences are equipped with canonical mixed Hodge structures,
and this sequence is not only an exact sequence of vector spaces but also an
exact sequence of mixed Hodge structures. This exact sequence is useful in that
the Hodge numbers of $H^n(\phi_tM)$ is given by the sum of $H^n(X_0)$ and
$H^n(\phi_tM)$.

\medskip\noindent%
\textit{Example.}
Consider the degeneration of a genus two curve to a genus one
curve with a node (see Figure~\ref{limitmhs}).
The weight filtration of three vector spaces are given as
\begin{align*}
&H^1(\psi_tM):~~~W_{-1}=\{0\},~W_0=\{\gamma_2^*\},~~~W_1=\{\gamma_2^*, \gamma_3^*, \gamma_4^*\},~~W_2=\{\gamma_1^*,\gamma_2^*,\gamma_3^*, \gamma_4^*\}     \nonumber \\
&H^1(X_0):~~~W_{-1}=\{0\},~W_0=\{\gamma_2^*\},~~~W_1=\{\gamma_2^*,\gamma_3^*,\gamma_4^*\}     \nonumber \\
&H^1(\phi_tM):~~W_0=W_{1}=\{0\},~~W_2=\{\gamma_1^*\}
\end{align*}
The morphisms are morphism of MHS of type $(0,0)$.

\medskip\noindent%
\textbf{Local invariant cycle theorem.} Consider weight one part of the
variation of MHS, and we consider the limit MHS of this part. There is a quite
useful facts for us: the cohomology groups of the singular variety $X_0$ is
given by the monodromy invariant part of the nearby cycle $X_t$ (the ``local
invariant cycle theorem'')%
\footnote{%
  We warn the reader that the ``local invariant cycle theorem'' in mathematics
  has a crucial hypothesis that the family is proper and the total space is
  nonsingular. While our family is by fiat non-proper, it is defined by a
  \textbf{tame} polynomial, and the theorem still works (one proves this using
  the exact sequence~\eqref{eq:okk}). If the polynomial is not tame, the local
  invariant cycle theorem does not hold in general.}. This fact has important
physical consequences as we explain now. First, for the polynomial relevant for
4d $\mathcal{N}=2$ theory, the maximal size of the Jordan block of the monodromy
operator on the nearby fiber $\psi_tM$ is two! We focus on $\lambda=1$ part of
the monodromy operator, and the nilpotent part takes the following form
\begin{equation}
N_{\lambda=1}=\left[\begin{array}{cccccc}
0&~&~&~&~&~\\
&\ldots&~&~&~&~\\
~&~&0&~&~&~\\
~&~&~&J_2&~& \\
&~&~&~& \ldots&\\
&~&~&~& &J_2\\
\end{array}\right]
\end{equation}
Here $J_2=\left[\begin{array}{cc} 0&1\\ 0&0 \end{array}\right] $. According to
local invariant cycle theorem, the invariant vector of $N_{\lambda=1}$ is the
cohomology group of the singular variety. This implies that the invariant vector
belonging to Jordan block of size $1$ always comes from the singular variety. For
the Jordan block of size two, one basis belongs to vanishing cohomology, and one
basis belongs to the cohomology of the singular variety. For a $J_2$ block, the
weight filtration has weight zero and two: the weight two (zero) part belongs to
vanishing (singular variety) cohomology. We now have
\begin{itemize}
\item The vanishing cohomology with $\lambda=1$ is always paired with a
  cohomology of the singular variety, and the Jordan block has size two. The
  Hodge numbers are $h^{1,1}$ (for vanishing cohomology), and $h^{0,0}$ (for
  $X_0)$%
  \footnote{We consider the curve case to fix the idea, and the general case is
    similar.}.
\item Now since $\lambda=1$ part of the vanishing cohomology is related to mass
  parameter of the interacting theory, the above result can be interpreted physically as follows: the
  $U(1)$ flavor symmetry of the interacting theory is gauged.
\end{itemize}
Finally, we make an observation of weight two part of variation of MHS, and look
at its limit MHS: the monodromy action on it is trivial and so its Hodge type is
always $h^{1.1}$. These cycles can be either vanishing or is supported on
singular variety.

\subsection{Computations}

In this subsection, we discuss how to compute explicitly the Mixed Hodge structure
for three spaces associated with singular variety, and show how to find the low
energy theory from these structures. First, let us review what we have learned:

\begin{enumerate}
\item Firstly, we can compute the Hodge filtration of the nearby cycle by using
  the method of last section. The monodromy action is often not easy to compute,
  so it is not easy to compute the Hodge numbers for nearby cycle.
\item The MHS for the singular variety can be computed using resolution of
  singularity.
\item If the local singularities are all isolated singularities, we can compute
  the MHS using the computers.
\end{enumerate}
Now if the polynomial is tame, using~\eqref{eq:okk} we
can then find the MHS on nearby cycle by using the results from the second and third items listed above. Some examples will be given in next
subsection.

\medskip\noindent%
\textbf{Newton filtration and MHS for vanishing cohomology.} Although we can
compute the MHS for the vanishing cohomology associated with a local singularity
using a computer software, here we will review a combinatorial method using
the so-called Newton filtration of the associated Newton polyhedron
\cite{saito1988exponents}.

Let $f=\sum c_m x^m$ be a germ of a function with an isolated singularity. Its
support is $\operatorname{Supp}f=\{m\in \mathbb{N}^{n+1}:c_m\neq0\}$. Let
$\Gamma_{+}(f)$ be the Newton polyhedron with respect to infinity, i.e., the
convex hull of
$\bigcup_{m\in\operatorname{Supp}f}(m+\mathbb{R}_{+}^{n+1})\subset\mathbb{R}^{n+1}$.
Let
$\Gamma(f)$ be the Newton boundary, i.e., the union of compact faces $\sigma$ of
$\Gamma_{+}(f)$. We further require that $f$ is Newton non-degenerate and
convenient.

Let's consider a homogeneous function
$h:\mathbb{R}^{n+1}\to \mathbb{R}_{+},$,
\(h(\lambda a)= \lambda h(a)\)
such that $h(\Gamma)=1$. $\Gamma$ defines a decreasing filtration on
\(\mathcal{O}:=\mathbb{C}[\![x_0,\ldots,x_n]\!]\),
\begin{equation}
N^a{\cal O}=\{g(x)\in {\cal O}:h(\operatorname{supp}g)\geq a \},
\end{equation}
This is called the Newton filtration.
Newton filtration induces a filtration $\bar{\nu}$ on the Jacobian algebra
$J_f$. It also induces a filtration on $\Omega_{X,0}^{n+1}$ which is defined as
\begin{equation}
\nu(\omega)=\nu(gx_0\ldots x_n)-1,~~~\omega=g(x)\mathrm{d}x_0\wedge\ldots \wedge \mathrm{d}x_n
\end{equation}
Finally, we can define an order on the Brieskorn lattice
$M_0=\Omega_{X,0}^{n+1} /df\wedge d\Omega_{X,0}^{n-1}$:
\begin{equation}
\bar{\nu}([\omega])=\max \{\nu(\eta):[\eta]=[\omega]\},
\end{equation}
and a filtration on $M_0$:
\begin{equation}
N^a M_0=\{[\omega]\in M_0: \bar{\nu}([\omega])\geq a \}.
\end{equation}
Similarly we have a filtration on $\Omega_f=M_0/tM_0=\Omega_{X,0}^{n+1}/df\wedge \Omega_{X,0}^n$.
Given a vector space \(H\) with a filtration $F^\bullet$, one can define the
Poincaré  polynomial
\begin{equation}
P_{H,F^\bullet}(t)=\sum g_\alpha t^\alpha
\end{equation}
with $g_\alpha= dim F^\alpha/F^{>\alpha}$. The relation with the Poincaré
polynomial of the singularity spectrum is
\begin{equation}
\mathrm{Sp}(f)=\sum_{i=1}^\mu t^{\alpha_i}=P_{\Omega_f, V^\bullet}(t)
\end{equation}

The Poincare polynomial on $\Omega_f$ has following simple description:
\begin{equation}
\tilde{P}_{\Omega_f, V^\bullet}=t P_{\Omega_f, V^\bullet}(t)=\sum_\sigma (-1)^{n-dim~\sigma}(1-t)^{k(\sigma)} P_{A_\sigma(t)}+(-1)^{n+1}.
\end{equation}
Here we sum over the faces of $\Gamma(f)$, and $k(\sigma)$ is the dimension of the minimal coordinate plane containing $\sigma$. $P_{A(\sigma)}$ is given as
\begin{equation}
P_{A_\sigma}(t)={1\over (1-t^{\omega_1})\ldots (1-t^{\omega_k})}
\end{equation}
Here $A_\sigma$ is the cone over $\sigma$, and it is isomorphic to a free algebra; $\omega_1,\ldots, \omega_k$ are weights.

\subsection{Low energy physics at special vacua}

Next we discuss how to find the low energy theory for the special vacuum by
using the MHS associated with singular fiber. In this section, we assume that
the polynomial is tame. The physical interpretations of three MHS are:
\begin{itemize}
\item The abelian gauge theory is described by the MHS of the singular variety:
  the weight one part gives the abelian gauge theory; the weight two part are
  flavor central charge; and weight zero part describes the gauging of the flavor symmetry of the 
  theory associated with singularity.
\item The interacting theory is described by the MHS of local singularities. The
  singularity spectrum contains important information for the theory: the middle
  spectra is identified with mass parameter.
\item The abelian gauge theory and interacting theory is coupled through gauging
  the abelian flavor symmetry of the interacting theory: this is reflected by
  the Hodge numbers: each $h^{0,0}$ of the singular variety is coupled with one
  $h^{1,1}$ from the singularity.
\end{itemize}
So we can find the low energy theory by computing these MHS.

\medskip\noindent%
\textit{Example 1.} Consider the polynomial map $f=x^3+y^7+2y^3$ ($(x,y)$ are
$\mathbb{C}$ variables.). This can be thought of as turning on relevant
deformation of $(A_2, A_6)$ theory which is described by the polynomial
$f=x^3+y^7$. The critical point of the hypersurface $f=0$ is given by the
equations $f={\partial f\over \partial x}={\partial f\over \partial y}=0$, and
the only critical point is the origin. We shall now determine the IR theory
using MHS.

\begin{figure}[H]
\begin{center}
\tikzset{every picture/.style={line width=0.75pt}} %set default line width to 0.75pt

\begin{tikzpicture}[x=0.55pt,y=0.55pt,yscale=-1,xscale=1]
%uncomment if require: \path (0,479); %set diagram left start at 0, and has height of 479

%Straight Lines [id:da6117605533307109]
\draw    (102,86) -- (102,263) ;
\draw [shift={(102,83)}, rotate = 90] [fill={rgb, 255:red, 0; green, 0; blue, 0 }  ][line width=0.08]  [draw opacity=0] (10.72,-5.15) -- (0,0) -- (10.72,5.15) -- (7.12,0) -- cycle    ;
%Straight Lines [id:da1361861893435674]
\draw    (62,243) -- (219,243) ;
\draw [shift={(222,243)}, rotate = 180] [fill={rgb, 255:red, 0; green, 0; blue, 0 }  ][line width=0.08]  [draw opacity=0] (10.72,-5.15) -- (0,0) -- (10.72,5.15) -- (7.12,0) -- cycle    ;
%Straight Lines [id:da33927932876938804]
\draw    (102,103) -- (162,243) ;
%Shape: Circle [id:dp5490410333471059]
\draw  [color={rgb, 255:red, 0; green, 0; blue, 0 }  ,draw opacity=1 ][fill={rgb, 255:red, 0; green, 0; blue, 0 }  ,fill opacity=1 ] (118,163.25) .. controls (118,161.46) and (119.46,160) .. (121.25,160) .. controls (123.04,160) and (124.5,161.46) .. (124.5,163.25) .. controls (124.5,165.04) and (123.04,166.5) .. (121.25,166.5) .. controls (119.46,166.5) and (118,165.04) .. (118,163.25) -- cycle ;
%Shape: Circle [id:dp9686067017785835]
\draw  [color={rgb, 255:red, 0; green, 0; blue, 0 }  ,draw opacity=1 ][fill={rgb, 255:red, 0; green, 0; blue, 0 }  ,fill opacity=1 ] (119,184.25) .. controls (119,182.46) and (120.46,181) .. (122.25,181) .. controls (124.04,181) and (125.5,182.46) .. (125.5,184.25) .. controls (125.5,186.04) and (124.04,187.5) .. (122.25,187.5) .. controls (120.46,187.5) and (119,186.04) .. (119,184.25) -- cycle ;
%Shape: Circle [id:dp038674915151676315]
\draw  [color={rgb, 255:red, 0; green, 0; blue, 0 }  ,draw opacity=1 ][fill={rgb, 255:red, 0; green, 0; blue, 0 }  ,fill opacity=1 ] (119,205.25) .. controls (119,203.46) and (120.46,202) .. (122.25,202) .. controls (124.04,202) and (125.5,203.46) .. (125.5,205.25) .. controls (125.5,207.04) and (124.04,208.5) .. (122.25,208.5) .. controls (120.46,208.5) and (119,207.04) .. (119,205.25) -- cycle ;
%Shape: Circle [id:dp428530821427505]
\draw  [color={rgb, 255:red, 0; green, 0; blue, 0 }  ,draw opacity=1 ][fill={rgb, 255:red, 0; green, 0; blue, 0 }  ,fill opacity=1 ] (119,223.25) .. controls (119,221.46) and (120.46,220) .. (122.25,220) .. controls (124.04,220) and (125.5,221.46) .. (125.5,223.25) .. controls (125.5,225.04) and (124.04,226.5) .. (122.25,226.5) .. controls (120.46,226.5) and (119,225.04) .. (119,223.25) -- cycle ;
%Shape: Circle [id:dp4875391072897268]
\draw  [color={rgb, 255:red, 0; green, 0; blue, 0 }  ,draw opacity=1 ][fill={rgb, 255:red, 0; green, 0; blue, 0 }  ,fill opacity=1 ] (139,223.25) .. controls (139,221.46) and (140.46,220) .. (142.25,220) .. controls (144.04,220) and (145.5,221.46) .. (145.5,223.25) .. controls (145.5,225.04) and (144.04,226.5) .. (142.25,226.5) .. controls (140.46,226.5) and (139,225.04) .. (139,223.25) -- cycle ;
%Shape: Circle [id:dp75982358922676]
\draw  [color={rgb, 255:red, 0; green, 0; blue, 0 }  ,draw opacity=1 ][fill={rgb, 255:red, 0; green, 0; blue, 0 }  ,fill opacity=1 ] (137,204.25) .. controls (137,202.46) and (138.46,201) .. (140.25,201) .. controls (142.04,201) and (143.5,202.46) .. (143.5,204.25) .. controls (143.5,206.04) and (142.04,207.5) .. (140.25,207.5) .. controls (138.46,207.5) and (137,206.04) .. (137,204.25) -- cycle ;
%Straight Lines [id:da6800009776579647]
\draw    (460,84) -- (460,261) ;
\draw [shift={(460,81)}, rotate = 90] [fill={rgb, 255:red, 0; green, 0; blue, 0 }  ][line width=0.08]  [draw opacity=0] (10.72,-5.15) -- (0,0) -- (10.72,5.15) -- (7.12,0) -- cycle    ;
%Straight Lines [id:da4888718342165943]
\draw    (420,241) -- (577,241) ;
\draw [shift={(580,241)}, rotate = 180] [fill={rgb, 255:red, 0; green, 0; blue, 0 }  ][line width=0.08]  [draw opacity=0] (10.72,-5.15) -- (0,0) -- (10.72,5.15) -- (7.12,0) -- cycle    ;
%Straight Lines [id:da2428351552664998]
\draw    (460.5,181) -- (520,241) ;
%Shape: Circle [id:dp40469605226578165]
\draw  [color={rgb, 255:red, 0; green, 0; blue, 0 }  ,draw opacity=1 ][fill={rgb, 255:red, 208; green, 2; blue, 27 }  ,fill opacity=1 ] (457.25,181) .. controls (457.25,179.21) and (458.71,177.75) .. (460.5,177.75) .. controls (462.29,177.75) and (463.75,179.21) .. (463.75,181) .. controls (463.75,182.79) and (462.29,184.25) .. (460.5,184.25) .. controls (458.71,184.25) and (457.25,182.79) .. (457.25,181) -- cycle ;
%Shape: Circle [id:dp846257484184525]
\draw  [color={rgb, 255:red, 0; green, 0; blue, 0 }  ,draw opacity=1 ][fill={rgb, 255:red, 208; green, 2; blue, 27 }  ,fill opacity=1 ] (516.75,241) .. controls (516.75,239.21) and (518.21,237.75) .. (520,237.75) .. controls (521.79,237.75) and (523.25,239.21) .. (523.25,241) .. controls (523.25,242.79) and (521.79,244.25) .. (520,244.25) .. controls (518.21,244.25) and (516.75,242.79) .. (516.75,241) -- cycle ;
%Shape: Circle [id:dp3467494612017792]
\draw  [color={rgb, 255:red, 0; green, 0; blue, 0 }  ,draw opacity=1 ][fill={rgb, 255:red, 208; green, 2; blue, 27 }  ,fill opacity=1 ] (456.75,101) .. controls (456.75,99.21) and (458.21,97.75) .. (460,97.75) .. controls (461.79,97.75) and (463.25,99.21) .. (463.25,101) .. controls (463.25,102.79) and (461.79,104.25) .. (460,104.25) .. controls (458.21,104.25) and (456.75,102.79) .. (456.75,101) -- cycle ;

% Text Node
\draw (229,245.4) node [anchor=north west][inner sep=0.75pt]    {$x$};
% Text Node
\draw (89,60.4) node [anchor=north west][inner sep=0.75pt]    {$y$};
% Text Node
\draw (587,243.4) node [anchor=north west][inner sep=0.75pt]    {$x$};
% Text Node
\draw (447,58.4) node [anchor=north west][inner sep=0.75pt]    {$y$};

\end{tikzpicture}

\end{center}
\caption{Left: The Newton polytope for the polynomial $f=x^3+y^7+2y^3$, it is the the convex hull of the support of $f$ and $zero$; Right: The Newton polytope for the local singularity at origin of polynomial $f=x^3+y^7+2y^3$, and this is the convex hull of the support of $f$ and infinity.}
\label{irtheory}
\end{figure}
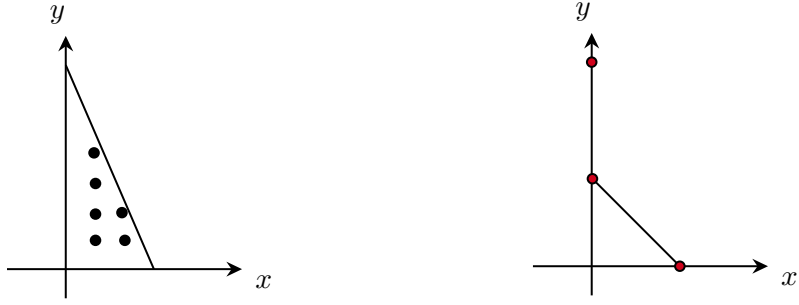

First we can compute the MHS for the smooth fiber (use $t$ to denote the
coordinate $\mathbb{C}$, and $t=0$ denotes the singular fiber). The Newton
polythedron for the polynomial $f$ is given in figure \ref{irtheory}, this is the
convex hull for the support of $f$ and zero. Using the formula in \ref{hodgecurve}
($h^{1,0}$ is equal to the number of internal points of the polyhedron), so we
find that the Hodge numbers of the smooth fiber are $h^{1,0}=h^{0,1}=6$. The
Hodge filtration of nearby cycle is then given by
\begin{equation}
F_{\mathrm{nearby}}^0\subset F_{\mathrm{nearby}}^1
\end{equation}
with $\dim F_{\mathrm{nearby}}^0=12$, \(\dim F_{\mathrm{nearby}}^1=6\).
There is no weight two part. This gives the MHS structure for the smooth fiber
(in fact this is a pure Hodge structure).

The limit MHS for the vanishing cohomology could be found using the Newton
polyhedron for the local singularity at the origin. The polytope is the convex
hull of the support of $f$ and infinity, see Figure~\ref{irtheory}. The limit
MHS is computed as $h_{\mathrm{van}}^{1,0}=h_{\mathrm{van}}^{0,1}=1$ and $h_{\mathrm{van}}^{1,1}=2$. The
Hodge and weight filtration for vanishing cohomology is then
\begin{align}
&F_{\mathrm{van}}^0\subset F_{\mathrm{van}}^1 \nonumber\\
&0=W_0\supset W_1 \supset W_2
\end{align}
We have $\dim F_{\mathrm{van}}^0=4$, \(\dim F_{\mathrm{van}}^1=3\),
and $\dim W_1=2$, \(\dim W_2=4\). The IR theory is a SCFT whose SW geometry is
given by $f=x^3+y^3$. The spectral pairs are
\begin{equation}
((1/2,1),1),((0,1),2) ,((-1/2,1),1)
\end{equation}
From the spectrum, we see that this theory has a $U(1)\times U(1)$ flavor
symmetry, which is actually enhanced to $SU(3)$ flavor symmetry. The Hodge
numbers are $h^{1,0}=h^{0,1}=1$ and $h^{1,1}=2$.

Let's now use above information to get the MHS for the singular variety $X_0$.
The crucial input is that the Hodge type for the nearby cycle is the same as the
smooth fiber. First, since $H^1(X_t)=12$ and $H^1_{\mathrm{van}}=4$, we find
$H^1(X_0)=8$, and it has Hodge filtration for $X_0$:
\begin{equation}
F_{\mathrm{sing}}^0\subset F_{\mathrm{sing}}^1
\end{equation}
with $\dim F_{\mathrm{sing}}^0=8$, \(\dim F_{\mathrm{sing}}^1=3\)
(Since $\dim F^1 H^1(X_t)=6$, \(\dim F^1 H^1_{\mathrm{van}}=3\)).
Using $F^1=\oplus_{p\geq 1} h^{p,q}$, we find $h^{1,0}(X_0)=h^{0,1}(X_0)=3$.
This implies that there is a $U(1)^3$ abelian gauge theory described by the
cohomology of singular variety. Finally, we have $h^{0,0}(X_0)=2$.
$h^{0,0}(X_0)$ is paired with $h^{1,1} (X_{\mathrm{van}})$ of the vanishing
cohomology.

Now we can derive the full IR theory: The theory associated with vanishing
cohomology is a SCFT ( $(A_1, D_4)$ theory), and this theory has a $SU(3)$
flavor symmetry. The two $U(1)$ subgroups of the flavor symmetry are gauged.
There is also a decoupled $U(1)^3$ abelian gauge theory (The weight one part of
the $H^1(X_0)$). The theory can be schematically written as
\begin{figure}[H]
\begin{center}
\tikzset{every picture/.style={line width=0.75pt}} %set default line width to 0.75pt
\begin{tikzpicture}[x=0.55pt,y=0.55pt,yscale=-1,xscale=1]
%uncomment if require: \path (0,479); %set diagram left start at 0, and has height of 479

%Straight Lines [id:da006387412741975851]
\draw    (316,174) -- (341.5,156) ;
%Straight Lines [id:da5779941314917529]
\draw    (316,185) -- (339.5,198) ;
%Shape: Circle [id:dp26478439059933345]
\draw   (340.5,150) .. controls (340.5,142.82) and (346.32,137) .. (353.5,137) .. controls (360.68,137) and (366.5,142.82) .. (366.5,150) .. controls (366.5,157.18) and (360.68,163) .. (353.5,163) .. controls (346.32,163) and (340.5,157.18) .. (340.5,150) -- cycle ;
%Shape: Circle [id:dp8459623487277086]
\draw   (339.5,198) .. controls (339.5,190.89) and (345.26,185.13) .. (352.38,185.13) .. controls (359.49,185.13) and (365.25,190.89) .. (365.25,198) .. controls (365.25,205.11) and (359.49,210.88) .. (352.38,210.88) .. controls (345.26,210.88) and (339.5,205.11) .. (339.5,198) -- cycle ;
%Flowchart: Or [id:dp7452006775594375]
\draw   (378,174.5) .. controls (378,170.36) and (381.25,167) .. (385.25,167) .. controls (389.25,167) and (392.5,170.36) .. (392.5,174.5) .. controls (392.5,178.64) and (389.25,182) .. (385.25,182) .. controls (381.25,182) and (378,178.64) .. (378,174.5) -- cycle ; \draw   (378,174.5) -- (392.5,174.5) ; \draw   (385.25,167) -- (385.25,182) ;

% Text Node
\draw (302,169) node [anchor=north west][inner sep=0.75pt]   [align=left] {T};
% Text Node
\draw (347,140) node [anchor=north west][inner sep=0.75pt]   [align=left] {1};
% Text Node
\draw (347,188) node [anchor=north west][inner sep=0.75pt]   [align=left] {1};
% Text Node
\draw (414,163.4) node [anchor=north west][inner sep=0.75pt]    {$U( 1)^{3}$};

\end{tikzpicture}
\end{center}
\end{figure}

Here $T$ is a rank one SCFT with flavor symmetry $SU(3)$,
and a $U(1)\times U(1)$ subgroup of flavor symmetry is gauged.

\medskip\noindent%
\textit{Example 2.} Consider the polynomial $f=z+{1\over z}+x^3-3a^2 x$, and
this polynomial describes the SW geometry of pure $SU(3)$ super Yang-Mills
theory, here $z$ is a $\mathbb{C}^*$ variable and $x$ is a $\mathbb{C}$
variable. Here we turn on one parameter deformation parameterized by $a$. The
critical points of $f$ is given by
\begin{equation}
{\partial f\over \partial z}=0,~~{\partial f \over \partial x}=0
\end{equation}
We find four solutions $z=\pm 1,~~x=\pm a$. The critical values are
\begin{align}
& c_{(1,a)}=2-2a^3,~~c_{(1,-a)}=2+2a^3,    \\
& c_{(-1,a)}=-2-2a^3,~~c_{(-1,-a)}=-2+2a^3.
\end{align}
By choosing $a$ appropriatly, the critical values will merge. We have two choices:

\begin{enumerate}[wide]
\item
  First, if $a=0$, then the curve takes the form
\begin{equation}
f=z+{1\over z}+x^3
\end{equation}
and the critical value is $\pm 2$, and there is only one singular point
$(x,z)=(0,1)$ on $f=2$. Near the local singularity, the curve takes the form
\begin{equation}
f^{'}=(z+1)^2/z+x^3=y^2+x^3
\end{equation}
The IR theory associated with local singularity is nothing but the $(A_1, A_2)$
theory. Let's now compute the IR theory at this vacuum using MHS. The Hodge
structure of the smooth fiber is $\dim F_{\mathrm{near}}^0=4$,
\(\dim F_{\mathrm{near}}^1=2\). The Mixed Hodge structure
for the vanishing cohomology is
$\dim F_{\mathrm{van}}^0=2$, \(\dim F_{\mathrm{van}}^1=1\). This implies that the
MHS for the singular variety is $\dim F_{\mathrm{sing}}^0=2$,
\(\dim F_{\mathrm{sing}}^1=1\). The IR theory is then
\begin{equation}
T \oplus U(1)
\end{equation}
Here $T$ is the $(A_1, A_2)$ theory, and there is free abelian $U(1)$ gauge
theory in the IR too. These two sectors are decoupled.

\item Secondly, we can take $a^3=1$, i.e., $a=1$. Then $c_{(1,1)}=c_{(-1,-1)}$.
  The critical value is zero, and the singular fiber has two critical points:
  $(x,z)=(1,1)$ and $(x,z)=(-1,-1)$, and each local singularity is a $A_1$
  singularity. So we have $\dim F_{\mathrm{van}}^0=2, \dim F_{\mathrm{van}}^1=2$
  ($h^{1,1}_{\mathrm{van}}=2$). This implies that $\dim F_{\mathrm{sing}}^0=2$,
  \(\dim F_{\mathrm{sing}}^1=0\) ($h^{0,0}_{\mathrm{sing}}=2$).
  The IR theory is just two copies of $U(1)$
  gauge theory coupled with a hypermultiplet:
\begin{figure}[H]
\begin{center}
\tikzset{every picture/.style={line width=0.75pt}} %set default line width to 0.75pt

\begin{tikzpicture}[x=0.35pt,y=0.35pt,yscale=-1,xscale=1]
%uncomment if require: \path (0,479); %set diagram left start at 0, and has height of 479

%Shape: Circle [id:dp26478439059933345]
\draw   (242.5,178.75) .. controls (242.5,168.39) and (250.89,160) .. (261.25,160) .. controls (271.61,160) and (280,168.39) .. (280,178.75) .. controls (280,189.11) and (271.61,197.5) .. (261.25,197.5) .. controls (250.89,197.5) and (242.5,189.11) .. (242.5,178.75) -- cycle ;
%Flowchart: Or [id:dp7452006775594375]
\draw  [color={rgb, 255:red, 208; green, 2; blue, 27 }  ,draw opacity=1 ] (380,180) .. controls (380,168.95) and (388.39,160) .. (398.75,160) .. controls (409.11,160) and (417.5,168.95) .. (417.5,180) .. controls (417.5,191.05) and (409.11,200) .. (398.75,200) .. controls (388.39,200) and (380,191.05) .. (380,180) -- cycle ; \draw  [color={rgb, 255:red, 208; green, 2; blue, 27 }  ,draw opacity=1 ] (380,180) -- (417.5,180) ; \draw  [color={rgb, 255:red, 208; green, 2; blue, 27 }  ,draw opacity=1 ] (398.75,160) -- (398.75,200) ;
%Shape: Square [id:dp3221836552689288]
\draw   (320,160) -- (360,160) -- (360,200) -- (320,200) -- cycle ;
%Straight Lines [id:da47638399733540804]
\draw    (280,180) -- (320,180) ;
%Shape: Circle [id:dp3682227391771551]
\draw   (440,178.75) .. controls (440,168.39) and (448.39,160) .. (458.75,160) .. controls (469.11,160) and (477.5,168.39) .. (477.5,178.75) .. controls (477.5,189.11) and (469.11,197.5) .. (458.75,197.5) .. controls (448.39,197.5) and (440,189.11) .. (440,178.75) -- cycle ;
%Shape: Square [id:dp6849151297260949]
\draw   (517.5,160) -- (557.5,160) -- (557.5,200) -- (517.5,200) -- cycle ;
%Straight Lines [id:da9518735421457183]
\draw    (477.5,180) -- (517.5,180) ;

% Text Node
\draw (255,169) node [anchor=north west][inner sep=0.75pt]  [font=\large] [align=left] {1};
% Text Node
\draw (336,168) node [anchor=north west][inner sep=0.75pt]  [font=\large] [align=left] {1};
% Text Node
\draw (452.5,169) node [anchor=north west][inner sep=0.75pt]  [font=\large] [align=left] {1};
% Text Node
\draw (533.5,168) node [anchor=north west][inner sep=0.75pt]  [font=\large] [align=left] {1};
\end{tikzpicture}
\end{center}
\end{figure}
\end{enumerate}

\newpage
\section{UV theory}
\label{sec:5}
Recall that the SW geometry is given by a polynomial map
$F:\mathbb{C}^{*n}\times\mathbb{C}^m\times\mathbb{C}^{\mu_1}\rightarrow\mathbb{C}^{\mu_1}\times\mathbb{C}$.
We have shown how to find the low energy
theory of generic vacua and special vacua by using the MHS and limit MHS for the
fiber $F^{-1}(\lambda_\alpha, t=0)$, where $\lambda_\alpha$ are the Coulomb
branch parameters.

Recall the vacua structure of pure \(SU(2)\) theory in Figure~\ref{}. The
methods we explained in previous sections can be used to solve the IR physics
for $u=\pm1$ and a generic $u$. There is yet another special vacuum at
$u=\infty$, which gives rise to the UV theory.
The purpose of this section is to show can we define a MHS for
the $t\to\infty$ limit ($t$ is the coordinate on $\mathbb{C}$), from which we
can derive the information of the UV theory.

Similar to the finite $t$ situation, there is also a limit MHS when
$t\to \infty$ \cite{sabbah2006hypergeometric}: there is a limit Hodge filtration
associated to the variation of Hodge structure around $t=\infty$, and the weight
filtration is given by the monodromy group around $t=\infty$ point.
One can also define the spectrum of this MHS, etc. It is in general, however, not
easy to compute the MHS in this limit. In the following, we will explain how to
use global Gauss--Manin system and the Bireskorn module to compute the limit MHS
at infinity.

\medskip\noindent%
\textbf{Gauss-Manin system and its Fourier-Laplace transform.}
Let \(U\) be a nonsingular affine algebraic variety of dimension \(n+1\). Let
\(f: U \to \mathbb{A}^1\) be a regular function. Then the (total) Gauss--Manin
system \(f_+\mathcal{O}_U\) of \(f\) is a complex of sheaves whose cohomology
sheaves are regular holonomic D-module. A representative of it
is the complex
\begin{equation*}
\cdots \to \Omega^{n}_U(U)[\tau] \xrightarrow{\mathrm{d}_U - \tau \mathrm{d}_{U}f} \Omega^{n+1}_U(U)[\tau].
\end{equation*}
The last item being placed on the cohomology degree \(0\).

In the above complex, \(\tau\) is merely a placeholder. Let us describe how the
Weyl algebra \(\mathbb{C}[t,\partial_t]\) acts. Here $t$ is the coordinate for
$\mathbb{A}^1$, and $\partial_t$ is the derivation which will give the
Gauss--Manin connection. A typical element in the complex is written
as \(\tau^{m}\omega\) where
\(\omega\) is a differential form on \(U\) and \(m\) is a non-negative integer.
Then
\begin{equation}
  \label{eq:rule-gm-partial-t}
  \partial_t \cdot \tau^{m}\omega = \tau^{m+1}\omega ,
\end{equation}
and
\begin{equation}
  \label{eq:rule-gm-t}
  t \cdot \tau^m\omega =   - m \tau^{m-1}\omega + \tau^{m}f(x)\omega.
\end{equation}

In many cases (e.g., when \(f\) is tame), the fibers only have interesting
cohomology in dimension \(n\).
Thus we shall look at the \(\mathbb{C}[t,\partial_t]\)-module
\(M=H^{0}f_+\mathcal{O}_U\)
(or its Fourier--Laplace transform on \(\check{\mathbb{A}}^1\), which will be
explained later). We call
\(f_+\mathcal{O}_U\) the \emph{total} Gauss--Manin system and we shall call its
zeroth cohomology sheaf \(M\) the \emph{Gauss--Manin system}.

In the $t$ plane, there are many singular fibers (and atypiical fibers)
including the one hidden at $t=\infty$. These
singularities produce regular singularities of the Gauss--Manin system \(M\).
We can similarly define limit MHS for the nearby and vanishing cycles%
\footnote{Strictly speaking, the vanishing cycle of \(M\) at infinity is not
  specified since we did not specify the singular fiber at infinity. What we
  meant here is rather the vanishing cycle of the \emph{microlocalization} of
  \(M\) at infinity. We shall not explain the technical details here.}
for the fiber at $t=\infty$. To compute this MHS, it is better to look the
picture after the Fourier-Laplace (FL) transformation.

We first consider a microlocalization of $M[\partial_t^{-1}]$
(forcing $\partial_t$ action to be invertible). The Fourier-Laplace
transform module $G$ of $M[\partial_t^{-1}]$ is now a
$\mathbb{C}[\tau, \tau^{-1}][\partial_\tau]$ module, the actions being
$\tau=\partial_t$, and $\partial_\tau=-t$.
The D-module $G$, has only two singular points on $\tau$ plane, one at $\tau=0$
which is still a regular singularity; the other one is at $\theta=0$ (here
$\theta={1\over \tau}$), and the singularity is typically an order two irregular
singularity.

Given a $\mathbb{C}[t]$ module $M_0$
(which generates $M$) of $M$,
one can produce a lattice $G_0$ of the Laplace transform $G$.
The construction is following: Let $M_0'$ be the image of
$M_0$ in $M[\partial_t^{-1}]$ under the morphism
$M\to M[\partial_t^{-1}]$.
Let $G_0=C[\theta]\cdot M_0^{'}$ be the $C[\theta]$ module generated by
$M_0^{'}$, then $G_0$ is a $\mathbb{C}[\theta]$ module.
It is a \textbf{lattice} of $G$ in the following sense:
$\mathbb{C}[\theta, \theta^{-1}]\otimes_{C[\theta]} G_0=G$.
\(G_0\) is also stable under $\theta^2 \partial_\theta$.

\medskip\noindent%
\textbf{Spectral numbers of the pair $(G, G_0)$.}
\label{sec:orga972328}
Let us consider the D-module $G$ and its lattice $G_0$ on the $\tau$ plane.
On the one hand, $G$ has regular singularity at $\tau=0$, and so we
can define a Kashiwara--Malgrange filtration.
The existence of lattice implies that we can define a filtration as follows
$G_k=\theta^{-k}G_0$. The combinations of these two filtrations make it possible
to define the notion of spectral pairs.

We now define the limit MHS and spectral numbers form the pair $(G,G_0)$.
Let us recall the formalism of Kashiwara--Malgrange filtration.
Let \(G\) be a \(\mathbb{C}[\tau,\partial_{\tau}]\)-module with a regular
singularity at \(\tau=0\) and regular elsewhere. Let
\begin{equation*}
G_{(\alpha)}=\bigcup_{n=1}^{\infty}\mathrm{Ker}(\tau\partial_{\tau}+\alpha)^n.
\end{equation*}
Then \(G_{(\alpha)}\) is called a \emph{root space} of \(G\). It is known that
\(G_{(\alpha)}\) is a finite dimensional vector space over \(\mathbb{C}\). Define
the (\emph{increasing}) Kashiwara--Malgrange filtration to be
\begin{equation*}
V_{\beta}M = \sum_{\alpha\leq \beta} G_{(\alpha)}.
\end{equation*}
Here the sum is taken in the following sense: it is generated by
\(\bigoplus_{\alpha\leq\beta} G_{(\alpha)}\) over \(\mathbb{C}[\tau]\), but is
contained in the infinite product \(\prod_{\alpha\leq\beta}G_{(\alpha)}\). Note
that \(\tau V_{\beta} \subset V_{\beta-1}\), and
\(\partial_{\tau}V_{\beta}\subset V_{\beta+1}\). We shall assume that all the
\(\alpha\) appeared above are real numbers.
For each \(\alpha\), we define its \emph{multiplicity} by
\begin{equation*}
\nu_\alpha = \dim \frac{V_{\alpha}G \cap G_0}{V_{\alpha}G \cap \theta G_0 + V_{<\alpha}G \cap G_0}.
\end{equation*}
The \emph{spectrum} of the pair \((G,G_0)\) is then the set of pairs
\begin{equation*}
(\alpha, \nu_{\alpha})
\end{equation*}
where \(\alpha\) is a real number, \(m_{\alpha} > 0\) is the multiplicity of
\(\alpha\). By assumption we have excluded numbers with zero multiplicity from
the spectrum.

\medskip%
The above definition needs the tameness of \(f\) to define the correct spectrum. In
fact, it is possible to define spectrum for any regular function
\(f: U \to \mathbb{A}^1\) in terms of limit Hodge theory. The present definition
has the virtue of being relatively easy to calculate.

\medskip\noindent%
\textbf{Tame polynomial.} The above procedure works for the general regular
holonomic D module on $t$ plane. If the polynomial $f:U\rightarrow A^1$ is a
tame polynomial, one can write $G$ and $G_0$ explicitly:
\begin{equation*}
G=\Omega^n[\tau,\tau^{-1}]/(d-\tau df)\Omega^{n-1}(U)[\tau, \tau^{-1}]
\end{equation*}
and
\begin{equation*}
G_0=\Omega^{n}(U)[\theta]/(\theta d-df\wedge)\Omega^{n-1}(U)[\theta]
\end{equation*}
The fiber at $\theta=0$ is given by
\begin{equation*}
G_0/\theta G_0=\Omega^n(U)/df\wedge \Omega^{n-1}(U)
\end{equation*}
which is similar to the local singularity case. The spectrum numbers for a tame
polynomial has following important property \cite{sabbah2006hypergeometric}:
\begin{itemize}
\item The spectrum number ${{n-1}\over2}$ is related to the mass parameter.
\item The spectrum number is symmetric with respect to ${n-1\over2}$.
\item There is only one minimal spectral number $\alpha_0$.
\end{itemize}
These spectral numbers characterizes the limit MHS and encode the physical
information of $UV$ theory. Notice that these are exactly similar to the
property of spectrum number of local singularity.

\subsection{Newton filtration and spectral numbers}

Similar to the IR theory, it is possible to compute the spectrum using the
Newton filtration. The major difference here is that the Newton polytope is
now the convex hull of the $\operatorname{Supp} f$ and the \textbf{origin}.
Let $f=\sum c_m x^m$ be a polynomial map $f:\mathbb{C}^{n+1}\to \mathbb{C}$,
and $\operatorname{Supp} f=\{m\in \mathbb{N}^{n+1}:c_m\neq0\}$ be its support.
$\Gamma_{+}(f)$ be the Newton polyhedron with respect to zero,
and $\Gamma(f)$ be the Newton boundary, i.e., a union of
compact faces $\sigma$ of $\Gamma_{+}(f)$. We further require that $f$ is Newton
non-degenerate and convenient.

Let's consider a homogeneous function $h:\mathbb{R}^{n+1}\to \mathbb{R}_{+}$,
$h(\lambda a)=\lambda h(a)$, such that $h(\Gamma)=1$.
$h$ then defines an \textbf{increasing} filtration on $\mathcal{O}=\mathbb{C}[\![x]\!]$:
\begin{equation}
N^a{\cal O}=\{g(x)\in \mathbb{C}[x]:h(\operatorname{Supp})g\leq a \}.
\end{equation}

This is called the Newton filtration.
Newton filtration induces a filtration $\bar{\nu}$ on the Jacobian algebra
$Q_f$. It also induces a filtration on $\Omega_{X}^{n+1}$ which is defined as
\begin{equation}
\nu(\omega)=\nu(gx_0\ldots x_n)-1,~~~\omega=g(x)dx_0\wedge\ldots \wedge dx_n
\end{equation}

Finally, we can define an order on the Brieskorn lattice
$M_0=\Omega_{X}^{n+1} / df\wedge d\Omega_{X}^{n-1}$:
\begin{equation}
  \bar{\nu}([\omega])=\min \{\nu(\eta):[\eta]=[\omega]\},
\end{equation}
and an increasing filtration on $M_0$:
\begin{equation}
N^a M_0=\{[\omega]\in M_0: \bar{\nu}([\omega])\leq a \}.
\end{equation}
Similarly we have a filtration on $\Omega_f=M_0/tM_0=\Omega_{X}^{n+1}/df\wedge \Omega_{X}^n$.
Given a vector space H with  a filtration $F^\bullet$, one can define the Poincare polynomial
\begin{equation}
P_{H,F^\bullet}(t)=\sum g_\alpha t^\alpha
\end{equation}
with $g_\alpha= \dim F^\alpha/F^{<\alpha}$.

The relation with the Poincaré polynomial of the spectrum at $\infty$ is then
\begin{equation}
\mathrm{Sp}(f)=\sum_{i=1}^\mu t^{\alpha_i}=P_{\Omega_f, V^\bullet}(t)
\end{equation}
Here $\alpha_i$ are the spectrum numbers.

The Poincaré polynomial on $\Omega_f$ has following simple description:
\begin{equation}
  \tilde{P}_{\Omega_f, V^\bullet}=
  t P_{\Omega_f, V^\bullet}(t)=\sum_\sigma (-1)^{n-dim~\sigma}(1-t)^{k(\sigma)} P_{A_\sigma(t)}+(-1)^{n+1}.
\end{equation}
Here we sum over the faces of $\Gamma(f)$, and $k(\sigma)$ is the dimension of
the minimal coordinate plane containing $\sigma$. $P_{A(\sigma)}$ is given as
\begin{equation}
P_{A_\sigma}(t)={1\over (1-t^{\omega_1})\ldots (1-t^{\omega_k})}
\end{equation}
Here, $A_\sigma$ is the cone over $\sigma$,
and it is isomorphic to a free algebra $A_\sigma$,
which is given by $\mathbb{C}[y_1,\ldots, y_k]$,
with $y_k$ the generators and are monomials of the coordinate
$x_i$; $\omega_1,\ldots, \omega_k$ are weights of $y_i$
which can be found from the weight on $\sigma$.

\medskip\noindent%
\textit{Example.}
Consider the polynomial $f(x,y)=x^{5 }+x^{2}y^{2}+y^2$.
The associated Newton polytope is given in Figure~\ref{figure-19}.
According to the formula in \ref{}, the smooth fiber has $h^{1,0}=h^{0,1}=3$
(the number of interior of lattice points),
and $h^{1,1}=2$ (the number of lattice points on $\Gamma(f)$).

$\Gamma(f)$ has five faces $\sigma_1, \sigma_2, \sigma_3, \sigma_{13},
\sigma_{23}$. Various data for the cones are collected in table.~\ref{default}.
The Poincare polynomial is given by
\begin{align}
& \tilde{P}={(1-t)^2\over (1-t^{1/2})(1-t^{1/2})}+{(1-t)^2\over (1-t^{1/2})(1-t^{1/5})} \nonumber \\
& -{(1-t)\over (1-t^{1/2})}-{(1-t)\over (1-t^{1/5})}-{(1-t)^2\over (1-t^{1/2})}+1 \nonumber\\
&= 2t+t^{1/2}+t^{7/10}+t^{9/10}+t^{11/10}+t^{13/10}+t^{3/2}
\end{align}
So the spectrum is
$(-{1\over2},-{3\over10},-{1\over10},0,0,{1\over10},{3\over10},{1\over2})$,
which is symmetric with respect to 0.

\begin{figure}[H]
\begin{center}
\tikzset{every picture/.style={line width=0.75pt}} %set default line width to 0.75pt
\begin{tikzpicture}[x=0.45pt,y=0.45pt,yscale=-1,xscale=1]
%uncomment if require: \path (0,479); %set diagram left start at 0, and has height of 479

%Shape: Axis 2D [id:dp2130006356073162]
\draw  (290,300.1) -- (580,300.1)(319,85) -- (319,324) (573,295.1) -- (580,300.1) -- (573,305.1) (314,92) -- (319,85) -- (324,92)  ;
%Straight Lines [id:da09519891652814394]
\draw [color={rgb, 255:red, 208; green, 2; blue, 27 }  ,draw opacity=1 ]   (318.75,260.75) -- (359.25,260.75) ;
%Straight Lines [id:da031831827687212044]
\draw [color={rgb, 255:red, 208; green, 2; blue, 27 }  ,draw opacity=1 ]   (360.75,260.75) -- (418.5,299.75) ;
%Shape: Circle [id:dp29589456503237677]
\draw  [fill={rgb, 255:red, 0; green, 0; blue, 0 }  ,fill opacity=1 ] (358,260.75) .. controls (358,259.23) and (359.23,258) .. (360.75,258) .. controls (362.27,258) and (363.5,259.23) .. (363.5,260.75) .. controls (363.5,262.27) and (362.27,263.5) .. (360.75,263.5) .. controls (359.23,263.5) and (358,262.27) .. (358,260.75) -- cycle ;
%Shape: Circle [id:dp969715718548126]
\draw  [fill={rgb, 255:red, 0; green, 0; blue, 0 }  ,fill opacity=1 ] (378,280.75) .. controls (378,279.23) and (379.23,278) .. (380.75,278) .. controls (382.27,278) and (383.5,279.23) .. (383.5,280.75) .. controls (383.5,282.27) and (382.27,283.5) .. (380.75,283.5) .. controls (379.23,283.5) and (378,282.27) .. (378,280.75) -- cycle ;
%Shape: Circle [id:dp5755414451740095]
\draw  [fill={rgb, 255:red, 0; green, 0; blue, 0 }  ,fill opacity=1 ] (339,260.75) .. controls (339,259.23) and (340.23,258) .. (341.75,258) .. controls (343.27,258) and (344.5,259.23) .. (344.5,260.75) .. controls (344.5,262.27) and (343.27,263.5) .. (341.75,263.5) .. controls (340.23,263.5) and (339,262.27) .. (339,260.75) -- cycle ;
%Shape: Circle [id:dp19415777282379132]
\draw  [fill={rgb, 255:red, 0; green, 0; blue, 0 }  ,fill opacity=1 ] (359,280.75) .. controls (359,279.23) and (360.23,278) .. (361.75,278) .. controls (363.27,278) and (364.5,279.23) .. (364.5,280.75) .. controls (364.5,282.27) and (363.27,283.5) .. (361.75,283.5) .. controls (360.23,283.5) and (359,282.27) .. (359,280.75) -- cycle ;
%Shape: Circle [id:dp8021759822787611]
\draw  [fill={rgb, 255:red, 0; green, 0; blue, 0 }  ,fill opacity=1 ] (340,279.75) .. controls (340,278.23) and (341.23,277) .. (342.75,277) .. controls (344.27,277) and (345.5,278.23) .. (345.5,279.75) .. controls (345.5,281.27) and (344.27,282.5) .. (342.75,282.5) .. controls (341.23,282.5) and (340,281.27) .. (340,279.75) -- cycle ;
%Curve Lines [id:da6055544611878965]
\draw    (390,265) .. controls (429.6,235.3) and (452.05,263.43) .. (491.31,234.88) ;
\draw [shift={(492.5,234)}, rotate = 503.13] [color={rgb, 255:red, 0; green, 0; blue, 0 }  ][line width=0.75]    (10.93,-3.29) .. controls (6.95,-1.4) and (3.31,-0.3) .. (0,0) .. controls (3.31,0.3) and (6.95,1.4) .. (10.93,3.29)   ;

% Text Node
\draw (304,240) node [anchor=north west][inner sep=0.75pt]   [align=left] {1};
% Text Node
\draw (359,236.4) node [anchor=north west][inner sep=0.75pt]    {$2$};
% Text Node
\draw (425.5,279.15) node [anchor=north west][inner sep=0.75pt]    {$3$};
% Text Node
\draw (503,215.4) node [anchor=north west][inner sep=0.75pt]    {$\Gamma ( f)$};

\end{tikzpicture}

\end{center}
\caption{}
\label{figure-19}
\end{figure}

\begin{table}[htp]
\begin{center}
\begin{tabular}{|c|c|c|c|} \hline
Cone & Generators & Weights & $k_\sigma$ \\ \hline
$\sigma_1$& $(0,1)$ & ${1\over 2}$ & 1 \\ \hline
$\sigma_2$ & $(1,1)$ & ${1\over2}$ & 2\\ \hline
$\sigma_3$ & $(1,0)$& ${1\over 5}$ & 1\\ \hline
$\sigma_{12}$ & $(0,1),(1,1)$&$({1\over 2},{1\over2})$ & 2\\ \hline
$\sigma_{23}$&$(1,0),(1,1)$&$({1\over 5},{1\over 2})$ &2 \\ \hline
\end{tabular}
\end{center}
\caption{The Newton information of \(f(x,y)=x^5+x^2y^2+y^2\)}
\label{default}
\end{table}%

\medskip\noindent%
\textbf{Under diagram deformation.}
For the curve case ($n=1$), we can compute the spectrum using the under-diagram
deformation: these points satisfy the following condition $m+(1,1)$ belongs to
the interior of the polytope, and the spectrum for such lattice points are
\begin{equation}
\alpha(x^mdx^m)=\nu(x^m x_0\ldots x_n)-1
\end{equation}
These give the negative part of the spectrum. Since the spectrum is symmetric respect to $0$ for $n=1$, we can find the full spectrum.

\medskip\noindent%
\textit{Example.} Consider the example in Figure~\ref{figure-19}, there are three
lattice points such that $m+(1,1)$ is in the interior of the polytope. The
spectrum number from them is $(-{1\over 2}, -{3\over 10}, -{1\over 10})$, which
agrees with the computation using Poincaré polynomial.

\subsection{Variance of spectrum numbers}
The spectral numbers of MHS play an important role in finding the Coulomb branch
information of the interacting theory (either IR or UV theory). Here we would
like to point out an inequality of the variance of spectral numbers,
which might have interesting implication on the property of 4d $\mathcal{N}=2$
theory.

First, for the spectral numbers associated with an isolated hypersurface
singularity (the map is $f:\mathbb{C}^{n+1}\to \mathbb{C}$), the variance is conjectured to
satisfy following inequality \cite{hertling2002frobenius}:

\begin{equation}\label{eq:conjUV}
{1\over \mu}\sum_{i=1}^\mu(\alpha_i-{n-1\over2})^2 \leq {\alpha_\mu -\alpha_1\over 12}
\end{equation}
In the case of quasi-homogeneous singularity, we have the equality which is also proven. Since the
quasi-homogeneous singularity gives the SCFT, the above equality gives a quite
interesting constraint on the Coulomb branch spectrum of a SCFT. Using the
relation between the spectral number and the Coulomb branch spectrum
($\alpha_i=\alpha_1(1-\Delta_i)+1,~n=3$), we get the following equality for the
Coulomb branch spectrum of a SCFT:
\begin{equation}\label{eq:conjUV}
{1\over 2 r+f}\sum_{i=1}^r (\Delta_i-1)^2={\Delta_{\max}(\Delta_{\max}-1)\over 12}
\end{equation}
Here $r$ is the rank of the 4d theory, $f$ is the rank of flavor symmetry.
$\Delta_{max}$ is the maximal Coulomb branch scaling dimension.

There is also a dual conjecture, which states that for a tame polynomial, the
spectral numbers at $t=\infty$ satisfies the inequality:
\begin{equation}
{1\over \mu}\sum_{i=1}^\mu(\alpha_i-{n-1\over2})^2 \geq {\alpha_\mu -\alpha_1\over 12}
\end{equation}
In the case of quasi-homogeneous singularity, we also have the equality.

Since the local singularity describes IR free (or conformal) theory, while the
$t\to \infty$ limit describes the UV complete theory, we might think of the above
variance on the spectral number as a method to measure whether the theory is IR
free, UV free, or conformal, namely, the variance of the spectral number is a
``$\beta$'' function.

\medskip\noindent%
\textit{Example.} Let's see the consequence on the Coulomb branch spectrum of a
SCFT if we assume the equalities in \ref{eq:conjUV} hold.
To begin with, assume that the Coulomb branch moduli is just one dimensional,
and the rank of flavor symmetry is $f$. Then we have
\begin{equation}
{\Delta-1\over 2+f}={\Delta \over 12} \to \Delta={12\over 10-f}
\end{equation}
We have the following solutions:

\begin{center}
  \begin{tabular}{|c||c|c|c|c|c|c|c|}
    \hline
    \(f\) & 0 & 1 & 2 &  4 & 6 & 7 & 8\\
    \hline
    \(\Delta\) & ${6\over 5}$ & ${4\over 3}$ & ${3\over2}$ & 2 & 3 & 4 & 6 \\
    \hline
  \end{tabular}
\end{center}
It is intriguing that these agree with the data of rank one theory studied in
the literature.

We have to emphasize that not all SCFT can be described by a quasi-homogeneous
polynomial. However, the above computations imply that maybe there is a more
general constraint on the variance of Coulomb branch spectrum of a
$\mathcal{N}=2$ SCFT.

\newpage
\section{General base}
\label{sec:6}

\subsection{Low energy effective theory: non-tame case}
In previous sections we have focused on the ``constant deformation'' of the SW
geometry associated with a tame polynomial.
The virtue of tameness is at least of two-fold:
\begin{itemize}
\item its critical points are all isolated, and
\item the vanishing cycles are all localized at the finite
  distance, thus we do not need to deal with ``hidden'' singularities.
\end{itemize}
In practice, one has to consider the SW geometries defined by non-tame
polynomials, as they could arise from situations when one takes some physical
parameters to be zero. Some of the analysis discussed in previous sections broke
down. In this section we discuss how to deal with this obstacle by turning on
another parameter.

To begin with, Let us explain why there could be vanishing cycles localized at
infinity in the non-tame case. To fix ideas, we consider the situation when
$U=\mathbb{C}^2$, and the SW geometry is defined by a family of curves:
$f(x,y)-t=0$. Consider the projective completion of this family
\begin{equation}
F_{t}(x,y,z)=f_d(x,y)+f_{d-1}z+\ldots+tz^d=0
\end{equation}
where $(x,y,z)$ are homogeneous coordinates of the complex projective plane
$\mathbb{P}^{2}$,
and $d$ is the degree of $f$. The line at infinity is $z=0$, and points at
infinity are defined by the equation
\begin{equation}
f_d(x,y)=0
\end{equation}
A point at infinity is singular if and only if it satisfies the equation
\begin{equation}
{\partial f_d \over \partial x}={\partial f_d \over \partial y}=f_{d-1}=f_d=0.
\end{equation}
Let \(p\) be such a point. Then the localization of the polynomial near \(p\)
has a well-defined local Milnor number.
There is a dense Zariski open subset of \(\mathbb{C}\), such that for \(t\) in
this open, the sum of the Milnor numbers \(\mu^{\mathrm{gen}}\) at infinity is a
constant. We say \(t\) is a ``jump value'', if the number
\begin{equation}\label{eq:vanishing-cycle-non-tame}
\nu_{t}=\mu_{t}-\mu^{\mathrm{gen}},
\end{equation}
is bigger than zero.
Here $\mu_{t}$ is the sum of Milnor numbers of points at infinity lying in
\(F_{t}(p) = 0\).

When \(t\) is a jump value, we say there is a vanishing cycle
hidden at infinity.

\medskip\noindent%
\textit{Example.}
Consider the polynomial $f=xy^2-y$. Then we have $F=xy^2-y z^2-tz^3$. The points
at infinity (for each \(t\)) are $(0,1,0)$ and $(1,0,0)$. For each \(t\), the
singular point is $p=(1,0,0)$. The singularity type of $p$ at $t\neq 0$ is
$y^2-z^3$, and so the Milnor number is $2$. When \(t=0\), the singularity type
is $y^2-yz^2$ and so the Milnor number is $3$. Thus we say there is a vanishing
cycle hidden at infinity when $t=0$.

\medskip%
Assuming $f$ has only isolated critical points and $f$ is non-degenerate,
the dimension of $H^{1}(f^{-1}(t))$ and the local Milnor numbers satisfy the
following.
\begin{enumerate}
\item Denote by \(m\) the dimension of first cohomology of the generic fiber.
  Define
  \begin{equation}
    \mu = \sum_{x\in \Sigma} \mu_x, \quad \nu=\sum \nu_{t}
  \end{equation}
  Here $\Sigma$ is the set of critical points of $f$. We then have
  \begin{equation}
    m=\mu+\nu
  \end{equation}
\item Consider an arbitrary fiber $f^{-1}(t)$, and we have the relation
  \begin{equation}
    m=\dim(H^1(f^{-1}(t)))+(\mu_t+\nu_t)
  \end{equation}
  Here $\mu_s$ is the sum of Milnor numbers of the singular points on $f(t)=0$,
  and $\nu_t$ is the Milnor number of the vanishing cycles hidden at $\infty$ of the
  fiber $f(t)=0$.
\end{enumerate}

The second equation gives the relation between the cohomology of nearby fiber,
special fiber, and the vanishing cohomology on the special fiber which includes also 
the vanishing cycle at infinity.

The existence of vanishing cycles hidden at infinity has two
implications:
\begin{enumerate}
\item The total vanishing cycles are not supported at the singularity visible by
  the variety defining the SW geometry, and we need to compute it using the
  nearby cycles and the the property of singular fibers itself.
\item It is possible that the $t$ direction does not give us the correct
  ``generic fiber'', so we have to consider other directions to get the low
  energy physics.
\end{enumerate}

\medskip\noindent%
\textit{Example.} The first phenomenon can be explained by following polynomial
$f=x^n+z^k$, where $z$ is a $\mathbb{C}^*$ variable. The fiber at $t=0$ is
atypical, but the variety is smooth. The vanishing cycle can be thought of as
supported at the origin $(x,z)=(0,0)$ which is actually removed.

\medskip\noindent%
\textit{Example.}
The second phenomenon is more nontrivial.
Consider a theory whose SW geometry is described by
the polynomial
\begin{equation}\label{eq:xie-dan}
f=x^4+x^2y^5+a^2y^3
\end{equation}
Here $x,y$ are $\mathbb{C}$ variables. This describes a theory with flavor
symmetry $U(2)$, and $a$ is the mass parameter.

\begin{figure}[H]
\begin{center}
\tikzset{every picture/.style={line width=0.75pt}} %set default line width to 0.75pt

\begin{tikzpicture}[x=0.45pt,y=0.45pt,yscale=-1,xscale=1]
%uncomment if require: \path (0,479); %set diagram left start at 0, and has height of 479

%Shape: Axis 2D [id:dp47858939468805173]
\draw  (120,281) -- (325,281)(140,120) -- (140,293) (318,276) -- (325,281) -- (318,286) (135,127) -- (140,120) -- (145,127)  ;
%Straight Lines [id:da19314920562846916]
\draw    (180,181) -- (222,281) ;
%Shape: Circle [id:dp7492666853825705]
\draw  [fill={rgb, 255:red, 0; green, 0; blue, 0 }  ,fill opacity=1 ] (177,181) .. controls (177,179.34) and (178.34,178) .. (180,178) .. controls (181.66,178) and (183,179.34) .. (183,181) .. controls (183,182.66) and (181.66,184) .. (180,184) .. controls (178.34,184) and (177,182.66) .. (177,181) -- cycle ;
%Shape: Circle [id:dp696681044741166]
\draw  [fill={rgb, 255:red, 0; green, 0; blue, 0 }  ,fill opacity=1 ] (219,281) .. controls (219,279.34) and (220.34,278) .. (222,278) .. controls (223.66,278) and (225,279.34) .. (225,281) .. controls (225,282.66) and (223.66,284) .. (222,284) .. controls (220.34,284) and (219,282.66) .. (219,281) -- cycle ;
%Straight Lines [id:da9168427996211557]
\draw    (140,221) -- (180,181) ;
%Shape: Circle [id:dp05175395525638393]
\draw  [fill={rgb, 255:red, 0; green, 0; blue, 0 }  ,fill opacity=1 ] (137,221) .. controls (137,219.34) and (138.34,218) .. (140,218) .. controls (141.66,218) and (143,219.34) .. (143,221) .. controls (143,222.66) and (141.66,224) .. (140,224) .. controls (138.34,224) and (137,222.66) .. (137,221) -- cycle ;
%Shape: Circle [id:dp21806166811376237]
\draw  [fill={rgb, 255:red, 0; green, 0; blue, 0 }  ,fill opacity=1 ] (157,242) .. controls (157,240.34) and (158.34,239) .. (160,239) .. controls (161.66,239) and (163,240.34) .. (163,242) .. controls (163,243.66) and (161.66,245) .. (160,245) .. controls (158.34,245) and (157,243.66) .. (157,242) -- cycle ;
%Shape: Circle [id:dp9730626650831553]
\draw  [fill={rgb, 255:red, 0; green, 0; blue, 0 }  ,fill opacity=1 ] (177,261) .. controls (177,259.34) and (178.34,258) .. (180,258) .. controls (181.66,258) and (183,259.34) .. (183,261) .. controls (183,262.66) and (181.66,264) .. (180,264) .. controls (178.34,264) and (177,262.66) .. (177,261) -- cycle ;
%Shape: Axis 2D [id:dp16955173747415553]
\draw  (419,282) -- (624,282)(439,121) -- (439,294) (617,277) -- (624,282) -- (617,287) (434,128) -- (439,121) -- (444,128)  ;
%Straight Lines [id:da42896521271417676]
\draw    (479,182) -- (521,282) ;
%Shape: Circle [id:dp1401308055077064]
\draw  [fill={rgb, 255:red, 0; green, 0; blue, 0 }  ,fill opacity=1 ] (476,182) .. controls (476,180.34) and (477.34,179) .. (479,179) .. controls (480.66,179) and (482,180.34) .. (482,182) .. controls (482,183.66) and (480.66,185) .. (479,185) .. controls (477.34,185) and (476,183.66) .. (476,182) -- cycle ;
%Shape: Circle [id:dp7883250153403962]
\draw  [fill={rgb, 255:red, 0; green, 0; blue, 0 }  ,fill opacity=1 ] (518,282) .. controls (518,280.34) and (519.34,279) .. (521,279) .. controls (522.66,279) and (524,280.34) .. (524,282) .. controls (524,283.66) and (522.66,285) .. (521,285) .. controls (519.34,285) and (518,283.66) .. (518,282) -- cycle ;
%Straight Lines [id:da41601711537870734]
\draw    (439,282) -- (479,182) ;
%Shape: Circle [id:dp05269893174828799]
\draw  [fill={rgb, 255:red, 0; green, 0; blue, 0 }  ,fill opacity=1 ] (157,201) .. controls (157,199.34) and (158.34,198) .. (160,198) .. controls (161.66,198) and (163,199.34) .. (163,201) .. controls (163,202.66) and (161.66,204) .. (160,204) .. controls (158.34,204) and (157,202.66) .. (157,201) -- cycle ;
%Shape: Circle [id:dp9438294872584871]
\draw  [fill={rgb, 255:red, 0; green, 0; blue, 0 }  ,fill opacity=1 ] (456,242) .. controls (456,240.34) and (457.34,239) .. (459,239) .. controls (460.66,239) and (462,240.34) .. (462,242) .. controls (462,243.66) and (460.66,245) .. (459,245) .. controls (457.34,245) and (456,243.66) .. (456,242) -- cycle ;
%Shape: Circle [id:dp5321624823133164]
\draw  [fill={rgb, 255:red, 0; green, 0; blue, 0 }  ,fill opacity=1 ] (477,241) .. controls (477,239.34) and (478.34,238) .. (480,238) .. controls (481.66,238) and (483,239.34) .. (483,241) .. controls (483,242.66) and (481.66,244) .. (480,244) .. controls (478.34,244) and (477,242.66) .. (477,241) -- cycle ;
%Shape: Circle [id:dp6007981702680196]
\draw  [fill={rgb, 255:red, 0; green, 0; blue, 0 }  ,fill opacity=1 ] (496,242) .. controls (496,240.34) and (497.34,239) .. (499,239) .. controls (500.66,239) and (502,240.34) .. (502,242) .. controls (502,243.66) and (500.66,245) .. (499,245) .. controls (497.34,245) and (496,243.66) .. (496,242) -- cycle ;
%Shape: Circle [id:dp0020845489663905425]
\draw  [fill={rgb, 255:red, 0; green, 0; blue, 0 }  ,fill opacity=1 ] (497,262) .. controls (497,260.34) and (498.34,259) .. (500,259) .. controls (501.66,259) and (503,260.34) .. (503,262) .. controls (503,263.66) and (501.66,265) .. (500,265) .. controls (498.34,265) and (497,263.66) .. (497,262) -- cycle ;
%Shape: Circle [id:dp5689795382081864]
\draw  [fill={rgb, 255:red, 0; green, 0; blue, 0 }  ,fill opacity=1 ] (477,261) .. controls (477,259.34) and (478.34,258) .. (480,258) .. controls (481.66,258) and (483,259.34) .. (483,261) .. controls (483,262.66) and (481.66,264) .. (480,264) .. controls (478.34,264) and (477,262.66) .. (477,261) -- cycle ;
%Shape: Circle [id:dp7260910553853885]
\draw  [fill={rgb, 255:red, 0; green, 0; blue, 0 }  ,fill opacity=1 ] (456,261) .. controls (456,259.34) and (457.34,258) .. (459,258) .. controls (460.66,258) and (462,259.34) .. (462,261) .. controls (462,262.66) and (460.66,264) .. (459,264) .. controls (457.34,264) and (456,262.66) .. (456,261) -- cycle ;
%Shape: Circle [id:dp39075215779133643]
\draw  [fill={rgb, 255:red, 0; green, 0; blue, 0 }  ,fill opacity=1 ] (477,202) .. controls (477,200.34) and (478.34,199) .. (480,199) .. controls (481.66,199) and (483,200.34) .. (483,202) .. controls (483,203.66) and (481.66,205) .. (480,205) .. controls (478.34,205) and (477,203.66) .. (477,202) -- cycle ;
%Shape: Circle [id:dp7903849146749593]
\draw  [fill={rgb, 255:red, 0; green, 0; blue, 0 }  ,fill opacity=1 ] (477,220) .. controls (477,218.34) and (478.34,217) .. (480,217) .. controls (481.66,217) and (483,218.34) .. (483,220) .. controls (483,221.66) and (481.66,223) .. (480,223) .. controls (478.34,223) and (477,221.66) .. (477,220) -- cycle ;
%Shape: Circle [id:dp7403317482304204]
\draw  [fill={rgb, 255:red, 0; green, 0; blue, 0 }  ,fill opacity=1 ] (196,262) .. controls (196,260.34) and (197.34,259) .. (199,259) .. controls (200.66,259) and (202,260.34) .. (202,262) .. controls (202,263.66) and (200.66,265) .. (199,265) .. controls (197.34,265) and (196,263.66) .. (196,262) -- cycle ;
%Shape: Circle [id:dp2818489174519636]
\draw  [fill={rgb, 255:red, 0; green, 0; blue, 0 }  ,fill opacity=1 ] (177,202) .. controls (177,200.34) and (178.34,199) .. (180,199) .. controls (181.66,199) and (183,200.34) .. (183,202) .. controls (183,203.66) and (181.66,205) .. (180,205) .. controls (178.34,205) and (177,203.66) .. (177,202) -- cycle ;
%Shape: Circle [id:dp09352759314267112]
\draw  [fill={rgb, 255:red, 0; green, 0; blue, 0 }  ,fill opacity=1 ] (177,221) .. controls (177,219.34) and (178.34,218) .. (180,218) .. controls (181.66,218) and (183,219.34) .. (183,221) .. controls (183,222.66) and (181.66,224) .. (180,224) .. controls (178.34,224) and (177,222.66) .. (177,221) -- cycle ;
%Shape: Circle [id:dp7588186308472082]
\draw  [fill={rgb, 255:red, 0; green, 0; blue, 0 }  ,fill opacity=1 ] (177,241) .. controls (177,239.34) and (178.34,238) .. (180,238) .. controls (181.66,238) and (183,239.34) .. (183,241) .. controls (183,242.66) and (181.66,244) .. (180,244) .. controls (178.34,244) and (177,242.66) .. (177,241) -- cycle ;
%Shape: Circle [id:dp2993291314173172]
\draw  [fill={rgb, 255:red, 0; green, 0; blue, 0 }  ,fill opacity=1 ] (157,221) .. controls (157,219.34) and (158.34,218) .. (160,218) .. controls (161.66,218) and (163,219.34) .. (163,221) .. controls (163,222.66) and (161.66,224) .. (160,224) .. controls (158.34,224) and (157,222.66) .. (157,221) -- cycle ;
%Shape: Circle [id:dp46368470639567794]
\draw  [fill={rgb, 255:red, 0; green, 0; blue, 0 }  ,fill opacity=1 ] (195,241) .. controls (195,239.34) and (196.34,238) .. (198,238) .. controls (199.66,238) and (201,239.34) .. (201,241) .. controls (201,242.66) and (199.66,244) .. (198,244) .. controls (196.34,244) and (195,242.66) .. (195,241) -- cycle ;
%Shape: Circle [id:dp9455904386352185]
\draw  [fill={rgb, 255:red, 0; green, 0; blue, 0 }  ,fill opacity=1 ] (157,261) .. controls (157,259.34) and (158.34,258) .. (160,258) .. controls (161.66,258) and (163,259.34) .. (163,261) .. controls (163,262.66) and (161.66,264) .. (160,264) .. controls (158.34,264) and (157,262.66) .. (157,261) -- cycle ;
\end{tikzpicture}
\end{center}
\caption{Turning on the mass parameter}
\label{fig:xie-dan}
\end{figure}
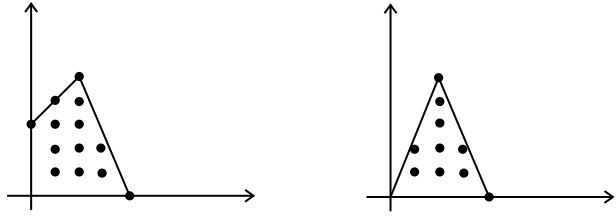

\begin{enumerate}[wide]
\item For $a\neq 0$, the dimension of first cohomology group in $t$ direction is
  \(20\): $H^{1}(f_{a\neq0}^{-1}(t)) \cong \mathbb{C}^{20}$. (One computes by
  using the combinatorial formula in \ref{sec:3.3}, the Newton polygon is the
  left hand polyhedron in Figure~\ref{fig:xie-dan}).
\item If $a=0$, then the generic fiber has a \(17\) dimensional cohomology group
  $H^{1}(f_{a=0}^{-1}(t)) \cong \mathbb{C}^{17}$. The Newton polyhedron is the
  left polyhedron in Figure~\ref{fig:xie-dan}).
\end{enumerate}

However, the correct low energy theory should only has rank $9$ (given by the
\(h^{1,1}\) of the ``generic fiber'', see the computation below). In order to get
the correct answer at the vacuum with $a=0$ and generic $t$, we should turn on the parameter
\(a\) and view the theory as a specialization of~\eqref{eq:xie-dan} when
\(a \to 0\).

\begin{figure}[H]
\begin{center}

\tikzset{every picture/.style={line width=0.75pt}} %set default line width to 0.75pt

\begin{tikzpicture}[x=0.45pt,y=0.45pt,yscale=-1,xscale=1]
%uncomment if require: \path (0,479); %set diagram left start at 0, and has height of 479

%Curve Lines [id:da787560965228935]
\draw    (298,235) .. controls (338,205) and (456,320) .. (496,290) ;
%Curve Lines [id:da29278855470752085]
\draw [color={rgb, 255:red, 208; green, 2; blue, 27 }  ,draw opacity=1 ]   (335,295) .. controls (375,265) and (365,155) .. (405,125) ;
%Shape: Circle [id:dp5843168468049498]
\draw  [fill={rgb, 255:red, 0; green, 0; blue, 0 }  ,fill opacity=1 ] (366,215) .. controls (366,213.34) and (367.34,212) .. (369,212) .. controls (370.66,212) and (372,213.34) .. (372,215) .. controls (372,216.66) and (370.66,218) .. (369,218) .. controls (367.34,218) and (366,216.66) .. (366,215) -- cycle ;
%Curve Lines [id:da39180145558561175]
\draw [color={rgb, 255:red, 80; green, 227; blue, 194 }  ,draw opacity=1 ]   (295,202) .. controls (335,172) and (453,287) .. (493,257) ;

% Text Node
\draw (473,274.4) node [anchor=north west][inner sep=0.75pt]    {$a$};
% Text Node
\draw (380,131.4) node [anchor=north west][inner sep=0.75pt]    {$t$};
% Text Node
\draw (375,196.4) node [anchor=north west][inner sep=0.75pt]    {$( 0,t_0)$};

\end{tikzpicture}
\end{center}
\caption{Different specializations}
\label{fig:different-para}
\end{figure}
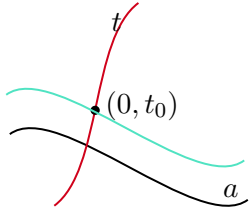

Recall that we are dealing with the cohomology of the curve
\begin{equation*}
x^4+x^2y^5+a^2 y^3-t=0.
\end{equation*}
If we still adapt the ``tame case'' strategy, we would have fixed $a$, and vary
$t$. See the red curve in Figure~\ref{fig:different-para}.
As we have said, this causes the problem if $a=0$,
Instead we turn on the deformation \(a\) and hold \(t=t_0\) as fixed.
Thus we should use the blue curve in Figure~\ref{fig:different-para} as our base
of deformation. Define
\begin{equation*}
  X = \{(x,y,a)  \in \mathbb{C}^{2} \times \mathbb{C}: x^4+x^2y^5+a^2 y^3-t=0\}.
\end{equation*}
Then the projection \((x,y,a) \mapsto a\) defines an algebraic function
\(g\colon X \to \mathbb{C}\). The correct low energy effective theory should be
read out by the vanishing cycle \(\phi_{g}M\), where \(M\) is the Gauss--Manin
system \(H^{0}g_{+}\mathcal{O}_{X}\) on \(\mathbb{C}\). Note that this time the
Gauss--Manin system is associated with the function \(g\) on the complex
analytic space \(X\) instead of that of \(f\).

The vanishing cycle can be calculated
using~\eqref{eq:vanishing-cycle-non-tame}, as follows.
The projective completion of the curve is
\begin{equation*}
x^4 z^3+ x^2 y^5+a^2y^3 z^4-t_0 z^7=0
\end{equation*}
When $a\neq 0$,
there are two singularities at $z=0$ patch,
and the local form are given by:
\begin{enumerate}[wide]
\item[(a)] $x^2+z^4=0$;
\item[(b)] $z^3+y^5=0$.
\end{enumerate}
The local Milnor numbers are respectively
$\mu_1=3$ and $\mu_2=8$. For $a=0$, the type (a) singularity at $z=0$ is changed
to $x^2-t_0 z^7=0$ (with Milnor number 6) and the type (b) is not changed.
So we see that there are three vanishing cycles hidden at infinity.
To really determine whether these extra cycles are vanishing are not,
we need to compute the monodromy and the asymptotical behavior of the period
integral over these three cycles. Instead of doing the explicit computation,
we can get the result using MHS. The fiber with $a=0$ is smooth,
and the Hodge numbers are $h^{1,0}=8,h^{1,1}=1$. For the nearby fibers ($a\neq 0$),
we have $h^{1,0}=9$ and $h^{1,1}=2$. This implies that cohomology associated
with the ``vanishing cycle at infinity'', we have $h^{1,1}=1$, $h^{1,0}=h^{0,1}=1$. This 
suggests that there might be trivial monodromy and the theory at $a=0, t_0$ is a generic vacuum. 

\medskip\noindent%
\textbf{Summary.} In previous sections, we treated exclusively tame
polynomials. The tameness implies that the nearby cycle treated there can be regarded as
the fiber of the generic vacuum. However, in the non-tame case, as the above
example illustrates, we need to turn on a ``generic'' parameter, and calculate
the vanishing cycle along the ``generic direction'', in order to
decide the low energy effective theory.

Finally, we would like to point out that it is also possible that the vanishing
cycle is supported not on isolated point, but on one dimensional locus, and this
happens if the singular locus is not isolated. For example, in the $t=0, a=0$
limit, the singular fiber has non-isolated singularity. The study of low energy
physics of this particular point is more complicated, and we leave it for future
study.

\subsection{Discriminant locus and global structure of the base}

We now discuss the global structure of the base, i.e.,
the full Coulomb branch parameter space.
The base is separated into generic vacua and special vacua.
The space of special vacua has a complicated structure given by a
stratification, which depends upon the type of low energy effective theory.
In our study, the space of special vacua are given by the locus where
the topology of the SW geometries change
(we recapitulate that this does not necessarily imply that the fibers become
singular, it could be that there are fibers which have vanishing cycles hidden
at infinity). In mathematics, the locus of special vacua is called the
discriminant locus of the moduli space, and its structure is not easy to
determine in general.

In this subsection we only briefly describe a class of examples where one can
analyze the discriminant locus explicitly, and leave the general study in a
future publication.

The models are the so-called $(A_1, G)$ (with $G=ADE$)
Argyres-Douglas theories. The most down-to-earth case is the
$(A_1, A_n)$ theory, whose SW geometry is given by
\begin{equation}
F=x^2+y^{n+1}+\sum_{i=1}^{n-1}\lambda_i y^{n-i};
\end{equation}
The discriminant locus of this model is given by the locus where $F$ becomes
singular, namely,
\begin{equation}
{\partial F\over \partial x}={\partial F\over \partial y}=F=0
\end{equation}
This still takes a complicated form.
However, it is shown in \cite{arnold2012singularity} that the base could be written as the orbifold
\begin{equation}
{\mathbb{C}^{n}/W},
\end{equation}
where $W$ is the Weyl group of group $A_{n}$.
We may describe it more explicitly.
First, we embed $\mathbb{C}^n$ into $\mathbb{C}^{n+1}$ by the equation
$x_1+\ldots+x_{n+1}=0$, and $W$ is generated by the Weyl reflection  $s_{ij}$
associated with hyperplane $H_{ij}$ defined by $x_i=x_j$:
\begin{equation}
s_v:x\to x-2(x,v)/(v,v)
\end{equation}
Here $v\in h$, and $(\cdot, \cdot)$ is the standard inner product.
Now the space $\mathbb{C}^n/W$ has following structure
\begin{enumerate}
\item The invariant polynomial of $\mathbb{C}^n/W$ is generated by
  $\sigma_1,\ldots, \sigma_n$ with degree $(2,\ldots, n+1)$, and it is freely
  generated so that as a complex analytic space it is isomorphic to
  $\mathbb{C}^n$. This is in agreement with the fact that the coordinate ring of
  the Coulomb branch of $\mathcal{N}=2$ theories is free generated.
\item The discriminant locus is given by the union of hyperplane $H_{ij}$.
\end{enumerate}

This gives a nice description of the global structure of the $(A_1, A_n)$
theory. For the general $(A_1, G)$ theory, the parameter space is given by
$\mathbb{C}^{\mu}/W$, here $\mu$ is the rank of $G$, and $W$ is the Weyl group
of type $G$. The description of the discriminant locus is the same as that of \(A\)
type.

\newpage
\section{Seiberg-Witten differential}
\label{sec:7}

The SW geometries $F:U\times \mathbb{C}^{\mu_1}\to \mathbb{C}^{\mu_1}\times
\mathbb{C}$ are parametrized by $\mathbb{C}^{\mu_1}$ which consists of coupling
constants and Coulomb branch moduli. If we fix the coupling constants, and vary
the Coulomb branch moduli, we should get a special K\"ahler structure.

The extra datum one needs in the definition of a SKG is a SW differential
$\lambda_{SW}$. The SW differential will be a section of the
cohomology bundle called primitive form, which is defined over the whole base
but is more or less determined by a constant deformation of a generic vaccum.
The SKG is obtained by the periods of the primitive form.

\subsection{Primitive form and SW differential}

As we have reviewed in section 2, the low energy effective action on the Coulomb
branch has a rigid special K\"ahler structure. We recall the main definition here.
A rigid special K\"ahler manifold is a complex manifold $M$, and on each chart
there are $n$ closed holomorphic 1-forms $U_\alpha dz^\alpha$:
\begin{equation}
  \partial_{\bar{\alpha}}U_\beta=0,
  \quad
  \partial_{\alpha} U_{\beta} - \partial_{\beta} U_{\alpha} =0.
\label{sk1}
\end{equation}
These functions satisfies following three conditions:
\begin{itemize}
\item the holomorphic function $U_\alpha$ satisfies:
\begin{equation}
  \langle U_\alpha, U_\beta \rangle =0,~~~g_{\alpha\bar{\beta}}=i\langle U_\alpha, \bar{U}_{\bar{\beta}} \rangle~\text{is the positive definite K\"ahler metric}.
\label{specials}
\end{equation}
Here the $2n$-dimensional vector $U$ is written as $U_\alpha=(e_\alpha^A,
h_\alpha^A),~~A=1,\ldots,n$, and the symplectic pairing is defined as
\begin{equation*}
 \langle U_\alpha, U_\beta \rangle =e_\alpha^A h_\beta^A-h_\alpha^A e_\beta^A,~~~ \langle U_\alpha, \bar{U}_{\bar{\beta}} \rangle =e_\alpha^A \bar{h}_{\bar{\beta}}^A-h_\alpha^A \bar{e}_{\bar{\beta}}^A
\end{equation*}
\end{itemize}

We'd like to derive a special K\"ahler geometry from the SW geometry.
This is achieved by using periods of integrals of a SW differential.
SW differential $\Omega$ is a section of cohomology bundle over the
parameter space.
Let's choose a basis of homology cycles as $(A_i, B_i), i=1,\ldots, r$ and
$L_i,i=1,\ldots, f$, and the nonzero intersection form takes the following form
\begin{equation*}
A_i\cdot B_j=\delta_{ij}
\end{equation*}
The holomorphic functions $U_\alpha^A$ is given by the period integral of SW
differential $\Omega$:
\begin{equation*}
h_\alpha^A=\int_A \partial_{\lambda_\alpha} \Omega,~~e_\alpha^B=\int_B \partial_{\lambda_\alpha} \Omega
\end{equation*}
Here $\lambda_\alpha$ parameterizes Coulomb branch moduli.  The period integral is holomorphic and so that the first condition in
\ref{sk1} is automatically satisfied, and the second equation is also automatically satisfied. The conditions in \ref{specials} put  following constraints on the SW differential:

\refstepcounter{equation}
\begin{quote}
  (\theequation)\label{okkk}
  $\partial_{\lambda_\alpha}\Omega$ gives a basis for the space
  $H^{{n+1\over 2},{n-1\over2}}$
  (this is given by the mixed Hodge structure, and $n=1$ or $n=3$).
\end{quote}

Then the condition in \ref{specials} is given by the first and second
Hodge--Riemann bilinear relation.

Let's give an example in the curve case. We
have a basis $\omega_\alpha$ of $H^{1,0}$,
\(\alpha=1,\ldots, n\)
(which is given by the derivative of the SW differential), and a symplectic
basis of one cycles
$(A_i,B_i)$, \(i=1,\ldots, n\). The period matrix is a
$g\times 2g$ matrix $[\int_{A_j}\omega_\alpha,\int_{B_j}\omega_\alpha]=[e^A_{\alpha}, h_{B\alpha}]$,
and the period matrix is given by
\begin{equation*}
\Pi_{AB}=h_{A\alpha}{(e^{-1}})^{\alpha}_B.
\end{equation*}
The first Hodge--Riemann bilinear relation is
\begin{equation*}
  \int \omega \wedge \eta =0, \quad \forall \omega, \eta \in H^{1,0}
\end{equation*}
This is because $\omega\wedge \eta$ is a $(2,0)$ form, and on the curve a
$(2,0)$ form is zero. The second Riemann bilinear relation is
\begin{equation*}
\sqrt{-1}\int_X \omega\wedge \bar{\omega} >0
\end{equation*}
here $\omega \in H^{1,0}$.
The first and second Riemann bi-linear relation implies that
\begin{equation*}
\Pi_{AB}=\Pi_{BA},\quad \text{Im}(\Pi_{AB})>0
\end{equation*}
These two conditions gives the condition listed in \ref{specials}.

The SW differential can not be uniquely fixed by above condition. We need to
impose one further condition on the asymptotical behavior of the period integral
of $\Omega$ over vanishing cycles. The condition is following

\begin{quote}
  Consider the periods of $\Omega$ over the vanishing homology classes of the $A_1$
  singularity $x_0^2+\ldots+x_n^2=0$. Physically, this gives a massless
  hypermultiplet. The periods should satisfy
  \begin{equation}
    \int_A \Omega = t+\ldots
    \label{sk3}
  \end{equation}
  Here $t$ is the coordinate of the base of Milnor fiberation. This requirement
  can be used to determine SW differential.
\end{quote}

\medskip\noindent%
\textbf{Primitive form.} To get a SW differential, one needs to find a section of
the cohomology bundle so that its derivative along the Coulomb branch moduli
gives the vector in the space $H^{n,0}$ which is defined using MHS. It turns
out that the primitive form would give the desired answer \cite{Li:2018rdd}. Here we give
a brief review with emphases on the use of MHS.

Fix the Coulomb branch parameters and assume $f$ is tame. We have defined a
Gauss-Manin system $M$ over the affine line parameterized by $t$ and a Brieskorn
lattice $M_0$. Both of them can be expressed in terms of spaces of differential
forms.

For our purposes, it is useful to consider their a Fourier--Laplace
transforms. The reader should recall the setting in \S\ref{sec:5}.
Thus we shall consider the D-module $G$ and its Bireskorn lattice
$G_0$, which are of the form
\begin{equation*}
G=\Omega^n[\tau,\tau^{-1}]/(d-\tau df)\Omega^{n-1}(U)[\tau, \tau^{-1}]
\end{equation*}
and
\begin{equation*}
G_0=\Omega^{n}(U)[\theta]/(\theta d-df\wedge)\Omega^{n-1}(U)[\theta]
\end{equation*}
Here $\tau$ is the dual coordinate, and $\theta= {1\over \tau}$.

It is shown in
\cite{saito_morihiko:structure-brieskorn-lattice,sabbah2006hypergeometric}
one can find a basis of $G_0$ satisfying following conditions
\begin{itemize}
\item The basis $v_1, \ldots, v_{\mu}$ satisfying following equations
\begin{equation*}
\theta^2\partial_\theta v= A_0 v+\theta A_1,
\end{equation*}
where \(A_0\) is nilpotent,
$A_1$ is a diagonalizable matrix, whose eigenvalues are the exponents
(which is equal to the minus one of the spectrum).
\item If we consider the full SW geometry
\begin{equation*}
F=f(x,y)+\sum_{\alpha=1}^{\mu_1} \lambda_\alpha \phi_\alpha
\end{equation*}
The above basis can be extended\footnote{At least in the examples we
  considered.}
to the basis over the base parameterized by
$\lambda_{\alpha}$.

\item Furthermore, one can choose $v_0$ on the full base so that
  $v_i=\partial_i v_0$, and here the derivative is over special coordinate (not
  the naive coordinate $\lambda_\alpha$).
\end{itemize}

The section $v_0$ is called primitive form and which gives the minimal spectrum. The
existence of it depends crucially on the MHS studied in the UV limit in section
\ref{sec:5}.

The upshot is the following: in our case, the spectrum is symmetric with respect
to ${n-1\over 2}$, and if we consider the derivative along the coordinates whose
spectrum is less than ${n-1\over 2}$, namely we consider
$v_0, \partial_{i} v_0, i=1, \ldots i$, then they span the Hodge space
$H^{{n+1\over 2}, {n-1\over2}}$! So the primitive form satisfy the condition
required by the Seiberg-Witten differential.

Given a primitive form $v_0$, one can form a sequence of elements by using
$\partial_t^{-1}$ action: $v_0^{(-i)}=\partial_t^{-i} v_0$, and the exponent of
which is given by $\alpha_0+i$. If the polynomial is just $A_1$ polynomial with
the form $f=x_0^2+\ldots+x_n^2$, then the exponent would be
$\alpha_0={n+1\over 2}-1$.
To ensure the condition listed in \ref{sk3}, we only need to make the following
identification:
\begin{equation*}
\alpha_0+i=1\to {n+1\over 2}-1+i=1 \to i={-n+3\over2}
\end{equation*}
and so the SW differential is given by $\partial_t^{{n-3\over2}} v_0$.

\medskip\noindent%
\textbf{Computations.}
Given a tame polynomial $f:U\to \mathbb{C}$.
It is shown in \cite{Li:2018rdd} that the primitive form takes the following form
\begin{align*}
& v_0= dx \wedge dy \wedge dz \wedge dw,~~~U=\mathbb{C}^4 \nonumber\\
&v_0={dx \over x} \wedge dy \wedge dz \wedge dw,~~U=\mathbb{C}^*\times \mathbb{C}^3
\end{align*}
If we fix the coupling constants, the primitive form is just a constant of above
form. To get a section of cohomological bundle, we consider Leray residue form
$\lambda_{SW}={v_0 \over df}$.

If $U$ is two dimensional, then $\lambda_{SW}=v_0^{(-1)}$ and we have
\begin{align*}
& \lambda_{SW}={v_0^{(-1)}\over df}= y dx,~~~U=\mathbb{C}^2 \nonumber\\
& \lambda_{SW}={v_0^{(-1)}\over df}=y{dx \over x},~~U=\mathbb{C}^*\times \mathbb{C}
\end{align*}

\subsection{Exact marginal deformations}

In the study of 4d $\mathcal{N}=2$ theories, one usually fixed the physical
parameters, i.e. the mass parameters, exact marginal deformations, relevant
couplings, etc.~and study the low energy physics on vacua parameterized by the
Coulomb branch moduli. There is a special Kähler structure on the Coulomb branch
moduli space and one can determine it by finding a SW geometry and a SW
differential. The SW differential can be found from the primitive form.

From our perspective, it is important to also consider the variations of all the
physical parameters including exact marginal deformations.
The primitive form does depend on these physical parameters, and so we get
generalized SW differential, which can be used to probe further the physical properties of the theory.

Here we focus on the space of exact marginal deformations.
This space has some interesting properties:
\begin{enumerate}[wide]
\item one can find weakly coupled gauge theory descriptions at the boundary of ${\cal M}$;
\item  One can define a metric on it;
\item There is a $tt^*$ equation on it. 
\end{enumerate}
The MHS studied in this paper is very useful in understanding the structure of conformal manifold.
Here we give an example of the non-trivial dependence of primitive form on the
exact marginal deformation.

\medskip\noindent%
\textit{Example.}
Consider following three class of examples
(see \cite{matsuo1998summary} for the study of these models):
\begin{align*}
& \tilde{E}_6:\quad w^2+yz^2-x(x-y)(x-\tau y)+ \sum_{i=0}^6 s_i P_i(x,y,z)  \nonumber\\
& \tilde{E}_7:\quad w^2+xy(x-y)(x-\tau y)+z^2  +\sum_{i=0}^7 s_i P_i(x,y,z)     \nonumber\\
&\tilde{E}_8:\quad w^2+x(x-y^2)(x-\tau y^2)+z^2+\sum_{i=0}^8 s_i P_i(x,y,z)
\end{align*}
The primitive form takes the form
$\Omega={1\over u(\tau)} dx \wedge dy \wedge dz \wedge d w$,
and $u(\tau)$ satisfies the differential equation
\begin{equation*}
\tau(\tau-1) {\partial^2 u\over \partial^2 \tau}-(2\tau-1) {\partial u\over \partial \tau}-{1\over 4}u=0
\end{equation*}
One can use the primitive form to compute two point function over the conformal
manifold. The above differential equation can be used to find the singularity on
conformal manifold: $\tau=0,1, \infty$, and so there should be three weakly
coupled gauge theory description, which is in agreement with the result in
\cite{Xie:2017vaf}.

\section{Conclusion}\label{sec:8}
The low energy physics at Coulomb branch of 4d $\mathcal{N}=2$ theories are very
rich and complicated as it involves strongly coupled dynamics of quantum field
theory. It is remarkable that one can completely solve the $SU(2)$ gauge theory by
finding a SW geometry and a SW differential \cite{Seiberg:1994rs,Seiberg:1994aj}. Since then, many SW
geometries were found, however, fewer studies were performed on these geometries
due to reasons listed in the introduction. In particular, the low energy physics
at the singular point of the Coulomb branch is rarely studied. In this paper, we
show that MHS can be used to determine the low energy physics at every vacuum of
$\mathcal{N}=2$ Coulomb branch theory.

A detailed summary of the use of MHS in determining low energy theory is as
follows.
\begin{enumerate}
\item For a generic vacuum where the low energy theory is abelian gauge theory,
  the cohomology of the associated fiber of SW geometry has a MHS: it carries
  \textbf{two} weights, where one weight describes electric-magnetic charge,
  while the other weight describes flavor charge.
\item For a special vacuum, one can associate three MHS: the MHS on the
  cohomology of singular fiber $X_0$, the vanishing cycle $\phi_t M$, and the
  nearby cycle $\psi_t M$. The MHS of $X_0$ gives the abelian gauge theory, and
  the MHS for the vanishing cycles gives the interacting theory. These two parts
  are coupled together by gauging the flavor symmetry of the interacting theory,
  and the coupling is reflected in the exact sequence between three MHS. This is
  the most important part of this paper.
\item It is also possible to assign a MHS from which one can obtain properties of
  the UV theory.
\item The MHS can be used to find the SW differential.
\end{enumerate}
We also explained the explicit computational devices for each of the physical
question listed above. Some more detailed computations would be given in
separate publications.

The method found in this paper can be generalized to study other supersymmetric field
theories: 2d Landau-Ginzburg models \cite{Xie:20212d}, 5d $(1,0)$ SCFT compactified on
circle \cite{Xie:20215d}, 6d $(1,0)$ theory compactified on torus \cite{Xie:20216d}. It is also
possible to use this method to study string compactifications, e.g., IIB string
theory on Calabi-Yau manifolds. The main new input is that it is now
possible to study the physics associated with the singular point on the moduli
space.

In this paper, we begin with a SW geometry and
use MHS to study the physical properties of the theory. It is also
possible to start with the general property of an $\mathcal{N}=2$ theory, and
see the constraint on the possible SW geometry. In general, the Coulomb branch
should have a stratification: the low energy theory is given by abelian gauge
theory at the stratum with largest dimension, and on the strata of lower
dimensions one has extra massless particles. At each point of a stratum, one
has a central charge function which locally has a flat connection. The crucial
physical constraint seems to be the gluing condition of these strata.
Mathematically, the Coulomb branch of $\mathcal{N}=2$ theories seems to give a
mathematical structure called mixed Hodge module, developed by Morihiko Saito \cite{saito1990mixed}.
We hope the study of mixed Hodge modules associated with 4d $\mathcal{N}=2$ theories would give a guideline of
classification of $\mathcal{N}=2$ theories. See also \cite{Argyres:2020nrr,Argyres:2020wmq} for the recent
attempt of classifying 4d $\mathcal{N}=2$ theories.

\section*{Acknowledgements}
We would like to thank Wenbin Yan and Jie Zhou for helpful discussions. D.~Xie
and D.~Zhang are supported by Yau mathematical sciences center at Tsinghua
University.

  %%%%%%%%%%%%%%%%%%%%%%%%%%%%%%%%%%%%%%%%%%%%%%%%%%%%%%%%%%%%%%%%%%%%%%%%%%%%%%%%%%%%%%%%%%%%%%%%%%%%%%%%%%%

\newpage
\appendix

\section{Resolution of singularity}
Consider a projective curve $F(x,y,z)=0$ inside $P^2$, here $(x,y,z)$ are homogeneous coordinates. The singular point of $F(x,y,z)$ is given by the solution of
the following equations
\begin{equation*}
F={\partial F\over \partial x}={\partial F\over \partial y}={\partial F\over \partial z}=0.
\end{equation*}

Let's review blow up of a point on the algebraic curve, and using blow-up repeatedly will resolve the singularity. First, we describe how to blow up the origin in $C^2$. We set
\begin{equation*}
X'= \{(x_0,x_1; z_0,z_1) \in C^2\times P^1 | x_0z_1=x_1z_0\}
\end{equation*}
Here $X^{'}$ is a hypersurface inside $C^2\times P^1$. There are two coordinate patches for $C^2\times P^1$ and therefore two coordinate patches for $X^{'}$:
\begin{align}
& U_0=\{(x;z)\in X^{'}| z_0 \neq 0 \} \nonumber \\
&U_1=\{(x;z)\in X^{'} |z_1 \neq 0 \}
\end{align}
with coordinates
\begin{align}
&v_0=x_0,~~u_0=z_1/z_0 \nonumber \\
&v_1=x_1,~~u_1=z_0/z_1
\end{align}
The two coordinate are bi-holomorphic to $C^2$ with coordinate change as
\begin{equation*}
u_1=u_0^{-1},~~v_1=u_0v_0
\end{equation*}
There is a projection of $\pi: X'\to C^2$ defined by
\begin{align}
& \pi:(u_0,v_0) \to (v_0, u_0v_0) \nonumber\\
& \pi:(u_1,v_1) \to (u_1v_1, v_1)
\end{align}
The important fact is that $\pi^{-1}(0)$ is given by a copy of $P^1$, while $\pi^{-1}(C^2/0)$ is bi-holomorphic to $C^2/0$. This blow-up process would reduce the singularity of an algebraic curve. In particular,
if two curves touch at a single point, then the blow-up process would separate two curves.

\textbf{Example}: Let's take $Y$ to be the affine curve defined by the equation $f(x_0, x_1)=x_0^p+x_1^q=0$, and we assume $p\leq q$.  Let's do a blow up at the origin of $C^2$, and the equation for the curve would be
\begin{align}
&Y_0^{'}= \pi^{-1}(Y) \cap U_0:~~v_0^p+(u_0v_0)^q=v_0^p(1+u_0^q v_0^{q-p})=0 \nonumber\\
& Y_1^{'}=\pi^{-1}(Y) \cap U_1:~~(u_1v_1)^p+v_1^q=v_1^p(u_1^p+v_1^{q-p})=0
\end{align}
\begin{enumerate}
\item If $p=q$, then $Y_0^{'}$ and $Y_1^{'}$ each consists of $p$ parallel lines. There are no singularities any more, and there are $p$ extra points added on $Y^{'}$.
\item If $q>p$, then $Y_0^{'}$ does not meet line $v
_0=0$, so it is smooth. We only need to consider $Y_1^{'}$:
\begin{enumerate}
\item  If $q-p<p$, then there is still a singularity, but the multiplicity is reduced.
\item If $q-p=p$, then this singularity can be removed by  blow-up.
\item If $q-p>p$, the singularity is also reduced.
\end{enumerate}
\end{enumerate}

\section{Computing cohomology group of affine curve}
Here we review how to compute first cohomology group $H^1(C)$ of an irreducible affine algebraic curve $C$. The basic idea is following:
\begin{enumerate}
\item Given an algebraic curve defined by the equation $f(x,y)=0$, we find its projective completion $\tilde{C}$ which is given by the equation $F(x,y,z)=0$, here $(x,y,z)$ are
the homogeneous coordinate for $P^2$. The affine curve is recovered by choosing the patch $z=1$, i.e. $f(x,y)=F(x,y,1)$. Several points are added at "infinity", namely the solution to equation $F(x,y,0)=0$, notice that
the solution $(x,y)$ and $\lambda (x,y), \lambda \neq 0$ are equivalent. These points are denoted as $n_{\infty}$, and the number of smooth points added at infinity are denoted as $m_{\infty}$.
\item The singular points of $\tilde{C}$ are computed by the solution of following equation
\begin{equation*}
{\partial F\over \partial x}={\partial F\over \partial y}={\partial F\over \partial z}=0.
\end{equation*}
The set of singular points are denoted as $\Sigma$.
\item Let's blow up the singular points of $\tilde{C}$, and the genus of the resolved curve is given as
\begin{equation*}
g={d(d-3)\over2}+1-{1\over2}\sum_i\nu_i(\nu_i-1)
\end{equation*}
here $d$ is the degree of $F$, and $\nu_i$ is the multiplicity at the $i$th singular point.
\item Repeating blow up process until we get a smooth curve $\tilde{C}_f$, and the genus of the resolved curve is denoted as $g_0$.
\item The resolution of singularity is given by the map $\pi: \tilde{C}_f\to \tilde{C}$, and
\begin{equation*}
m=\# \pi^{-1}(\Sigma)
\end{equation*}
\item Finally, we have the formula of the dimension of $H^{1}(C)$:
\begin{equation*}
\dim H^{1}(C)=2g_0+m+m_{\infty}-1
\end{equation*}
\end{enumerate}

\textbf{Example}: Consider following affine curve $C$
\begin{equation*}
f=x^5+x^3y^2+1=0.
\end{equation*}
We would like to compute the dimension of first cohomology group. First, we consider  its projective compactifcation $\tilde{C}$:
\begin{equation*}
F=x^5+x^3y^2+z^5=0.
\end{equation*}
here $(x,y,z)$ are the homogenous coordinate of $P^2$. In the patch $z=1$, we recover the original curve. We also add  several points to the original curve, which is
the intersection of the above equation with the line $H=(x,y,0)$ in $P^2$, and we have equation
\begin{equation*}
x^5+x^3y^2=0.
\end{equation*}
If $y=0$, $x$ has to be zero, so the solution is $(0,0,0)$ which is not allowed. So $y\neq 0$ and we can set $y=1$, and the equation becomes
\begin{equation*}
x^5+x^3=0
\end{equation*}
and there are three solutions $x=0, \pm i$. The point $[0,1,0]$ is a singular point, while the other two points are non-singular. We then have $m_{\infty}=2$.

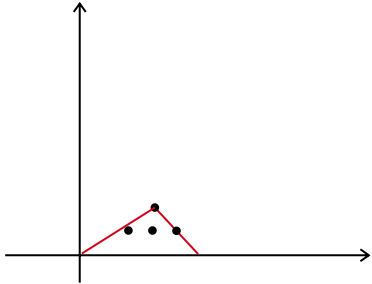
\begin{figure}[H]
\begin{center}

\tikzset{every picture/.style={line width=0.75pt}} %set default line width to 0.75pt

\begin{tikzpicture}[x=0.45pt,y=0.45pt,yscale=-1,xscale=1]
%uncomment if require: \path (0,479); %set diagram left start at 0, and has height of 479

%Shape: Axis 2D [id:dp2130006356073162]
\draw  (137.5,301) -- (440,301)(199.5,90) -- (199.5,324) (433,296) -- (440,301) -- (433,306) (194.5,97) -- (199.5,90) -- (204.5,97)  ;
%Straight Lines [id:da031831827687212044]
\draw [color={rgb, 255:red, 208; green, 2; blue, 27 }  ,draw opacity=1 ]   (262,261) -- (298,300) ;
%Shape: Circle [id:dp969715718548126]
\draw  [fill={rgb, 255:red, 0; green, 0; blue, 0 }  ,fill opacity=1 ] (259.25,261) .. controls (259.25,259.48) and (260.48,258.25) .. (262,258.25) .. controls (263.52,258.25) and (264.75,259.48) .. (264.75,261) .. controls (264.75,262.52) and (263.52,263.75) .. (262,263.75) .. controls (260.48,263.75) and (259.25,262.52) .. (259.25,261) -- cycle ;
%Shape: Circle [id:dp8021759822787611]
\draw  [fill={rgb, 255:red, 0; green, 0; blue, 0 }  ,fill opacity=1 ] (237.25,280.25) .. controls (237.25,278.73) and (238.48,277.5) .. (240,277.5) .. controls (241.52,277.5) and (242.75,278.73) .. (242.75,280.25) .. controls (242.75,281.77) and (241.52,283) .. (240,283) .. controls (238.48,283) and (237.25,281.77) .. (237.25,280.25) -- cycle ;
%Straight Lines [id:da20537324743380259]
\draw [color={rgb, 255:red, 208; green, 2; blue, 27 }  ,draw opacity=1 ]   (262,261) -- (201.25,299.6) ;
%Shape: Circle [id:dp5617160900209308]
\draw  [fill={rgb, 255:red, 0; green, 0; blue, 0 }  ,fill opacity=1 ] (257.25,280.25) .. controls (257.25,278.73) and (258.48,277.5) .. (260,277.5) .. controls (261.52,277.5) and (262.75,278.73) .. (262.75,280.25) .. controls (262.75,281.77) and (261.52,283) .. (260,283) .. controls (258.48,283) and (257.25,281.77) .. (257.25,280.25) -- cycle ;
%Shape: Circle [id:dp4528401155345678]
\draw  [fill={rgb, 255:red, 0; green, 0; blue, 0 }  ,fill opacity=1 ] (277.25,280.5) .. controls (277.25,278.98) and (278.48,277.75) .. (280,277.75) .. controls (281.52,277.75) and (282.75,278.98) .. (282.75,280.5) .. controls (282.75,282.02) and (281.52,283.25) .. (280,283.25) .. controls (278.48,283.25) and (277.25,282.02) .. (277.25,280.5) -- cycle ;

\end{tikzpicture}
\end{center}
\caption{The Newton polytope for the polynomial $f=x^5+x^3y^2+1$. There are two interior points, and two interior boundary points, so $H^1(X)=6$.}
\label{app1}
\end{figure}

The singularity of $\tilde{C}$ is given by the  solution of following equation:
\begin{equation*}
F={\partial F\over \partial x}={\partial F\over \partial y}={\partial F\over \partial z}=0
\end{equation*}
The solution of above equation is $[0,y,0]$, so there is only one singular point (point added in the infinity).  Near this point, we can set $y=1$, and get an affine equation
\begin{equation*}
x^5+x^3+z^5=0
\end{equation*}
and the singularity is at $x=z=0$, so $\nu=3$ (the lowest order  in the expansion around the singular point). After the blow up, we get a double point ($\nu=2$) and we need another blow up, so we find
\begin{equation*}
g={5-1)(5-2)\over 2}-{1\over 2}3\times (3-1)-{1\over 2}2\times (2-1)=2
\end{equation*}
and finally the number of points in $\pi^{-1}(\Sigma)$ is one, and we find
\begin{equation*}
\dim H^1(C)=6
\end{equation*}
This agrees with the formula using the combinatorial data of the corresponding Newton polytope, see figure. \ref{app1}. There are two internal lattice points, so $h^{1,0}=h^{0,1}=2$, and two lattice points on
the boundary of the Newton polytope inside first quadrant, so $dim H^{1}= 2\times 2+2=6$.

\bibliographystyle{JHEP}

\bibliography{ADhigher.bib}

\end{document}